\documentclass[a4paper,11pt]{article}
\pdfoutput=1 

\usepackage{jheppub} 

\usepackage[T1]{fontenc} 
\usepackage{amssymb}
\usepackage{xcolor}
\usepackage{caption}
\usepackage{subcaption}
\usepackage{float}
\usepackage[section]{placeins}
\usepackage{afterpage}
\usepackage{hyperref}
\usepackage{tikz-feynman}
\usepackage{lscape}
\usepackage{comment}
\usepackage{multirow}
\usepackage{makecell}
\usepackage{array,booktabs}

\title{\boldmath Comparison of spin-correlation and polarization variables of spin density matrix for top quark pairs at the LHC and New Physics implications}

\author[a,b,1]{Altan Cakir\note{Corresponding author.}}
\author[a]{and Orcun Kolay}

\affiliation[a]{Department of Physics Engineering, Istanbul Technical University, Maslak Istanbul, Turkey}
\affiliation[b]{Istanbul Technical University Artificial Intelligence, Data Science Research and Application Center, Maslak Istanbul, Turkey}

\emailAdd{altan.cakir@itu.edu.tr}
\emailAdd{orcun.kolay@itu.edu.tr}

\abstract{Precise determination of top-quark pairs is an essential tool for understanding the overall consistency of the standard model (SM) expectations, understanding limited New Physics (NP) models, through spin-spin correlation and polarization parameters, and has a critical impact on the analyses strategies at upcoming LHC programs. In this work, we review and discuss various state-of-the-art Monte Carlo (MC) methodologies as \textsc{MadGraph5}\_aMC@NLO, \textsc{Sherpa}, \textsc{Powheg-Box} and \textsc{Pythia8}, which are Matrix Element (ME)$/$Parton Shower (PS) matching generators including a complete set of spin correlation and polarization in top quark pair production with dileptonic final states. This is the first such study that not only compares the effects of different MC event generator approaches on spin density matrix elements and polarization parameters, but also investigates the effects of leading order (LO) and next-to-leading order (NLO) accuracy in QCD, and electroweak (EW) corrections via \textsc{Sherpa}. Moreover, as a continuation of the work, the prospects for possible NP scenarios through top-quark spin-spin correlation and polarization measurements for Supersymmetry (R parity conserved and violated models) and Dark Matter (top quarks associated mediator) models during upcoming LHC runs are briefly outlined.  We find that all SM MC predictions for the defined set of variables are generally consistent with the experimental data and theoretical predictions within the uncertainty variations. Besides, for the distributions of the $\cos\varphi$, laboratory-frame observables ($\cos\varphi_{lab}$ and $|\Delta\phi_{ll}|$) and the observables generated by parity (P) and charge-parity (CP) conserving interactions, we conclude some clues that the considered signals and beyond may well be separated from experimental data and located above the SM predictions.}


\begin{document} 
\maketitle
\flushbottom

\section{Introduction}
\label{sec:intro}

Top quark pair production is still one of the most critical process in collider physics today. As the heaviest known fundamental particle in the standard model (SM), the top-quark provides a unique opportunity to understand the critical behaviour of Yukawa coupling close to unity, making it a more important subject for beyond standard model (BSM) scenarios and in discussions of the stability of the SM universe. Top-quark properties is therefore an increasingly important input to many analyses in the following LHC runs. Understanding higher order effects together with electroweak corrections to top-quark processes play a crucial role in precision measurements, such as spin-correlation and polarization of top-quark pairs \cite{0a,0b,0c,0d,0e,0f,0g}, and many other unknown BSM signals, such as supersymmetry and effective field theory (EFT) models \cite{0h,0i}. More generally top quarks are unavoidable facts in BSM searches, both as signal objects and for backgrounds (e.g. as a source of leptons and b quarks).

Monte Carlo (MC) event generators are critical tools for the analysis and interpretation of LHC analyses in today research \cite{0j,0j1,0j2}. In particular, detailed aspects of the final states produced in individual processes, such as top-quark pair production, are to be required advanced calculations \cite{0k,0k1,0k2}. Top-quark pair production comprise for example the contribution of higher order matrix element calculations with multiple partons, the evaluation of parton shower and fragmentation, parton shower matching selection efficiencies, and the extrapolation of cross sections to the full phase space. In this manner, considering recent developments in higher-order perturbative corrections, in particular in QCD, but also in QED and in the electroweak sector, comparison studies of MC approaches are a critical subject that makes some precision measurements and data interpretation more visible. Based on a high level of calculation - studied with LHC data -  these contributions allow for both the realistic simulation of top-quark pairs precision measurements and the possible understanding of BSM signals. Thus, complex contributions of various  MC approaches to experimental objects and variables are a vital subject of collider-based particle physics studies.

In the following sections, we provide an overview of the up-to-date top-quark pairs hadron level production with various MC methodologies, above all, we will try to stress the main motivation of the extended discussion of a spin density matrix for the top pair system existing consistency and controversies concerning interpretation of the LHC measurements, as well as the sources of theory uncertainty. All in all, we review the precision measurement of spin density matrix of top quark-pairs; discuss the NP ambiguity; and the main strategies to measure spin-correleation and polarization top-quark pairs  will be presented. The interpretation of 22 different measurements and the theoretical uncertainties will be discussed while conclusion will contain some final remarks for future LHC runs.

\begin{figure}[t]
     \centering
     \begin{subfigure}[b]{0.3\textwidth}
         \centering
         \begin{tikzpicture}{diagram1}
         \begin{feynman}
         \vertex (i1) {\(\text{p}\)};
         \vertex [below=2.1cm of i1] (i2) {\(\text{p}\)};
         \vertex [right=1cm of i1] (m1);
         \vertex [right=2cm of i2] (m2);
         \vertex [blob, below=0.65cm of m1] (m3) {};
         \vertex [right=1.3cm of m1] (m4);
         \vertex [dot, below=0.5cm of m4] (m5) {};
         \vertex [dot, below=1cm of m5] (m6) {};
         \vertex [above right=0.1 and 2cm of m1] (o1) {\(\text{t}\)};
         \vertex [below=1cm of o1] (o2) {\(\tilde{\chi}_1^0\)};
         \vertex [below=0.8cm of o2] (o3) {\(\tilde{\chi}_1^0\)};
         \vertex [below=1cm of o3] (o4) {\(\overline{\text{t}}\)};
         \diagram* {
         (i1) -- (m3),
         (i1) -- (m3),
         (m5) -- [scalar, edge label'=\(\tilde{\text{t}}\)] (m3),
         (i2) -- (m3) -- [scalar, edge label'=\(\overline{\tilde{\text{t}}}\)] (m6),
         (m5) -- [fermion] (o1),
         (m5) -- [ghost] (o2),
         (m6) -- [ghost] (o3),
         (m6) -- [anti fermion] (o4),
};
         \end{feynman}
\end{tikzpicture}
\vspace*{2.1mm}
         \caption{}
         \label{fig:models_a}
     \end{subfigure}
     \hfill
     \begin{subfigure}[b]{0.3\textwidth}
         \centering
         \begin{tikzpicture}{diagram2}
         \begin{feynman}
         \vertex (i1) {\(\text{g}\)};
         \vertex [below=2cm of i1] (i2) {\(\text{g}\)};
         \vertex [right=2cm of i1] (m1);
         \vertex [right=2cm of i2] (m2);
         \vertex [below=1cm of m1] (m3);
         \vertex [right=1cm of m3] (m4);
         \vertex [right=2cm of m1] (o1) {\(\text{t}\)};
         \vertex [below=0.5cm of o1] (o2) {\(\chi\)};
         \vertex [below=1cm of o2] (o3) {\(\overline{\chi}\)};
         \vertex [right=2cm of m2] (o4) {\(\overline{\text{t}}\)};
         \diagram* {
         (i1) -- [gluon] (m1) -- [fermion] (o1),
         (i2) -- [gluon] (m2) -- [anti fermion] (o4),
         (m2) -- [fermion] (m3) -- [fermion] (m1),
         (m4) -- [scalar, edge label'=\(\phi/a\)] (m3),
         (m4) -- [fermion] (o2),
         (m4) -- [anti fermion] (o3),
};
         \end{feynman}
\end{tikzpicture}
\vspace*{2.1mm}
         \caption{}
         \label{fig:models_b}
     \end{subfigure}
     \hfill
     \begin{subfigure}[b]{0.3\textwidth}
         \centering
                  \begin{tikzpicture}{diagram3}
         \begin{feynman}
         \vertex (i1) {\(\text{p}\)};
         \vertex [below=2.1cm of i1] (i2) {\(\text{p}\)};
         \vertex [right=1cm of i1] (m1);
         \vertex [blob, below=0.65cm of m1] (m3) {};
         \vertex [right=1.5cm of m1] (m4);
         \vertex [dot, below=0.5cm of m4] (m5) {};
         \vertex [dot, below=1cm of m5] (m6) {};
         \vertex [dot, above right=0.5cm and 0.5cm of m5] (m7) {};
         \vertex [dot, below right=0.5cm and 0.5cm of m6] (m8) {};
         \vertex [above right=0.5cm and 1cm of m4] (o1) {\(\text{q}\)};
         \vertex [below=0.5cm of o1] (o2) {\(\text{q}\)};
         \vertex [below=0.5cm of o2] (o3) {\(\text{q}\)};
         \vertex [below=0.5cm of o3] (o4) {\(\text{t}\)};
         \vertex [below=0.5cm of o4] (o5) {\(\overline{\text{t}}\)};
         \vertex [below=0.5cm of o5] (o6) {\(\text{q}\)};
         \vertex [below=0.5cm of o6] (o7) {\(\text{q}\)};
         \vertex [below=0.5cm of o7] (o8) {\(\text{q}\)};
         \diagram* {
         (i1) -- (m3),
         (i1) -- (m3),
         (m5) -- [scalar, edge label'=\(\tilde{\text{t}}\)] (m3),
         (i2) -- (m3) -- [scalar, edge label'=\(\overline{\tilde{\text{t}}}\)] (m6),
         (m5) -- [fermion] (o4),
         (m5) -- [plain] (m7),
         (m7) -- [boson, edge label'=\(\tilde{\chi}_1^0\)] (m5),
         (m7) -- [plain] (o1),
         (m7) -- [plain] (o2),
         (m7) -- [plain] (o3),
         (m6) -- [anti fermion] (o5),
         (m6) -- [plain] (m8),
         (m6) -- [boson, edge label'=\(\tilde{\chi}_1^0\)] (m8),
         (m8) -- [plain] (o6),
         (m8) -- [plain] (o7),
         (m8) -- [plain] (o8),
};
         \end{feynman}
\end{tikzpicture}
         \caption{}
         \label{fig:models_c}
     \end{subfigure}
        \caption{Diagrams of the signal samples for stop quark pair production with decay into top quarks and stable neutralinos (a), direct DM production mediated by a scalar or pseudoscalar particle with associated top quark pair (b), and stop quark pair production with decay into top quarks and neutralinos decaying into light flavour quarks (c).}
        \label{fig:three_graphs}
\end{figure}
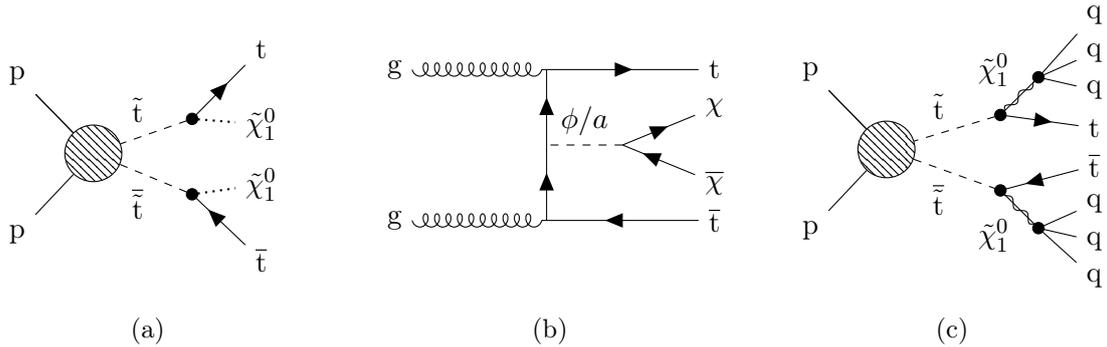

As it is known, even though the SM is a theory that has been proven many times by various tests, it has problems that it cannot solve, such as hierarchy problem and dark matter (DM). Supersymmetry (SUSY) as an extension of the SM could help to elucidate these both problems by introducing additional particles (e.g. bosonic superpartner of the top quark (stop, $\tilde{t}$)) that cancel large quantum corrections to the Higgs boson squared mass in the SM and feasible DM candidates. Due to the central role of the top quark in interaction with these hypothetical particles, collaborations like ATLAS and CMS try to measure top quark properties accurately to find these signals in the experiment. Searches for stop $\tilde{t}$ and DM candidate particles concentrate generally on high missing transverse energy $E_{T}^{miss}$ region stem from undetected lightest SUSY particles (LSP), as a decay product of $\tilde{t}$, and DM candidate particles. With these studies, most of the high $E_{T}^{miss}$ region, even low $E_{T}^{miss}$, was excluded by ATLAS and CMS Collaborations \cite{1a, 1b, 0e, 1d}. In the last step of the study, we will focus on the interpretation of SUSY and DM signals in low $E_{T}^{miss}$ region, which are studied with generic searching methods, by measuring precisely. The signals are based on simplified models, which are the similar topology with the models in the ref. \cite{1a}, of stop pair production and of top quark pair production in associated with fermionic DM candidate particles. In the SUSY signal, in figure \ref{fig:models_a}, a pair of stop $\tilde{t}$ decays an on-shell top quark $t$ and an LSP, in this case neutralino $\tilde{\chi}_{1}^{0}$, which is stable for this simplified model. In order to investigate the deviations of the signal from SM $t\bar{t}$, one mass point which is $175$ GeV for $\tilde{t}$ and $1$ GeV for $\tilde{\chi}_{1}^{0}$ will be used. An alternative dark matter signal in figure \ref{fig:models_b} include a scalar $\phi$ or pseudoscalar $a$ spin-0 mediator decayed into fermionic DM particles $\chi$ and an associated top quark pair. The coupling strengths of the DM-mediator and the mediator-SM fermions are assumed to be equal to one in the simplified model. Additionally, in this analysis, the scenario with $m_{\chi}=1$ GeV and $m_{\phi/a}=10$ GeV will be focused.

In the above models, we focused on a small mass point to ensure low $E_{T}^{miss}$. However, the additional model of R-Parity \cite{1e} violating (RPV) SUSY provides more mass points to study for low $E_{T}^{miss}$ due to the decaying $\tilde{\chi}_{1}^{0}$. The $\tilde{\chi}_{1}^{0}$ decays into three quarks with the help of a trilinear Yukawa coupling between quarks and superpartner of quarks (squark) \cite{1f}. In this study, the production chain includes a $\tilde{t}$ pair decaying into a $t$ and a $\tilde{\chi}_{1}^{0}$ as shown in figure \ref{fig:models_c}, and the $\tilde{\chi}_{1}^{0}$ goes into three light flavor quarks $q$, $\tilde{\chi}_{1}^{0}\rightarrow uds$, as considered in ref. \cite{1g}. For the RPV SUSY scenario, $\tilde{t}$ masses up to $670$ GeV have been excluded, by CMS Collaboration \cite{1g}, at $95\%$ confidence level with a maximum observed local significance of $2.8 \sigma$ for $m_{\tilde{t}}=400$ GeV.

This paper structured as follows: in the next section, we describe the MC data samples and the CMS data used in the analysis, with generator settings for all $t\bar{t}$ and signal samples. In section \ref{sec:3}, we introduce the set of observables of spin correlation and polarization in top quark pair production by using decomposition of spin density matrix. In section \ref{sec:4}, we present our results for the differential distributions of the observables. The results are divided into two parts as the comparison of different MC event generator configurations on the distributions of the observables of $t\bar{t}$ events and the investigation of behaviours of signals having different topology on the observables. Finally, our conclusions are given in section \ref{sec:5}.

\section{Event Samples and Data}
\label{sec:2}

A range of MC generators to provide predictions for top quark pair production and related BSM signal processes are used to investigate the effects of various approaches of MC generators on top pair spin correlation matrix and polarization variables and to assess the differences between BSM signals and top quark pairs.

A summary of the produced MC samples, describing various the SM top quark pair production processes and the possible signal scenarios, is shown in table \ref{tab:1}. The table also lists details of the simulation samples used, including the matrix element (ME) event generator with the order of the calculation and the central parton distribution function (PDF) set, the parton shower (PS) and hadronization model, and  the merging scheme used to remove the overlap between ME and PS.

\begin{table}[t]
\renewcommand{\arraystretch}{1.3}
\resizebox{\textwidth}{!}{
\centering
\begin{tabular}{p{2cm}p{2.8cm}p{3.3cm}p{2.5cm}p{2.5cm}p{2.1cm}}
 \hline\hline
 \textbf{Process} & \textbf{ME PDF} & \textbf{ME generator} & \textbf{Order} & \textbf{PS and} & \textbf{Merging}\\
 && \textbf{} & \textbf{(ME$/$PS)} & \textbf{hadronization} & \textbf{scheme}\\
 \hline
 $t\bar{t}+jets$&NNPDF3.0 NLO&MG5\_aMC@NLO&NLO$/$LL&\textsc{Pythia 8}&FxFx\\
 $t\bar{t}+jets$&NNPDF3.0 LO&MG5\_aMC@NLO&LO$/$LL&\textsc{Pythia 8}&MLM\\
 $t\bar{t}+jets$&NNPDF3.0 NLO&\textsc{Powheg-Box}&NLO$/$LL&\textsc{Pythia 8}&\textsc{Powheg}\\
 $t\bar{t}+jets$&NNPDF3.0 NLO&\textsc{Sherpa 2}&NLO($+$EW)$/$LL&\textsc{CSShower}&MEPS@NLO\\
 $t\bar{t}+jets$&NNPDF3.0 LO&\textsc{Sherpa 2}&LO$/$LL&\textsc{CSShower}&MEPS@LO\\
 \hline
 $\tilde{t}\tilde{t}\to t\tilde{\chi}_{1}^{0}\bar{t}\tilde{\chi}_{1}^{0}$&NNPDF3.1 LO&MG5\_aMC@NLO&LO$/$LL&\textsc{Pythia 8}&MLM\\
 $t\bar{t}+DM$&NNPDF3.1 LO& MG5\_aMC@NLO&LO$/$LL& \textsc{Pythia 8}& MLM\\
 \hline
\end{tabular}
}
\caption{\label{tab:1} Matrix Element (ME) event generator settings for $t\bar{t}$ and BSM signal samples used in this study. The generator versions, the order of QCD accuracy of associated predictions, the PDF sets used in ME calculations and the merging scheme are shown. }
\end{table}

In all background and signal samples $t\bar{t}$ events are produced with a top quark mass of $m_{t}=172.5$ GeV. The nominal $t\bar{t}$ sample (with up to two additional matched partons included in the ME calculations) is simulated using \textsc{MadGraph5}\_aMC@NLO (v2.6.7) \cite{b} at NLO QCD accuracy. To evaluate the effects of different perturbative orders of the calculation, we have been used the samples generated with \textsc{MadGraph5}\_aMC@NLO at LO accuracy as well. The parton showers are modelled using \textsc{Pythia} (v8.2.44) \cite{c} with the default showering tunes at leading logarithmic (LL) accuracy. The matrix elements at LO and NLO are merged with the parton shower using the MLM \cite{d} and the FxFx \cite{e} merging scheme, respectively. In order to determine state-of-the-art theoretical approaches between the different methods in MC generators, two additional generator setups are incorporated in the analysis chain. Firstly, $t\bar{t}$ sample is generated using the \textsc{Sherpa} (v2.2.8) \cite{f} generator with up to three extra partons at LO and NLO. The \textsc{Sherpa}, in production of $t\bar{t}$ events, employs \textsc{Amegic++} \cite{g1}, \textsc{Comix} \cite{g} and \textsc{OpenLoops} \cite{h,i,j} (for one-loop amplitudes) in matrix elements calculation which are merged with \textsc{CSShower} \cite{k} for parton showering (with LL accuracy) and hadronization by using the MEPS@NLO \cite{l,m} prescription. To match the ME and PS results, \textsc{Sherpa} and \textsc{MadGraph5}\_aMC@NLO make use of the same normalization method, which is MC@NLO \cite{n}. The other alternative $t\bar{t}$ sample is generated with \textsc{Powheg} (v2) \cite{o,p,q,r} including up to two extra matched partons at the ME level with NLO accuracy. The parton shower is employed by \textsc{Pythia} (v8.2.40) with $h_{damp}=272.72$ GeV parameter of \textsc{Powheg}.

For all SM samples listed, the initial state partons are modeled with the NNPDF3.0 PDF set \cite{s,t} implemented by LHAPDF \cite{u}. To determine the systematic uncertainties in predictions, uncertainties coming from PDF and scale variations are used as summed in quadrature. Parton distribution function uncertainties are evaluated by reweighting the $t\bar{t}$ samples generated by \textsc{MadGraph5}\_aMC@NLO at LO and \textsc{Sherpa} for all setups using generator weights related with each of the variations given by NNPDF3.0, CT10 \cite{v} and MMHT14 \cite{w} PDF sets. Renormalization and factorization scale uncertainties are calculated using variations by a factor of two around the central scales which are equal to $\frac{1}{2}\sum_{i=1}^{2}\sqrt{m_{t}^{2}+P_{T,i}^{2}}$ for \textsc{MadGraph5}\_aMC@NLO and $\sqrt{m_{t}^{2}+\frac{1}{2}\sum_{i=1}^{2}P_{T,i}^{2}}$ for \textsc{Sherpa} setup \cite{x}, where $P_{T,i}$ is the transverse momentum of top quarks.

Signal samples of SUSY and RPV SUSY are simulated at LO with up to three and two additional partons, respectively, with \textsc{MadGraph5}\_aMC@NLO (v2.4.2) interfaced to \textsc{Pythia} (v8.2.40) for parton showering and hadronization using the MLM merging scheme. For the SUSY sample in figure~\ref{fig:models_a}, the masses of the $\tilde{t}$ and the $\tilde{\chi}_{1}^{0}$ will be considered as $m_{\tilde{t}}=175$ GeV and $m_{\tilde{\chi}_{1}^{0}}=1$ GeV. On the RPV SUSY side, the masses of the $\tilde{t}$ and the $\tilde{\chi}_{1}^{0}$ are taken as $400$ GeV and $100$ GeV, respectively.

The other signal samples are top quark pairs associated with DM particles $\chi$ as shown in figure~\ref{fig:models_b} which is included scalar $\phi$ or pseudoscalar $a$ mediators. The DM signal samples are simulated at LO including up to one additional parton with \textsc{MadGraph5}\_aMC@NLO (v2.6.1) interfaced with the \textsc{Pythia} (v8.2.40) PS model using the MLM merging scheme. The masses of the $\chi$ and the $\phi/a$ will be considered as $m_{\chi}=1$ GeV and $m_{\phi/a}=10$ GeV. In all signal samples, NNPDF3.1 PDF set is used in ME calculation.

To assess the truth of the theoretical predictions, experimental measurements published by CMS Collaboration are directly used \cite{0i}. The CMS analysis unfolded the data to compare with parton level predictions at full phase space. In our analysis, the MC predictions have baseline selections compatible with the CMS analysis.

\section{Spin Correlations on Top Quark Pairs}
\label{sec:3}

The total cross section for the top pair production at accelerator energy $s$ can be written as
\begin{equation}
\label{eq:1}
\sigma(s,m_{t})= \sum_{i,j}\int_{0}^{1}\int_{0}^{1} \,dx_{1}\,dx_{2} \phi_{i,A}(x_{1},\mu_{F})\phi_{j,B}(x_{2},\mu_{F}) \hat{\sigma}_{ij\to t\bar{t}}(\frac{m_{t}^{2}}{\hat{s}},\mu_{R},\mu_{F},\alpha_{s}(\mu_{R}))
\end{equation}
by using factorization theorem. This theorem divides the total cross section into a short distance (hard) part $\hat{\sigma}_{t\bar{t}}$ and a long distance (soft) part, which is known as parton distribution function $\phi_{i,A/j,B}$ (PDF). The PDF determine the probability of the existence of a parton $i$ ($j$) in the hadron A (B) with a rate $x_{1}$ ($x_{2}$) of the longitudinal momentum of the hadron at the factorization scale $\mu_{F}$. The partonic cross section $\hat{\sigma}_{ij\to t\bar{t}}$ involves the spin information of top quark and anti-quark. However the spin information of a top quark is not detectable because of short life time of it. In spite of the fact that the spin of an unstable particle produced in high energy reactions is no straightforwardly noticeable quantity, it is extremely helpful to present the idea of a spin density matrix \cite{z} for the top pair system \eqref{eq:2}. The decay width of the top quark much smaller than its mass allows to use narrow width approximation (NWA) \cite{aa,ab}. Using NWA, the matrix element of the top pair production can be written as the trace of the production and the decay factors:

\begin{equation}
\label{eq:2}
|M|^{2} \propto Tr[\rho R \bar{\rho}]
\end{equation}
where $R$ is the spin density matrix of the on-shell top pair production and $\rho$ and $\bar{\rho}$ are the decay spin density matrices of the polarized top quark and antiquark, respectively. To look at the spin properties of the top quark and antiquark, spin density matrix can be decomposed in the spin space of $t$ and $\bar{t}$.

\begin{equation}
\label{eq:3}
R \propto \tilde{A}\mathbb{I}\otimes\mathbb{I}+\tilde{B}_{i}^{+}\sigma^{i}\otimes\mathbb{I}+\tilde{B}_{i}^{-}\mathbb{I}\otimes\sigma^{i}+\tilde{C}_{ij}\sigma^{i}\otimes\sigma^{j}
\end{equation}
Here, the first terms which are a 2x2 unit matrix \(\mathbb{I}\) and the Pauli matrices $\sigma^{i}$ in the tensor products refer to top spin spaces, and second terms refer to anti-top.

While the function $\tilde{A}$ is related to the total $t\bar{t}$ cross section and the top quarks kinematic properties, $\tilde{B}_{i}^{\pm}$ and $\tilde{C}_{ij}$ are 3-vectors that describe the polarizations of $t$ and $\bar{t}$ in each direction, and 3x3 matrix that determine the spin correlation between top quark and antiquark.

As mentioned earlier, decay of the top quarks before hadronization allows to study the spins of them by looking at the angular distribution of the decay products. Top quarks decay almost to a $W$ boson and a bottom quark. The final states of the decay coming from $W$ boson can be a quark-antiquark pair or a lepton and a neutrino. The angular distribution of the top decay products with respect to the spin alignment of the top is given by

\begin{equation}
\label{eq:4}
\frac{1}{\Gamma}\frac{d\Gamma}{d\cos\chi_{i}}=\frac{1}{2}(1+\alpha_{i}\cos\chi_{i})
\end{equation}
with $\chi_{i}$ the angle between the 3-momentum of decay product $i$ in the top rest frame and the direction of the top spin. $\alpha_{i}$ is the spin analyzing power of $i$. The spin analyzing power of the charged lepton is almost equal to $1$ which is maximum value. For a top antiquark, $\alpha_{i}$ is minus sign of the top quark one. It does not change much at NLO QCD \cite{ac}. To make use of high valued spin analyzing power, leptons have been used as a proxy for spins of top quarks and anti-quarks.

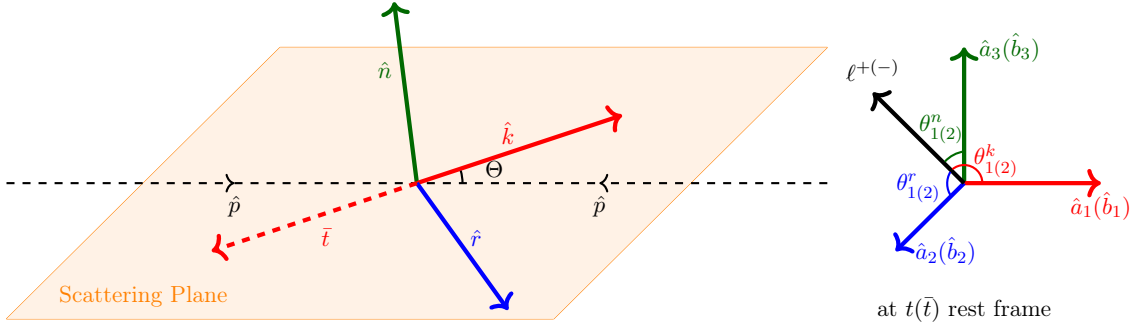
\begin{figure}[t]
\begin{tikzpicture}[scale=0.6, every node/.style={scale=0.8}]
\draw[orange] (3,-3) -- (9,3);
\draw[orange] (9,3) -- (-3,3);
\draw[orange] (-3,3) -- (-9,-3);
\draw[orange] (-9,-3) -- (3,-3);
\fill[orange!10] (3,-3) to (9,3)  to (-3,3)  to (-9,-3) to cycle;
\draw[thick, dashed] (-4,0) -- (4,0);
\draw[->, thick, dashed] (-9,0) -- (-4,0);
\draw[->, thick, dashed] (9,0) -- (4,0);
\definecolor{green1}{RGB}{0, 105, 0}
\draw[->, ultra thick, green1] (0,0) -- (-0.5,4);
\draw[->, ultra thick, red] (0,0) -- (4.5,1.5);
\draw[->, ultra thick, blue] (0,0) -- (2,-2.8);
\draw[->, ultra thick, red, dashed] (0,0) -- (-4.5,-1.5);
\node[orange] at (-6,-2.5) {Scattering Plane};
\node at (4,-0.5) {$\hat{p}$};
\node at (-4,-0.5) {$\hat{p}$};
\node[red] at (2,1.1) {$\hat{k}$};
\node[red] at (-2,-1.2) {$\bar{t}$};
\node[blue] at (1.3,-1.2) {$\hat{r}$};
\node[green1] at (-0.7,2.5) {$\hat{n}$};
\draw[thick] (1,0) arc (0:16:1);
\node at (1.7,0.31) {$\Theta$};
\draw[->, ultra thick, green1] (12,0) -- (12,3);
\draw[->, ultra thick, red] (12,0) -- (15,0);
\draw[->, ultra thick, blue] (12,0) -- (10.5,-1.5);
\node[green1] at (13,3) {$\hat{a}_{3}(\hat{b}_{3})$};
\node[red] at (15,-0.5) {$\hat{a}_{1}(\hat{b}_{1})$};
\node[blue] at (11.6,-1.5) {$\hat{a}_{2}(\hat{b}_{2})$};
\draw[->, ultra thick] (12,0) -- (10,2);
\node at (10,2.5) {$\ell^{+(-)}$};
\draw[green1, thick] (12,0.7) arc (98:138:0.7);
\draw[red, thick] (12.4,0) arc (0:130:0.4);
\draw[blue, thick] (11.7,0.25) arc (149:215:0.5);
\node[green1] at (11.5,1.2) {$\theta_{1(2)}^{n}$};
\node[red] at (12.7,0.5) {$\theta_{1(2)}^{k}$};
\node[blue] at (11,-0.1) {$\theta_{1(2)}^{r}$};
\node at (12,-2.8) {at $t(\bar{t})$ rest frame};
\end{tikzpicture}
\caption{\label{fig:Helicity_Basis}Coordinate system illustrating $t$ and $\bar{t}$ helicity spin basis. The $\hat{k}$ axis is described as the direction of flight of top quark measured in the $t\bar{t}$ CM. The directions $\hat{k}$, $\hat{r}$, $\hat{p}$ and $\bar{t}$ are in the scattering plane which the normal is the direction $\hat{n}$. The signs of the $\hat{r}$ and $\hat{n}$ axes can have inverse according to the angle $\Theta$ as shown in eq. \eqref{eq:6}. The angles between the direction of flight of the positively $\ell^{+}$ (negatively $\ell^{-}$) charged lepton in the top quark (anti-quark) rest frame and the axes $\hat{a}$ ($\hat{b}$) determined in eq. \eqref{eq:6} are used to calculate the coefficients in eqs. \eqref{eq:7}-\eqref{eq:10}.}
\end{figure}

The functions $\tilde{B}_{i}^{\pm}$ and $\tilde{C}_{ij}$ in equation \eqref{eq:3} can be written as a linear combination of coefficient functions associated with an orthonormal basis. The basis in this analysis was chosen as the helicity basis as it gives more spin correlation compared to other alternatives as discussed in ref. \cite{0c}. This basis can be shown in figure \ref{fig:Helicity_Basis}. $\hat{k}$ is described as helicity axis determined by the top quark direction in the $t\bar{t}$ center-of-mass (CM) frame. The direction $\hat{p}$ indicates the direction of one of the incoming partons in the same frame. Using the directions $\hat{k}$ and $\hat{p}$, the normal $\hat{n}$ to the scattering plane and the direction $\hat{r}$ perpendicular to $\hat{k}$ in the scattering plane are given by $\hat{n}=(\hat{p}\times\hat{k})/\sin(\Theta)$ and $\hat{r}=(\hat{p}-\hat{k}\cos(\Theta))/\sin(\Theta)$.

A redefinition of the $\hat{r}$ and $\hat{n}$ axes is necessitated owing to the Bose-Einstein symmetry of the initial $gg$ state \cite{0h}. The reference axes $\hat{a}$ and $\hat{b}$ are described with above regularization as
\begin{equation}
\label{eq:6}
\hat{a}\to \{\hat{k},sign(\cos\Theta)\hat{r},sign(\cos\Theta)\hat{n}\}, \,\,
 \hat{b}\to \{-\hat{k},-sign(\cos\Theta)\hat{r},-sign(\cos\Theta)\hat{n}\}
\end{equation}

Using the orthonormal basis, $\tilde{B}_{i}^{\pm}$ and $\tilde{C}_{ij}$ are decomposed as

\begin{equation}
\label{eq:coefficients_B}
\tilde{B}_{i}^{\pm}=b_{k}^{\pm}\hat{k}_{i}+b_{r}^{\pm}\hat{r}_{i}+b_{n}^{\pm}\hat{n}_{i},
\end{equation}
\begin{equation}
\label{eq:coefficients_C}
\begin{split}
&\tilde{C}_{ij}=c_{kk}\hat{k}_{i}\hat{k}_{j}+c_{rr}\hat{r}_{i}\hat{r}_{j}+c_{nn}\hat{n}_{i}\hat{n}_{j}\\
&+c_{rk}(\hat{r}_{i}\hat{k}_{j}+\hat{k}_{i}\hat{r}_{j})+c_{nr}(\hat{n}_{i}\hat{r}_{j}+\hat{r}_{i}\hat{n}_{j})+c_{kn}(\hat{k}_{i}\hat{n}_{j}+\hat{n}_{i}\hat{k}_{j})\\
&+c_{n}(\hat{r}_{i}\hat{k}_{j}-\hat{k}_{i}\hat{r}_{j})+c_{k}(\hat{n}_{i}\hat{r}_{j}-\hat{r}_{i}\hat{n}_{j})+c_{r}(\hat{k}_{i}\hat{n}_{j}-\hat{n}_{i}\hat{k}_{j})
\end{split}
\end{equation}

The coefficient functions $b_{i}^{\pm}$, $c_{ij}$ and $c_{i}$ which are functions of the partonic center of mass-energy $\hat{s}$ and $\cos\Theta$ are related to some discrete symmetries such as parity (P) and charge-parity (CP), and they are summarized in table \ref{tab:CP}.

\begin{table}[t]
\footnotesize
\renewcommand{\arraystretch}{1.3}
\resizebox{\textwidth}{!}{
\centering
\begin{tabular}{p{0.8cm}|p{4cm}p{2cm}p{3cm}p{2.5cm}}
 \hline\hline
 &\textbf{Observable} & \textbf{Measured} & \textbf{Coefficient} & \textbf{Symmetries}\\
 &&\textbf{coefficient}&\textbf{function}&\\
 \hline
				\multirow{10}{*}{\rotatebox{90}{\textbf{polarization}}}&$\cos\theta_{1}^{k}$ & $B_{1}^{k}$ & $b_{k}^{+}$ & P-odd, CP-even \\
				&$\cos\theta_{2}^{k}$ & $B_{2}^{k}$ & $b_{k}^{-}$ & P-odd, CP-even \\
				&$\cos\theta_{1}^{r}$ & $B_{1}^{r}$ & $b_{r}^{+}$ & P-odd, CP-even \\
				&$\cos\theta_{2}^{r}$ & $B_{2}^{r}$ & $b_{r}^{-}$ & P-odd, CP-even \\
				&$\cos\theta_{1}^{n}$ & $B_{1}^{n}$ & $b_{n}^{+}$ & P-even, CP-even \\
				&$\cos\theta_{2}^{n}$ & $B_{2}^{n}$ & $b_{n}^{-}$ & P-even, CP-even \\
				&$\cos\theta_{1}^{k*}$ & $B_{1}^{k*}$ &  $b_{k}^{+}$& P-odd, CP-even \\
				&$\cos\theta_{2}^{k*}$ & $B_{2}^{k*}$ & $b_{k}^{-}$ & P-odd, CP-even \\
				&$\cos\theta_{1}^{r*}$ & $B_{1}^{r*}$ & $b_{r}^{+}$ & P-odd, CP-even \\
				&$\cos\theta_{2}^{r*}$ & $B_{2}^{r*}$ & $b_{r}^{-}$ & P-odd, CP-even \\
				\hline
				\multirow{3}{*}{\rotatebox[origin=c]{90}{\thead{\textbf{diagonal} \\ \textbf{spin corr.}}}}&$\cos\theta_{1}^{k}\cos\theta_{2}^{k}$ & $C_{kk}$ & $c_{kk}$ & P-even, CP-even \\
				&$\cos\theta_{1}^{r}\cos\theta_{2}^{r}$ & $C_{rr}$ &  $c_{rr}$&P-even, CP-even  \\
				&$\cos\theta_{1}^{n}\cos\theta_{2}^{n}$ & $C_{nn}$ & $c_{nn}$ & P-even, CP-even \\
				\hline
				\multirow{6}{*}{\rotatebox[origin=c]{90}{\thead{\textbf{cross spin corr.}}}}&$\cos\theta_{1}^{r}\cos\theta_{2}^{k}+\cos\theta_{1}^{k}\cos\theta_{2}^{r}$ & $C_{rk}+C_{kr}$ & $c_{rk}$ & P-even, CP-even \\
				&$\cos\theta_{1}^{r}\cos\theta_{2}^{k}-\cos\theta_{1}^{k}\cos\theta_{2}^{r}$ & $C_{rk}-C_{kr}$ & $c_{n}$ & P-even, CP-odd \\
				&$\cos\theta_{1}^{n}\cos\theta_{2}^{r}+\cos\theta_{1}^{r}\cos\theta_{2}^{n}$ & $C_{nr}+C_{rn}$ & $c_{nr}$ & P-odd, CP-even \\
				&$\cos\theta_{1}^{n}\cos\theta_{2}^{r}-\cos\theta_{1}^{r}\cos\theta_{2}^{n}$ & $C_{nr}-C_{rn}$ & $c_{k}$ & P-odd, CP-odd \\
				&$\cos\theta_{1}^{n}\cos\theta_{2}^{k}+\cos\theta_{1}^{k}\cos\theta_{2}^{n}$ & $C_{nk}+C_{kn}$ & $c_{kn}$ & P-odd, CP-even \\
				&$\cos\theta_{1}^{n}\cos\theta_{2}^{k}-\cos\theta_{1}^{k}\cos\theta_{2}^{n}$ & $C_{nk}-C_{kn}$ & $-c_{r}$ & P-odd, CP-odd \\
				\hline
				&$\cos\varphi$ & $D$ & $-(c_{kk}+c_{rr}+c_{nn})/3$ & P-even, CP-even \\
				\hline
				\multirow{2}{*}{\rotatebox[origin=c]{90}{\thead{\textbf{lab.} \\ \textbf{frame}}}}&$\cos\varphi_{lab}$ & $A_{\cos\varphi}^{lab}$ & $-$ & $-$ \\
				&$|\Delta\phi_{ll}|$ & $A_{|\Delta\phi_{ll}|}$ & $-$ & $-$ \\
 \hline
\end{tabular}
}
\caption{\label{tab:CP}Observable quantities of the spin density matrix and their corresponding measured coefficients, coefficient functions and P and CP symmetry properties.}
\end{table}

According to table \ref{tab:CP}, the P- and CP-even coefficient functions ($c_{kk}$, $c_{rr}$, $c_{nn}$ and $c_{rk}$) can have large values due to the parity invariance of strong interaction. However, the presence of electroweak (EW) corrections can result in non-zero P-odd and CP-even coefficients ($c_{nr}$ and $c_{kn}$). On the other hand, the coefficients produced by CP-violating interactions are close to zero in the SM. The approximate CP invariance of the SM limits the top and anti-top quarks to have the same polarization, so $b_{i}^{+}-b_{i}^{-}$ has a value of zero\footnote{See ref. \cite{0h} for further details.}.

After the equations \eqref{eq:2} and \eqref{eq:3}, the normalized distribution for the two leptons final state can be written in the form of spin coefficients and direction of flight of the leptons. This can be simplified as the differential distribution of the cross-section. The cross section distribution for the double leptonic channel can be written as \cite{0i}

\begin{equation}
\label{eq:5}
\frac{1}{\sigma}\frac{d^{2}\sigma}{d\cos\theta_{1}^{i}d\cos\theta_{2}^{j}}=\frac{1}{4}(1+B_{1}^{i}\cos\theta_{1}^{i}+B_{2}^{j}\cos\theta_{2}^{j}-C_{ij}\cos\theta_{1}^{i}\cos\theta_{2}^{j})
\end{equation}
where $\theta_{1}^{i} (\theta_{2}^{j})$ is the angle between the direction of flight of the positively (negatively) charged lepton in the top quark (anti-quark) rest frame and a reference direction $\hat{a}$ ($\hat{b}$). The coefficients $B_{1,2}^{i,j}$ and $C_{ij}$ in \eqref{eq:5} reflect the properties of the spin density components in \eqref{eq:3}. The elements of $B_{1,2}$ and $C$ matrices have the same discrete symmetry properties with related coefficient functions, see in table \ref{tab:CP}. To extract each coefficient separately one can reduce the double differential cross section in equation \eqref{eq:5} to single differential distributions

\begin{equation}
\label{eq:7}
\frac{1}{\sigma}\frac{d\sigma}{d\cos\theta_{1}^{i}}=\frac{1}{2}(1+B_{1}^{i}\cos\theta_{1}^{i}),
\end{equation}

\begin{equation}
\label{eq:8}
\frac{1}{\sigma}\frac{d\sigma}{d\cos\theta_{2}^{i}}=\frac{1}{2}(1+B_{2}^{i}\cos\theta_{2}^{i}),
\end{equation}

\begin{equation}
\label{eq:9}
\frac{1}{\sigma}\frac{d\sigma}{d(\cos\theta_{1}^{i}\cos\theta_{2}^{i})}=\frac{1}{2}(1-C_{ii}\cos\theta_{1}^{i}\cos\theta_{2}^{i})\ln{\left(\frac{1}{|\cos\theta_{1}^{i}\cos\theta_{2}^{i}|}\right)},
\end{equation}

\begin{equation}
\label{eq:10}
\frac{1}{\sigma}\frac{d\sigma}{dx_{\pm}}=\frac{1}{2}\left(1-\frac{C_{ij}\pm C_{ji}}{2}x_{\pm}\right)\cos^{-1}|x_{\pm}|  \quad\quad(\mbox{for}\, i\neq j),
\end{equation}
where $x_{\pm}=\cos\theta_{1}^{i}\cos\theta_{2}^{j}\pm\cos\theta_{1}^{j}\cos\theta_{2}^{i}$. In addition to eqs. \eqref{eq:7}, \eqref{eq:8}, four further observables $\cos\theta_{1,2}^{i}$ with respect to modified axes $\hat{k}^{*}$ and $\hat{r}^{*}$ are calculated. Axes $\hat{k}^{*}$ and $\hat{r}^{*}$ for top quark are equal to $sign(\Delta|y|)\hat{k}$ and $sign(\Delta|y|)sign(\cos\Theta)\hat{r}$, respectively, while they are inverse for anti-top. Here $\Delta|y|=|y_{t}|-|y_{\bar{t}}|$ is the difference in absolute values of the rapidities of top and anti-top quarks in the laboratory frame. The observables in modified axes may sensitive to the NP contributions \cite{0h}.

The opening angle $\cos\varphi$, which is obtained by the inner product of the direction of flight of positively and negatively charged leptons in their own top quark rest frames, is taken into account as a useful spin correlation observable below. The analysis also involves two observables measured directly in the laboratory-frame. First one is $\cos\varphi_{lab}$, which is defined as $\cos\varphi$ but in the laboratory-frame. $|\Delta\phi_{ll}|$ is the other observable measured in the laboratory-frame. It is the absolute azimuthal difference between the two charged leptons measured in the laboratory-frame. The observables measured in the laboratory-frame cannot be obtained from spin correlation or polarization observables, but they could have high precision in measurement and computation.

\section{Results}
\label{sec:4}

In this section, the normalized differential distributions for the 22 observables mentioned in previous section are presented as a comparison of various MC approaches and experimental measurements published by the CMS Collaboration \cite{0i}. The physics objects in events predicted by MC generators are considered at generator-level. In that sense, objects come from MC history after decay but before hadronization. Last top quark in top quark decay chain as a top quark and first lepton after decay of W boson as a lepton are used in the calculations of observables. As leptons in dilepton events, prompt electrons and muons (not taus or decay products of taus) are taken into account.

Three different MC generators as \textsc{MadGraph5}\_aMC@NLO, \textsc{Sherpa} and \textsc{Powheg box} at LO and NLO QCD accuracy used in the analysis. As in the CMS analysis \cite{0i}, the analysis with MC generators is in full phase space. As the nominal $t\bar{t}$ sample, we have been chosen the sample simulated by \textsc{MadGraph5}\_aMC@NLO ME event generator at NLO interfaced with \textsc{Pythia8} for the parton shower and fragmentation.

\begin{figure}[t]
     \centering
         \includegraphics[width=.445\textwidth]{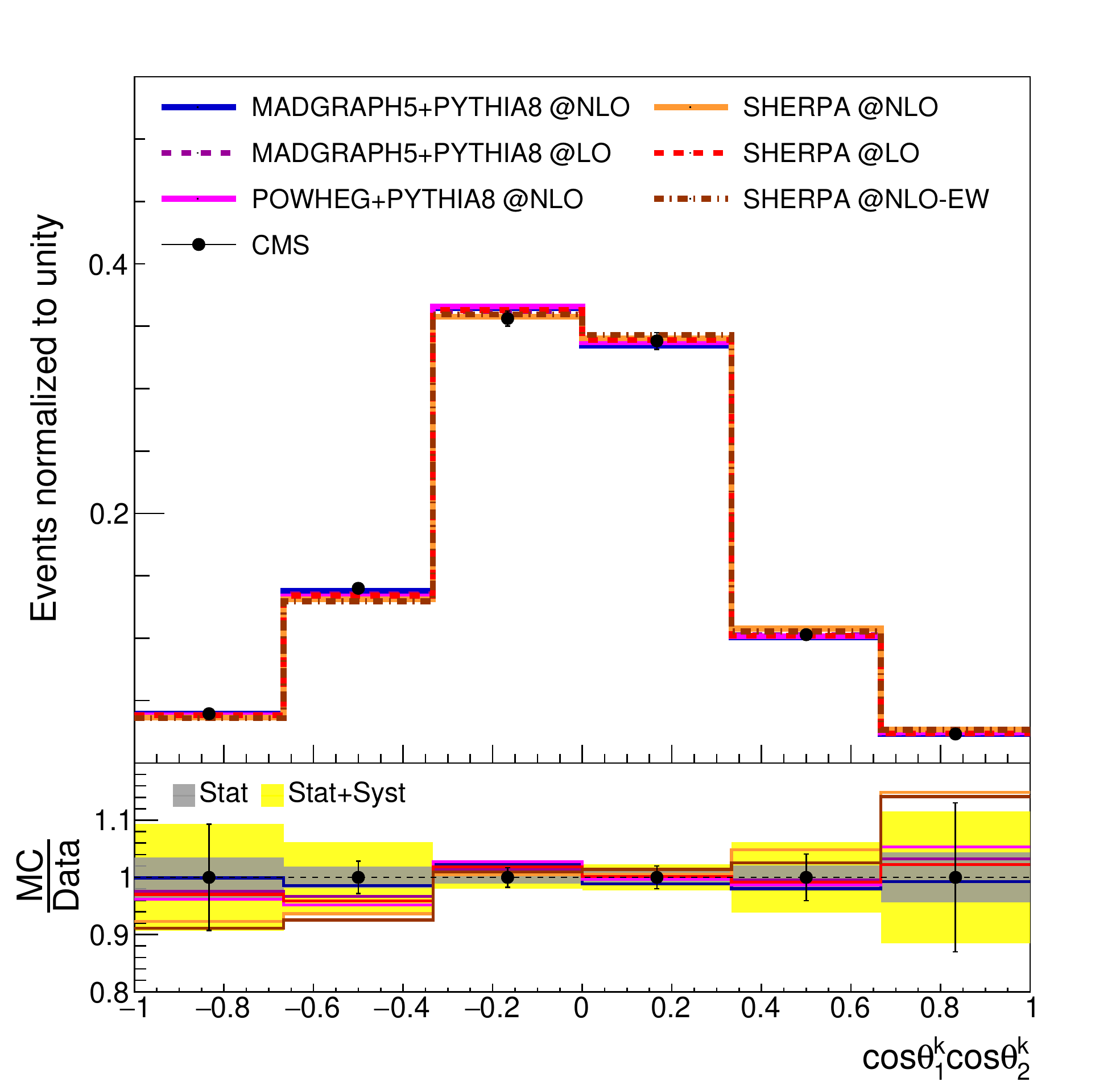}
     \hfill
         \includegraphics[width=.445\textwidth]{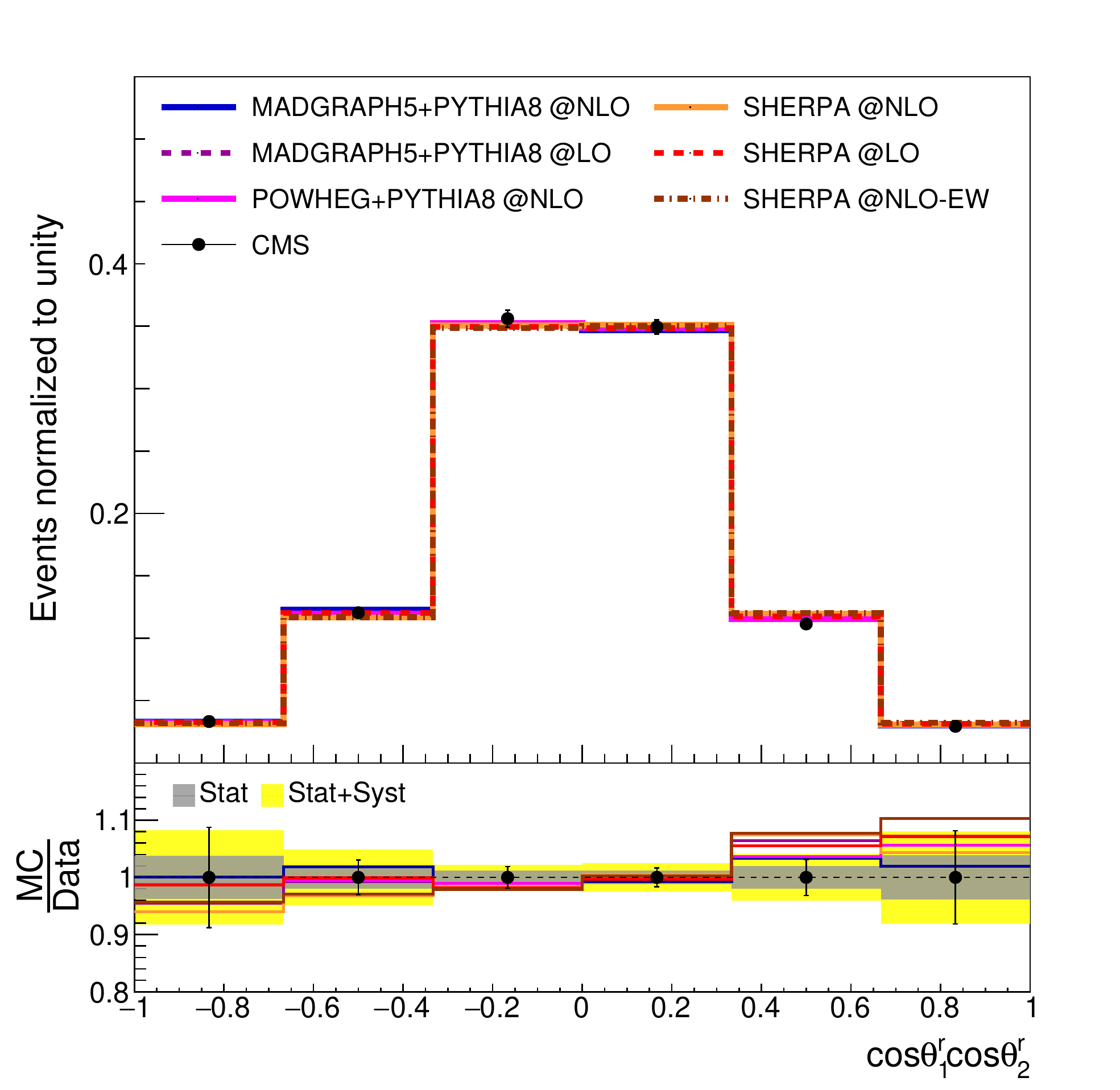}
     \hfill
         \includegraphics[width=.445\textwidth]{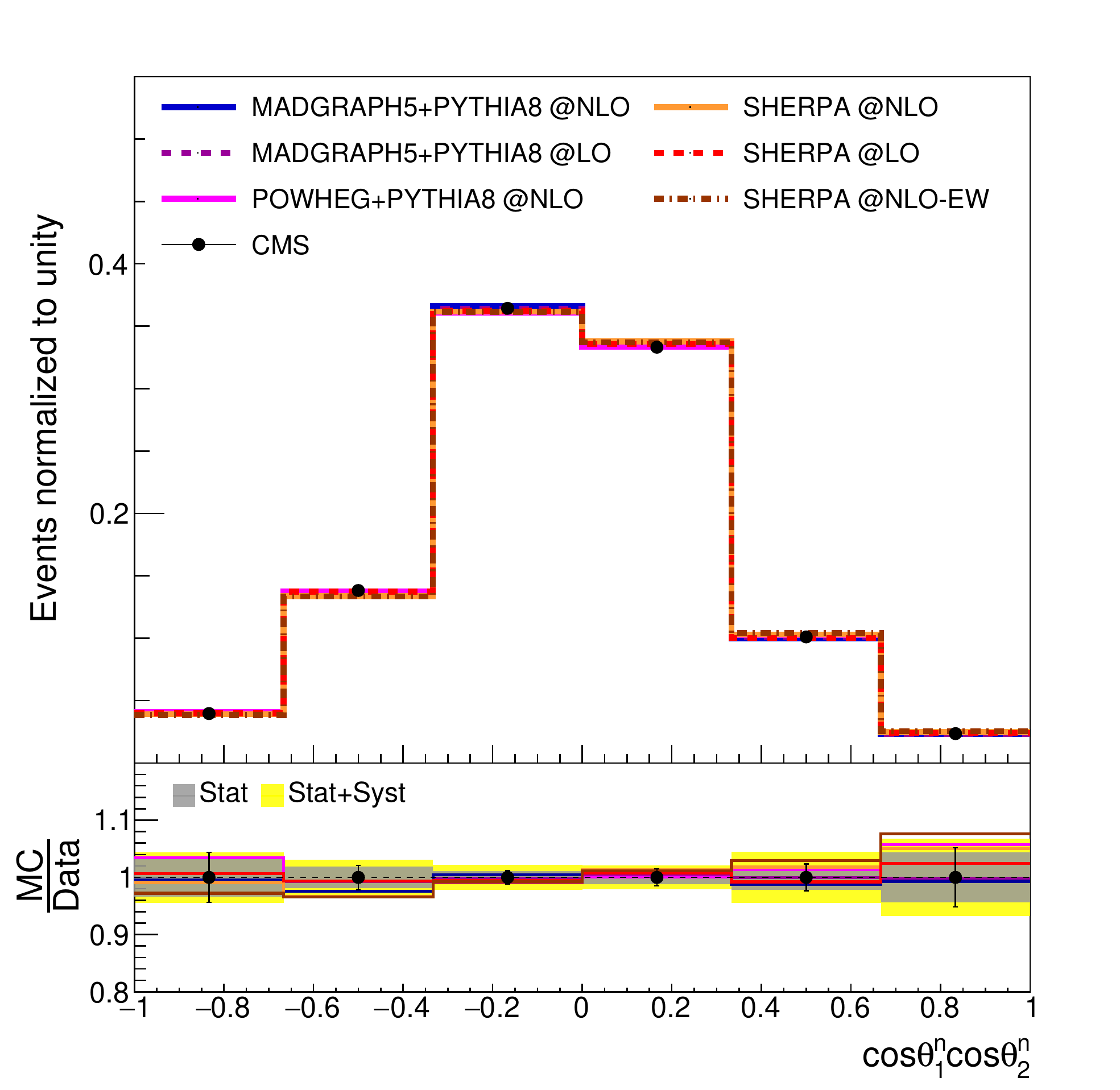}
        \caption{\label{fig:diagonal_spin}The normalized differential cross sections with respect to the diagonal spin correlation observables $\cos\theta_{1}^{i}\cos\theta_{2}^{i}$, $i=k,r,n$ compared to the CMS data and the predictions from \textsc{MadGraph5}\_aMC@NLO, \textsc{Sherpa} and \textsc{Powheg box} in full phase space. The ratio of the normalized differential cross sections predicted by the generator+parton shower configurations to the CMS data is shown in the lower panels. The statistical uncertainty band is shown in grey. The yellow uncertainty bands represent the overall uncertainty obtained by summing the systematic and statistical uncertainties in quadrature. The systematic uncertainty includes PDF and scale uncertainties.}
\end{figure}

\begin{figure}[t]
     \centering
         \includegraphics[width=.445\textwidth]{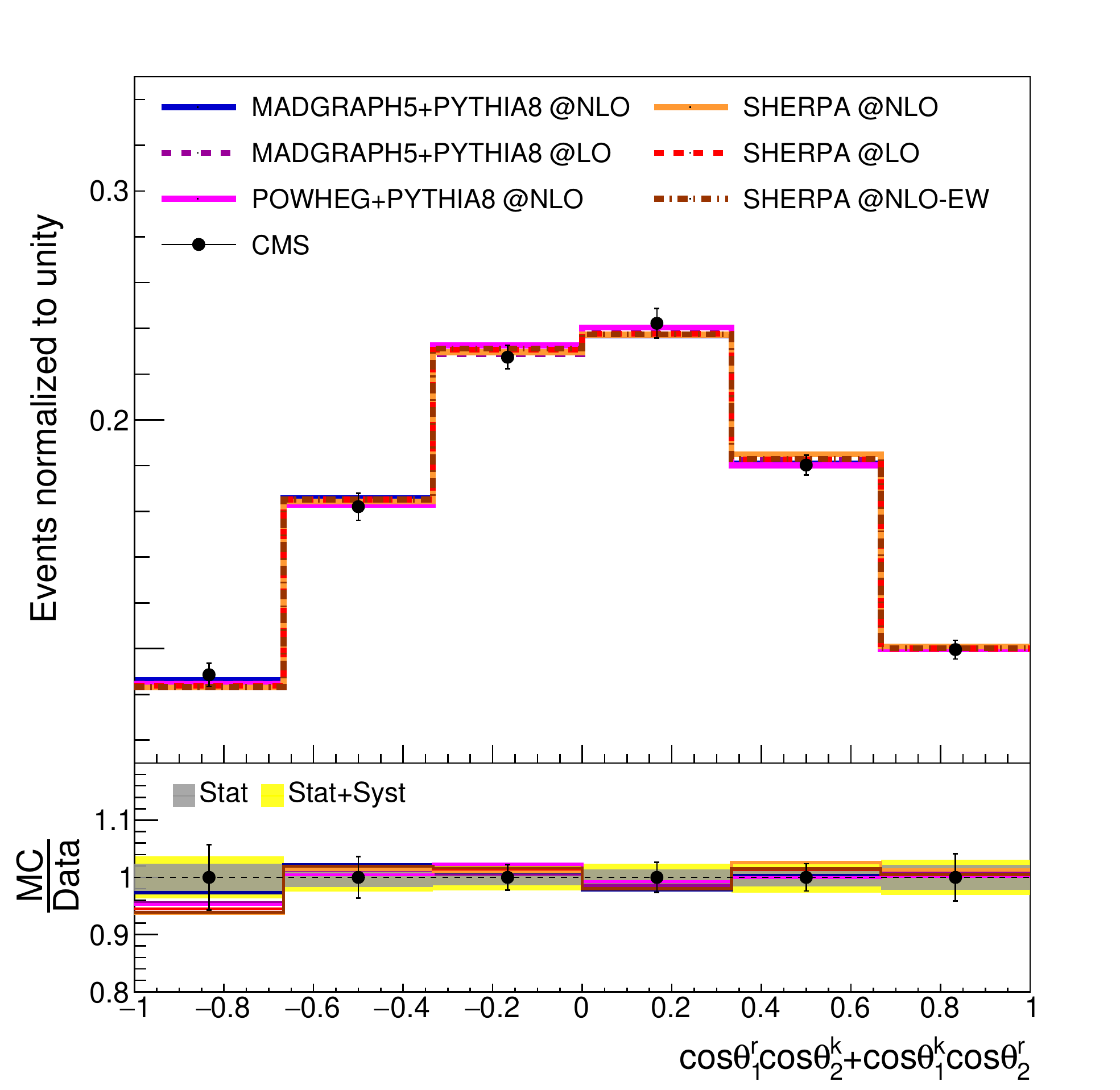}
     \hfill
         \includegraphics[width=.445\textwidth]{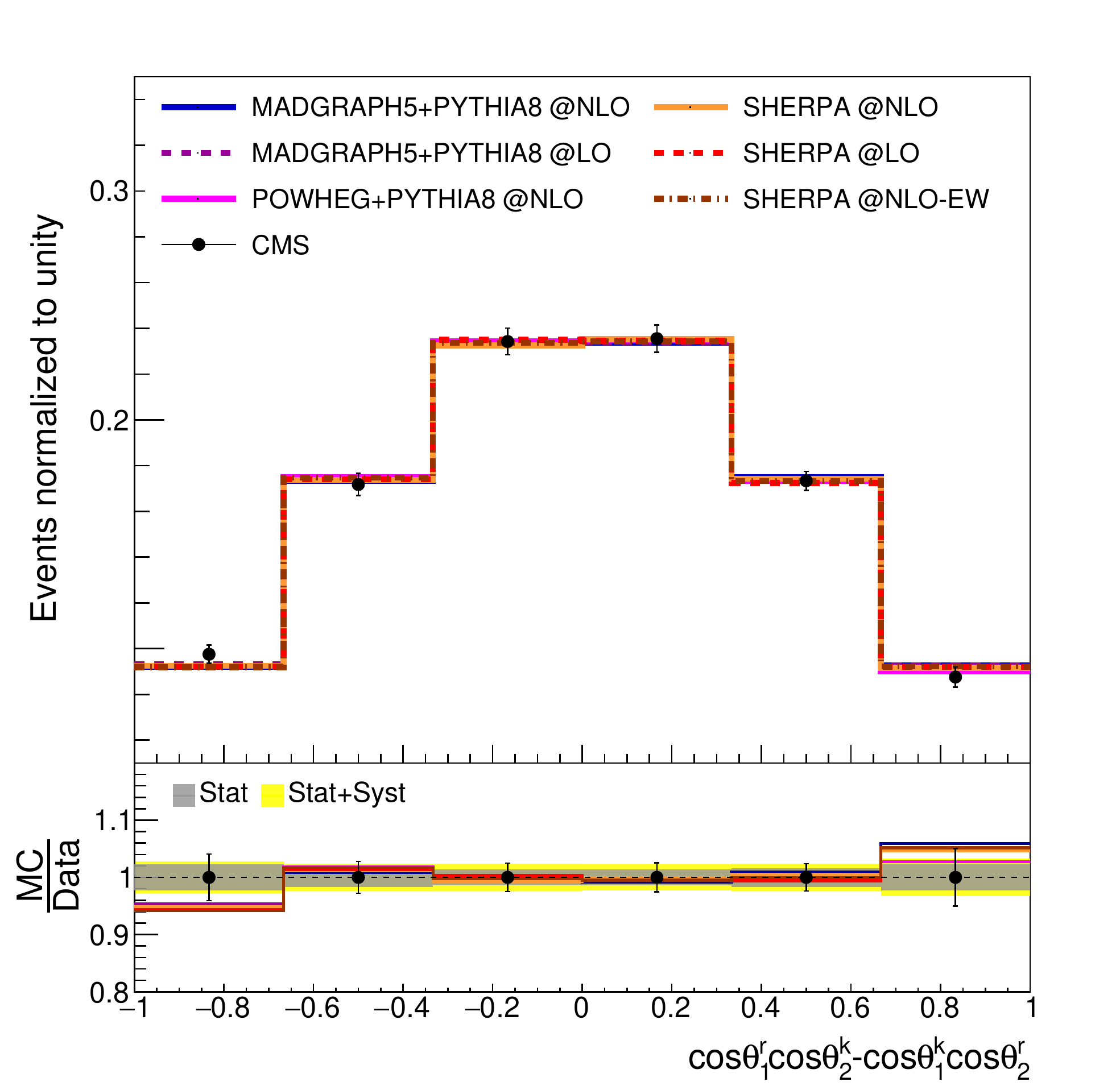}
     \hfill
         \includegraphics[width=.445\textwidth]{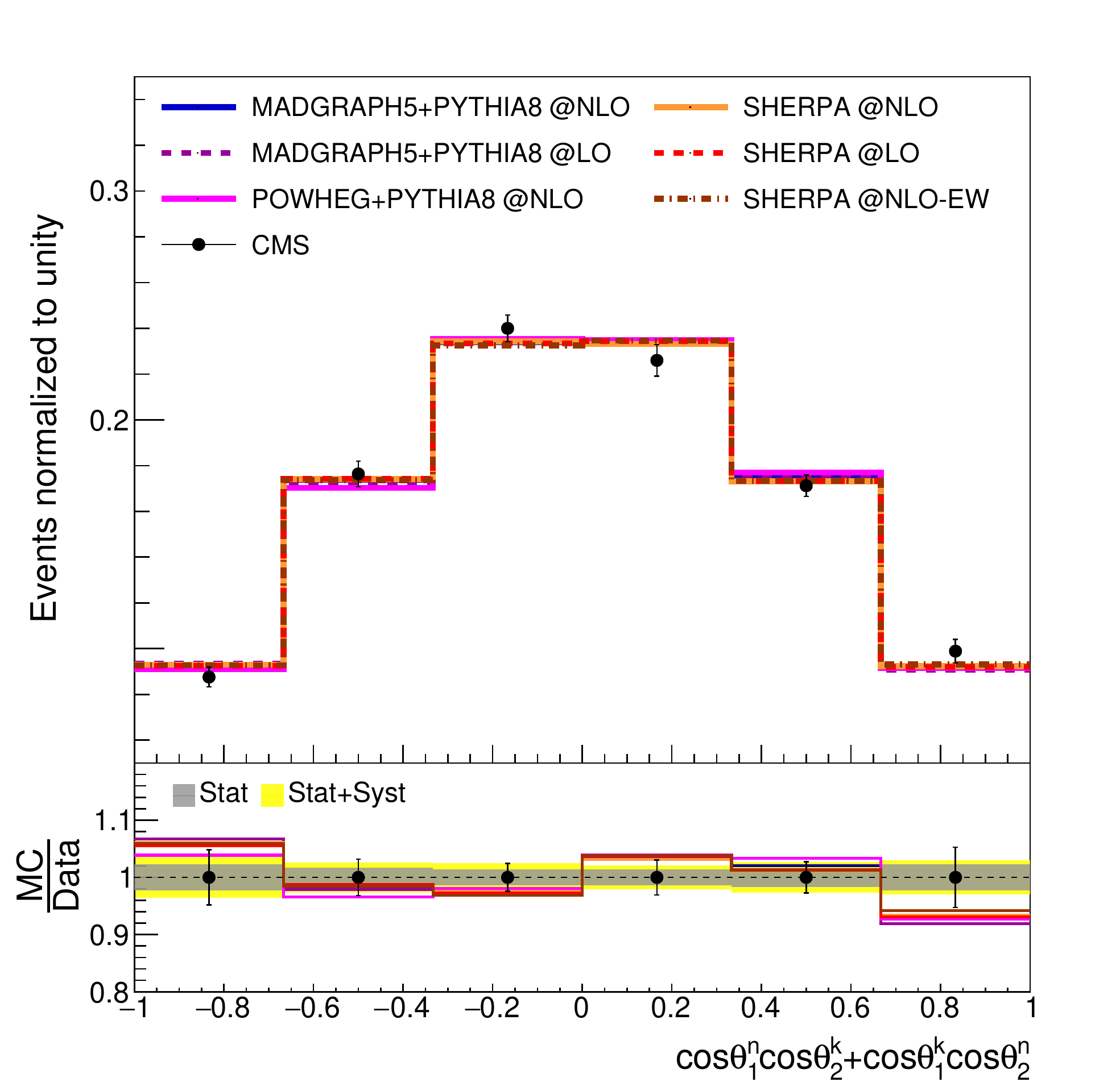}
      \hfill
         \includegraphics[width=.445\textwidth]{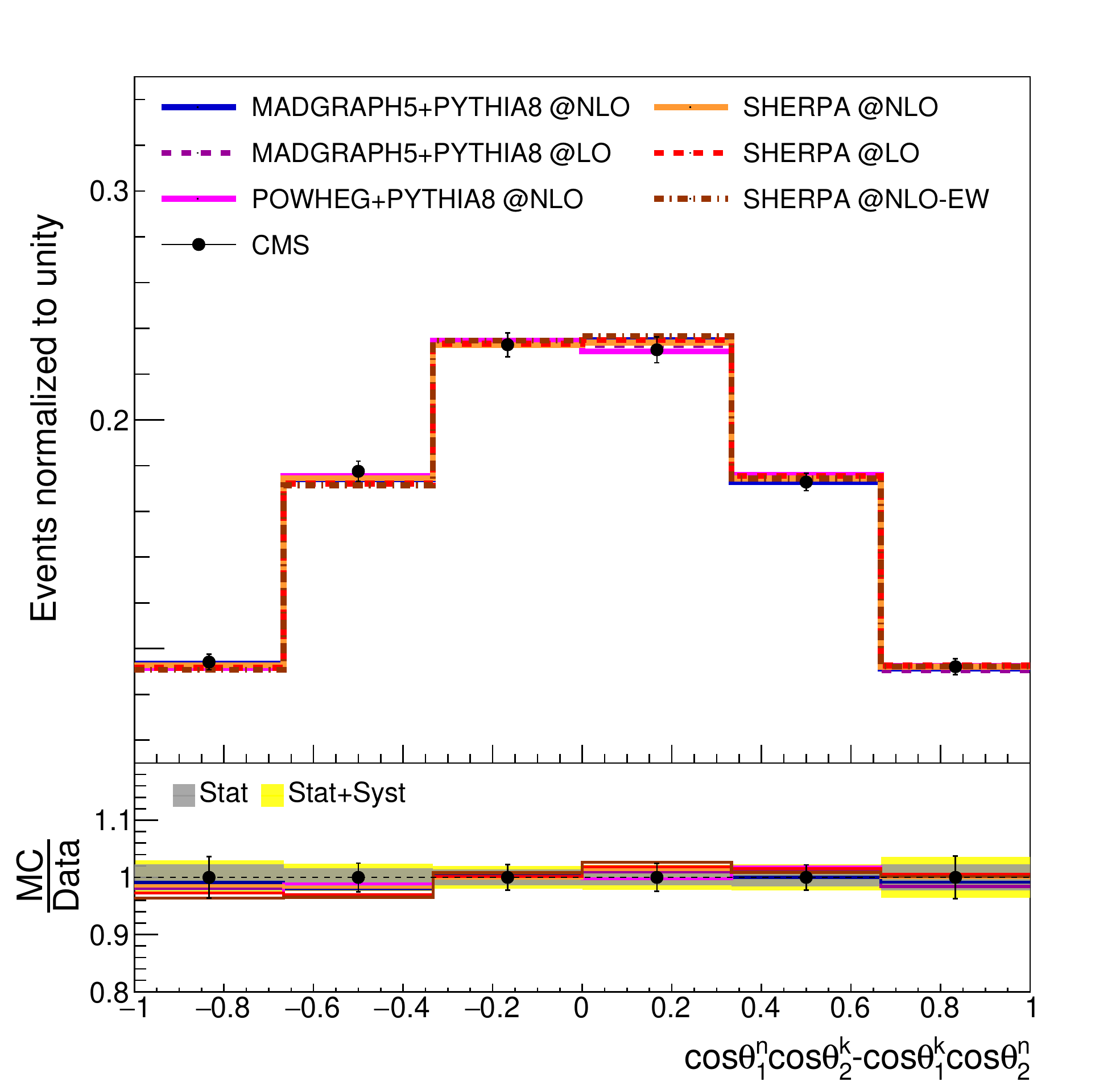}
     \hfill
         \includegraphics[width=.445\textwidth]{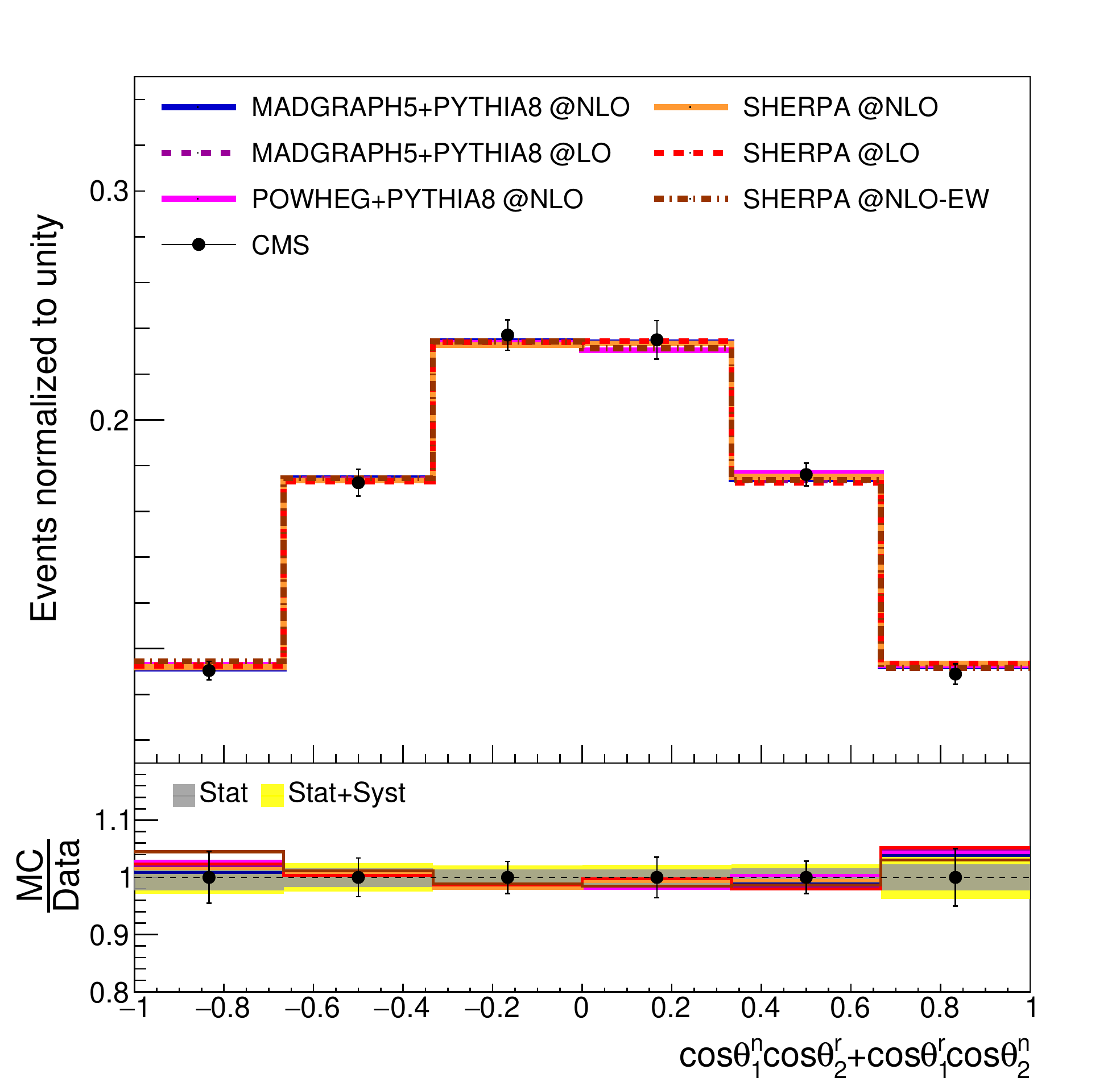}
     \hfill
         \includegraphics[width=.445\textwidth]{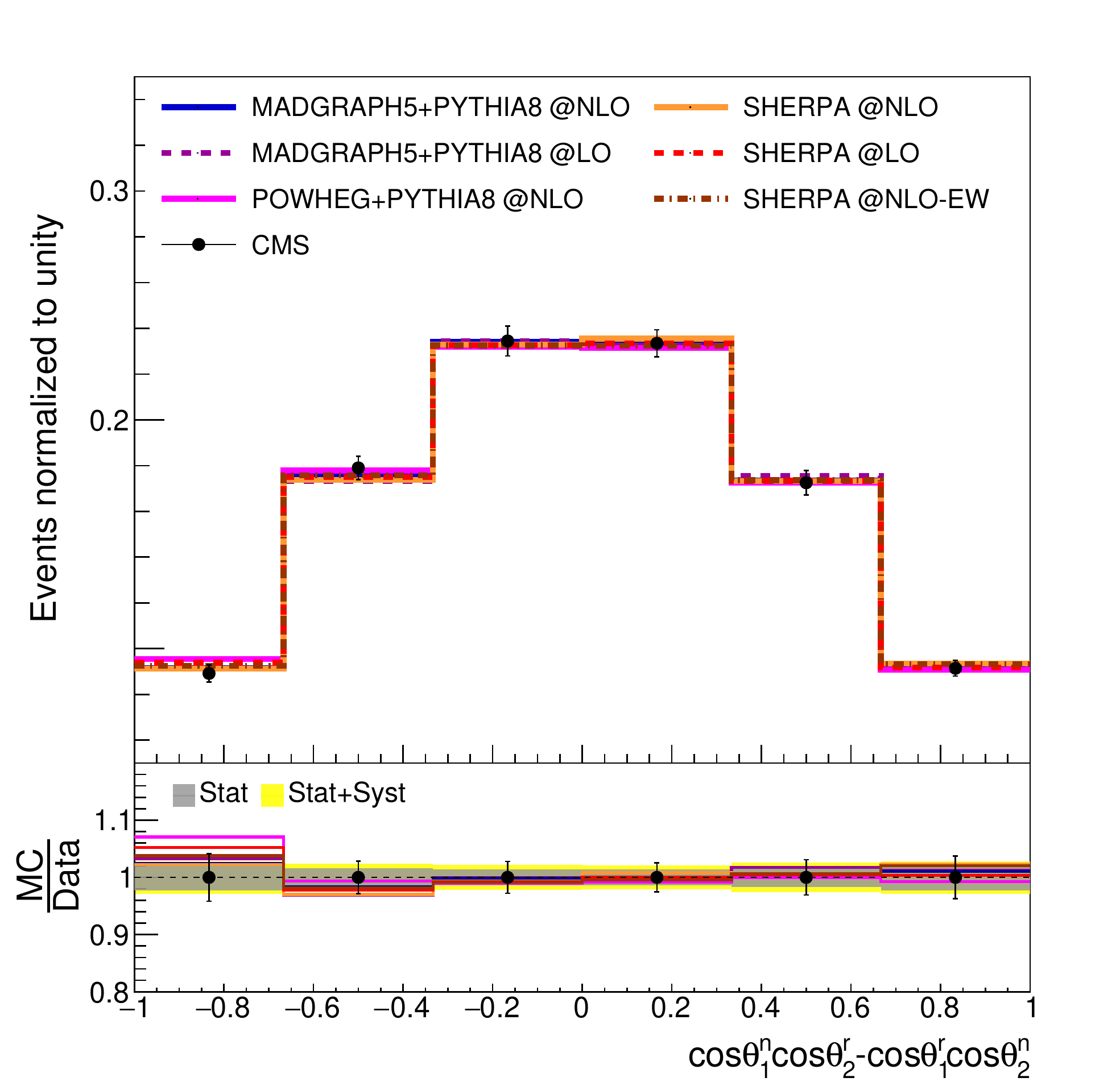}
        \caption{\label{fig:cross_spin} The normalized differential cross sections with respect to the cross spin correlation observables $\cos\theta_{1}^{i}\cos\theta_{2}^{j}\pm\cos\theta_{1}^{j}\cos\theta_{2}^{i}$, $i\neq j$. The ratio panels compare the MC predictions to the CMS data.}
\end{figure}

\begin{figure}[t]
     \centering
         \includegraphics[width=.445\textwidth]{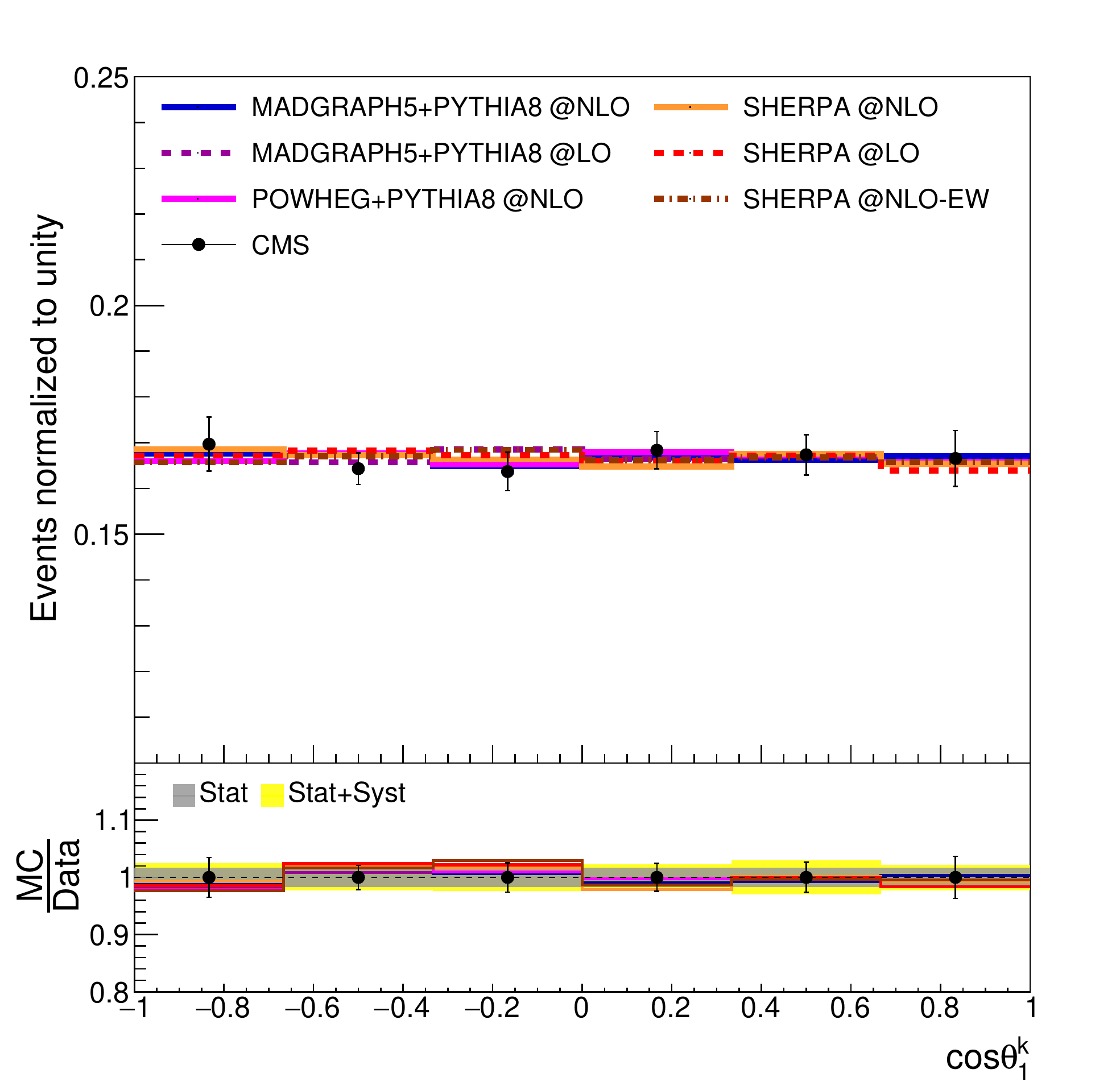}
         \hfill
         \includegraphics[width=.445\textwidth]{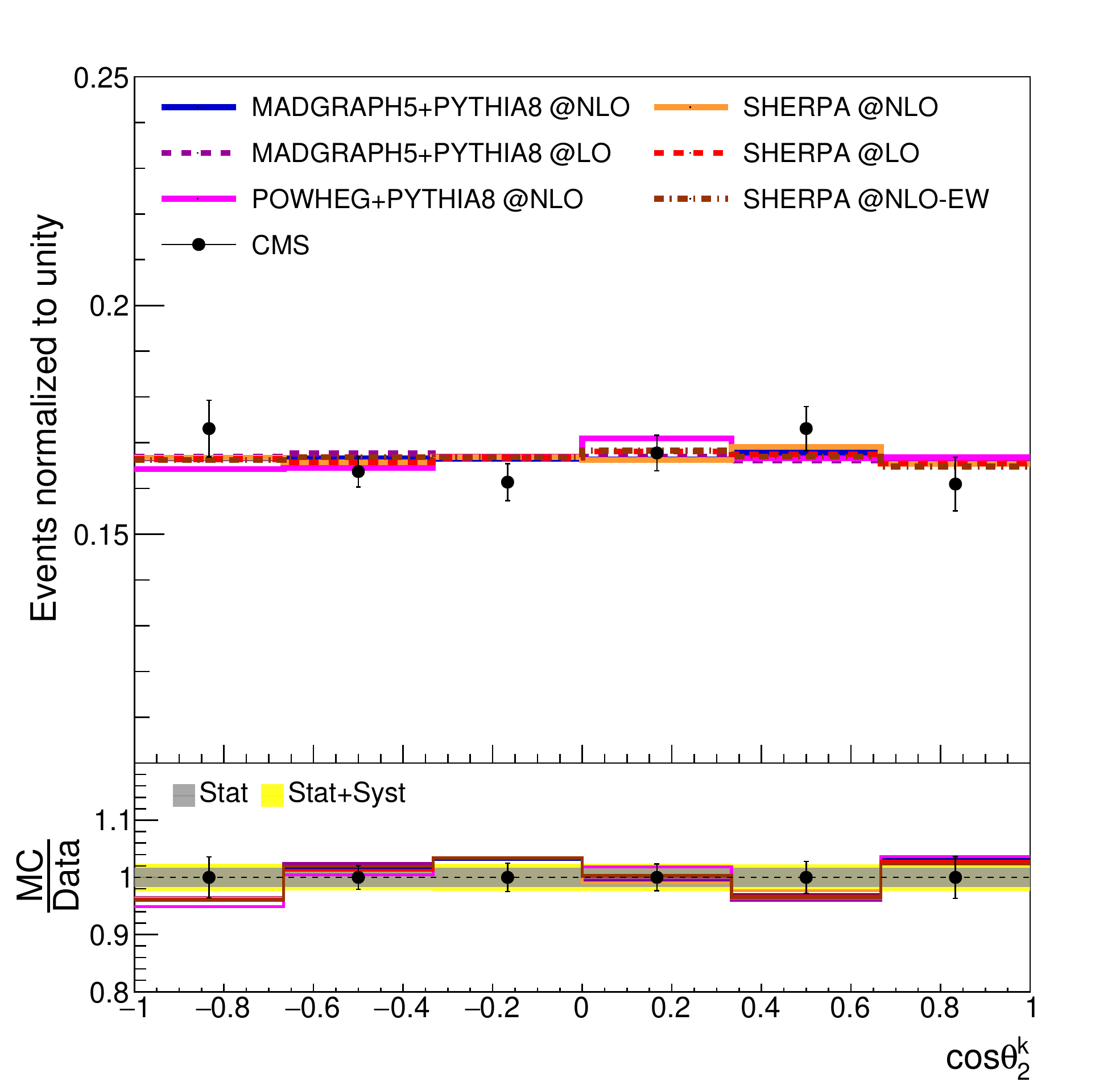}
     \hfill
         \includegraphics[width=.445\textwidth]{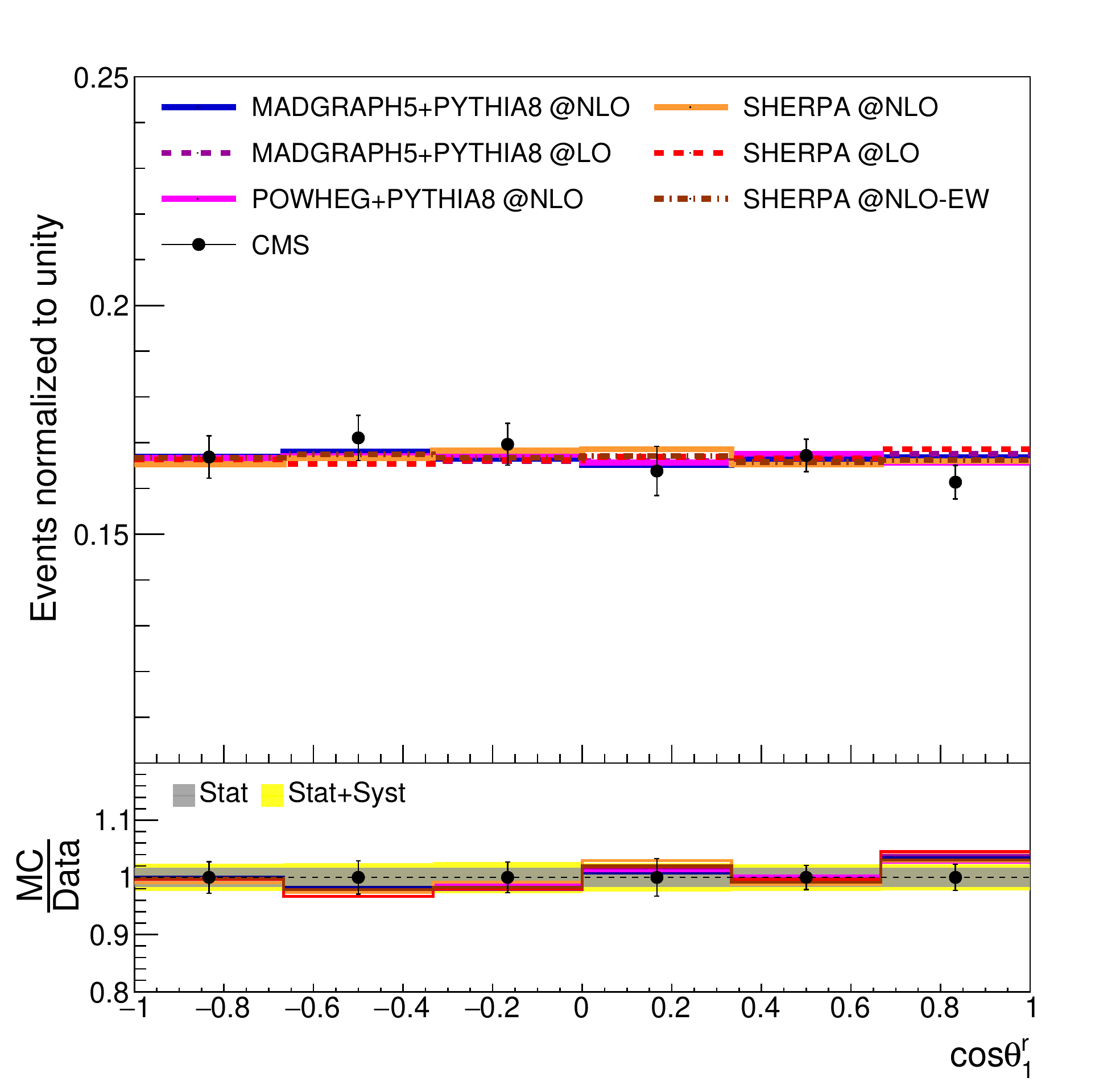}
         \hfill
         \includegraphics[width=.445\textwidth]{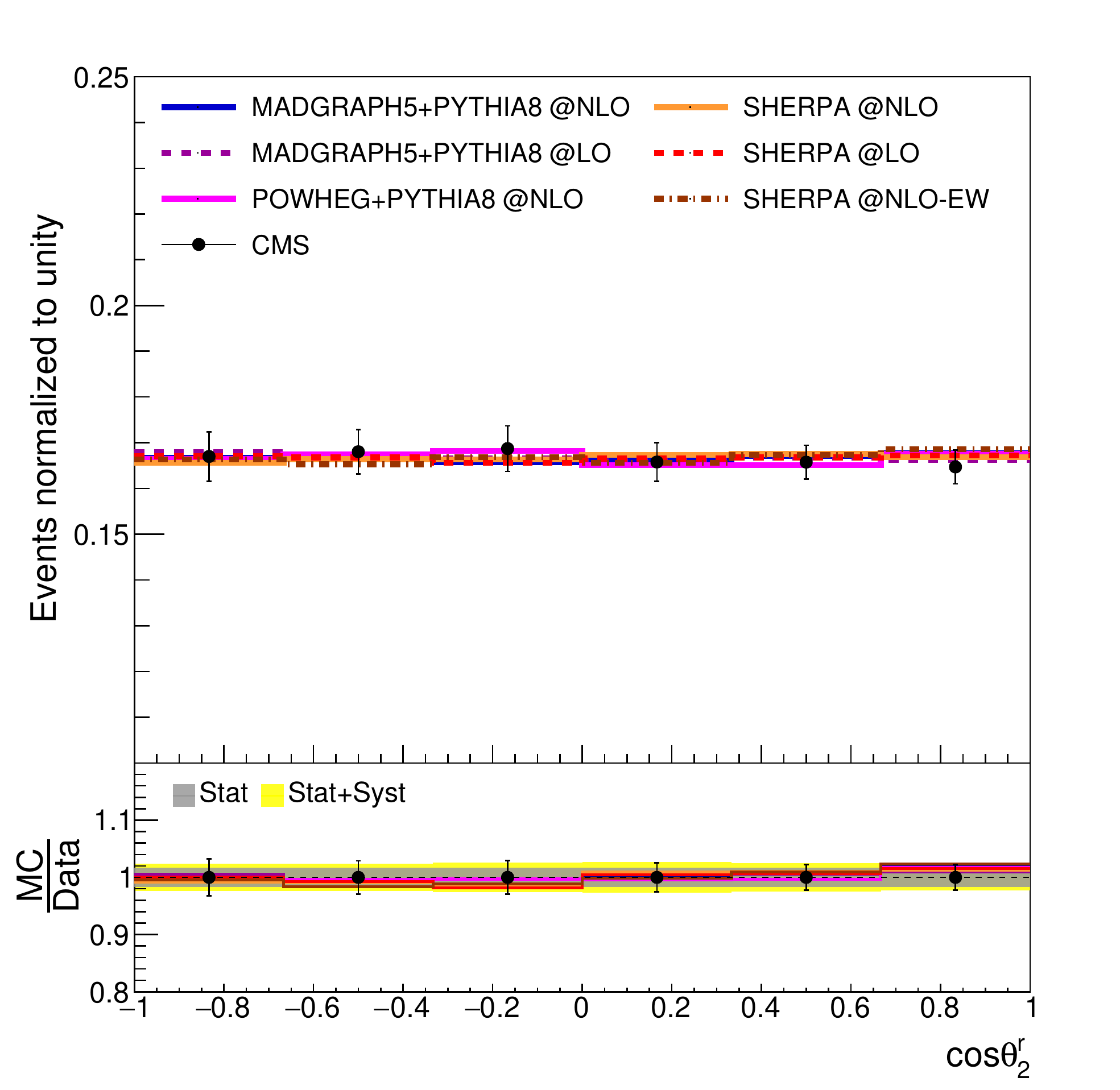}
     \hfill
         \includegraphics[width=.445\textwidth]{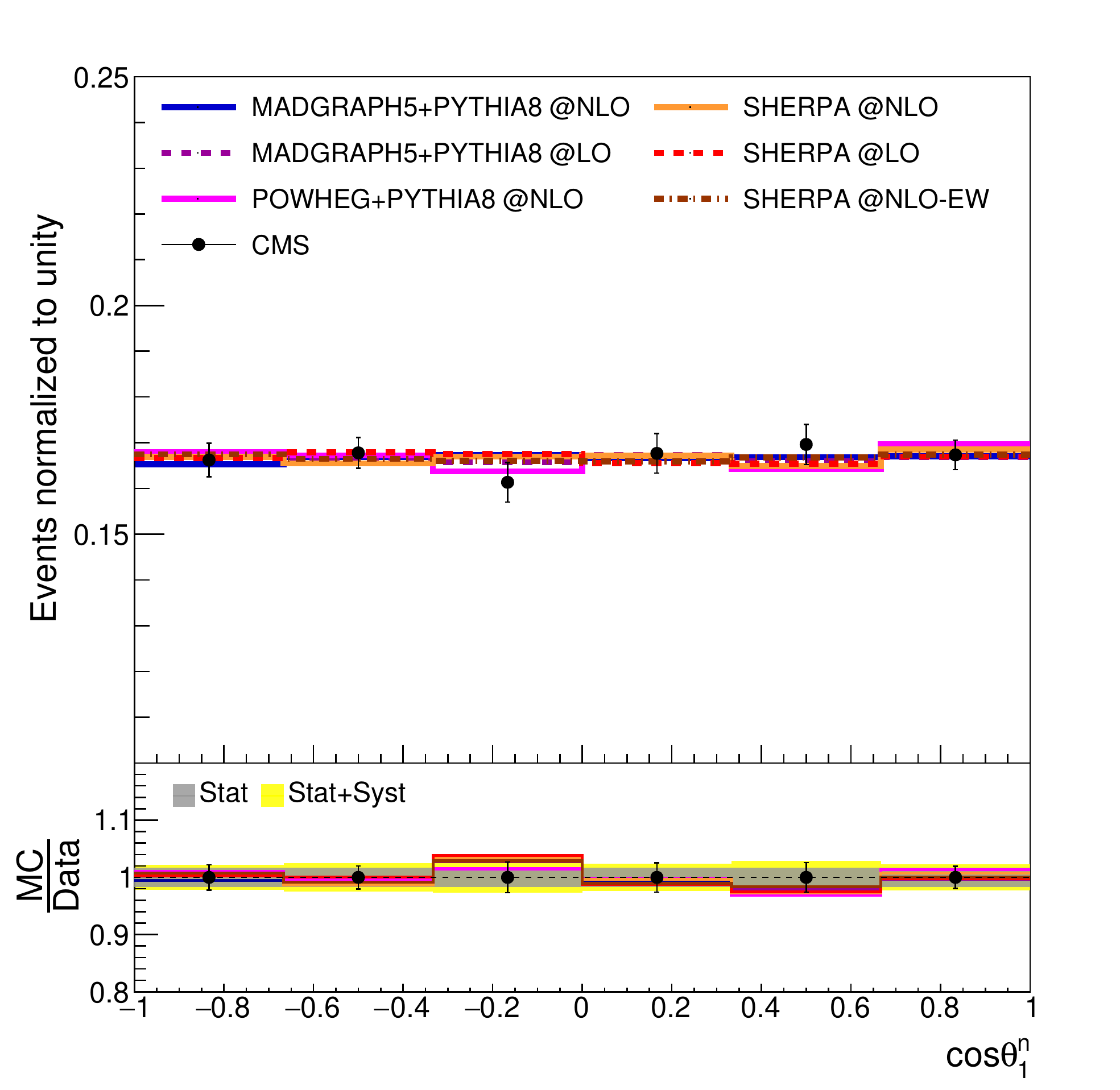}
          \hfill
         \includegraphics[width=.445\textwidth]{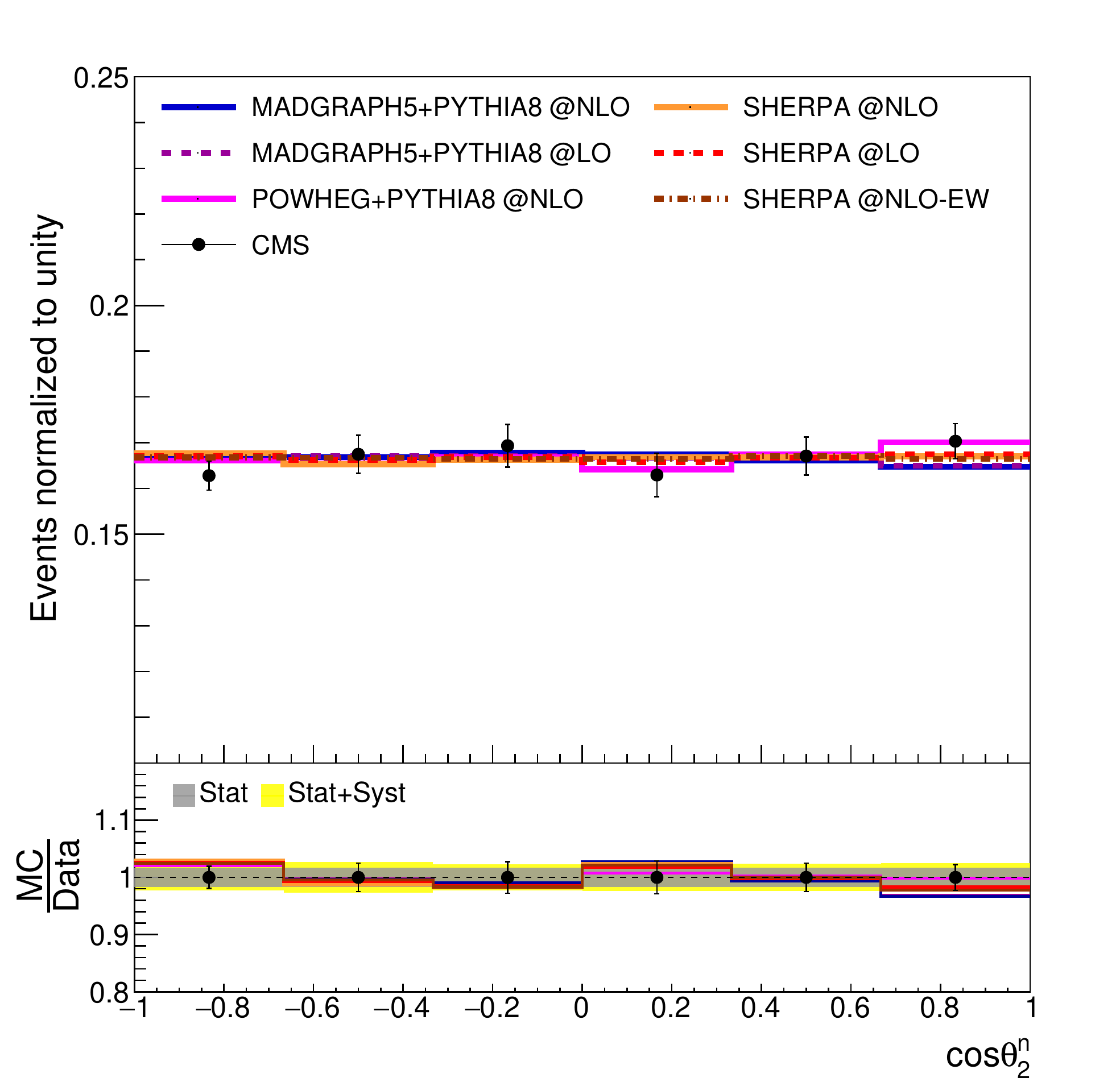}
        \caption{\label{fig:polarization} The normalized differential cross sections with respect to the observables $\cos\theta_{1}^{i}$ and $\cos\theta_{2}^{i}$, $i=k,r,n$. The ratio panels compare the MC predictions to the CMS data.}
\end{figure}

\begin{figure}[t]
     \centering
         \includegraphics[width=.445\textwidth]{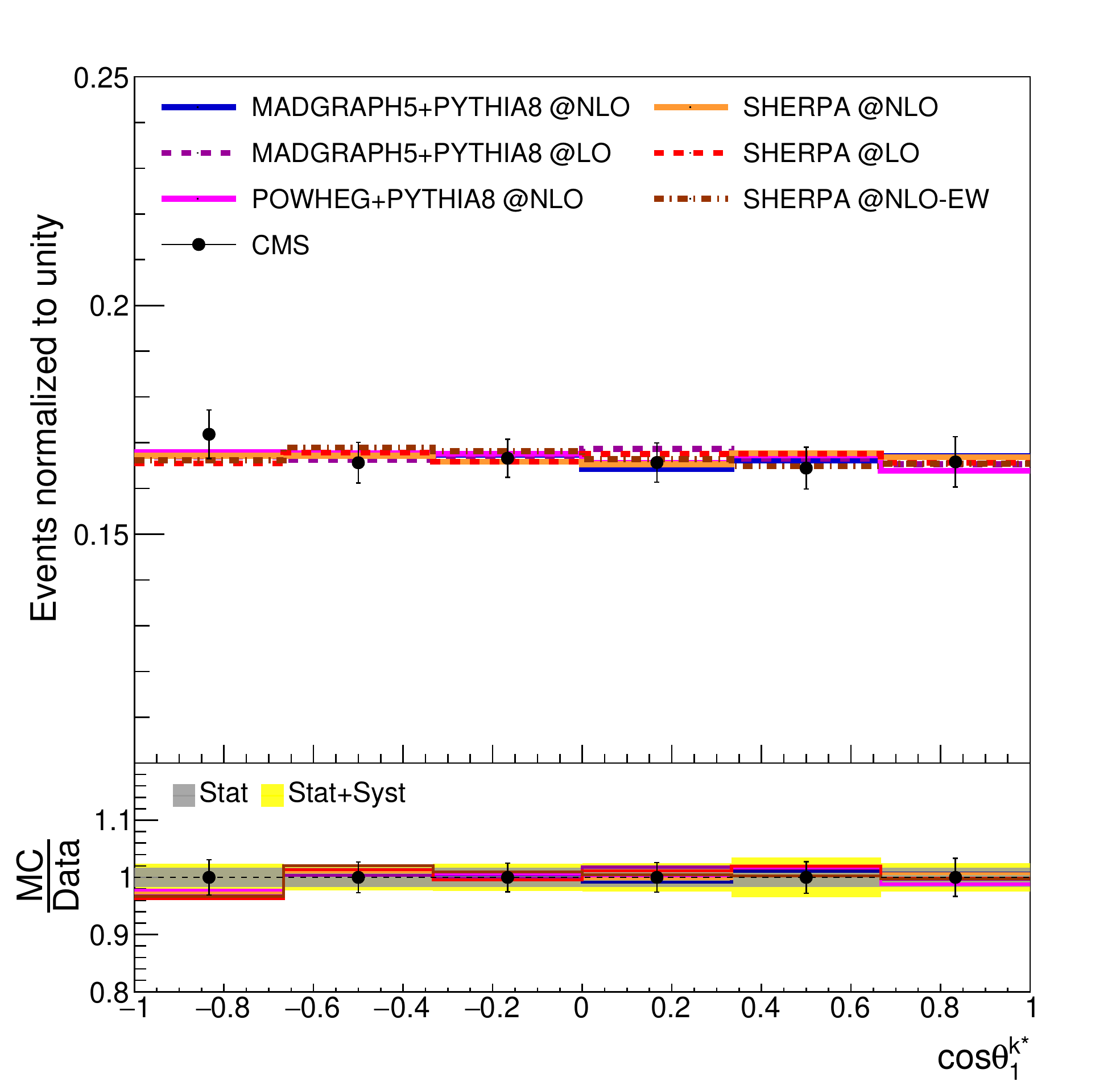}
     \hfill
         \includegraphics[width=.445\textwidth]{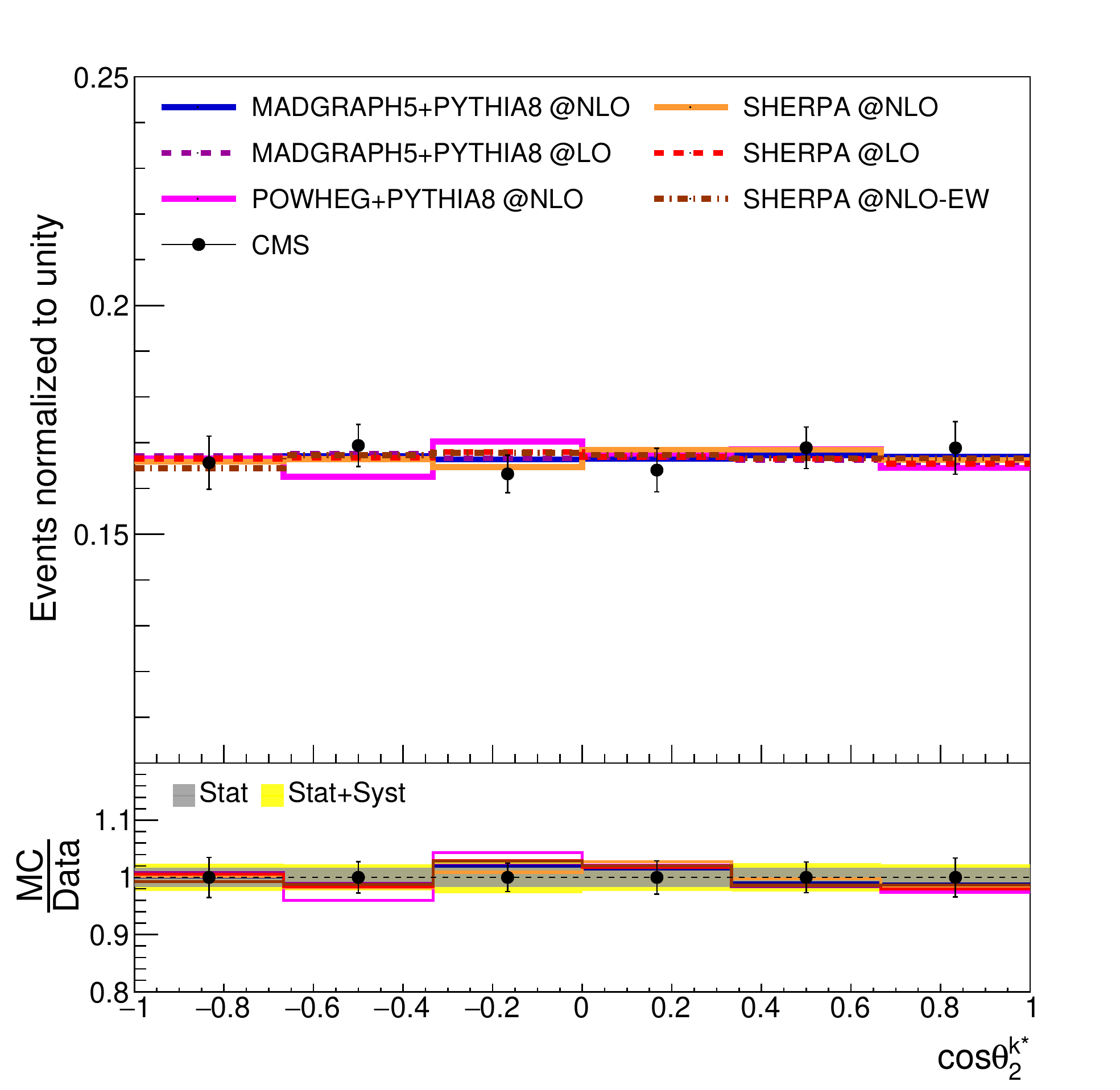}
     \hfill
         \includegraphics[width=.445\textwidth]{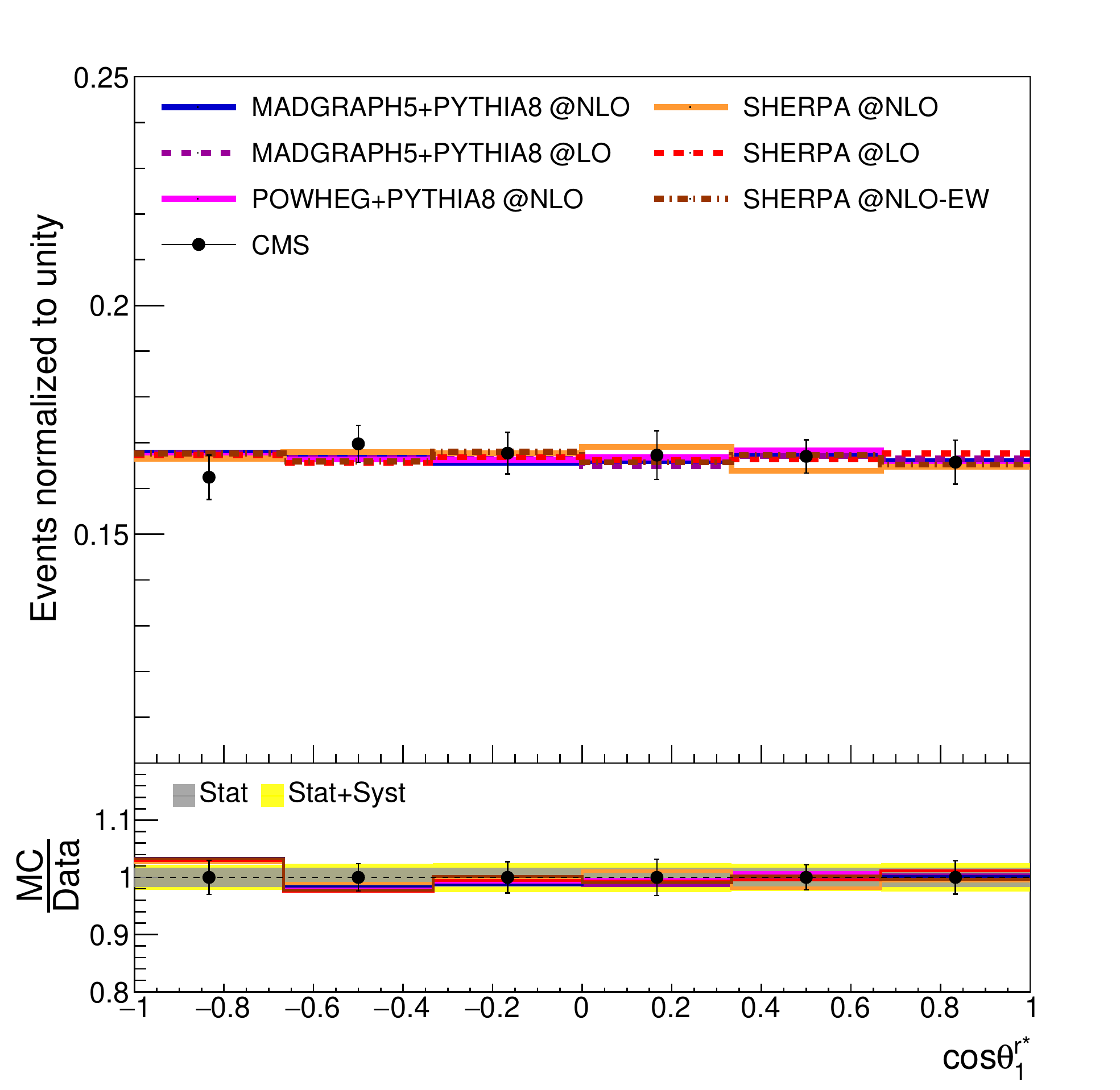}
     \hfill
         \includegraphics[width=.445\textwidth]{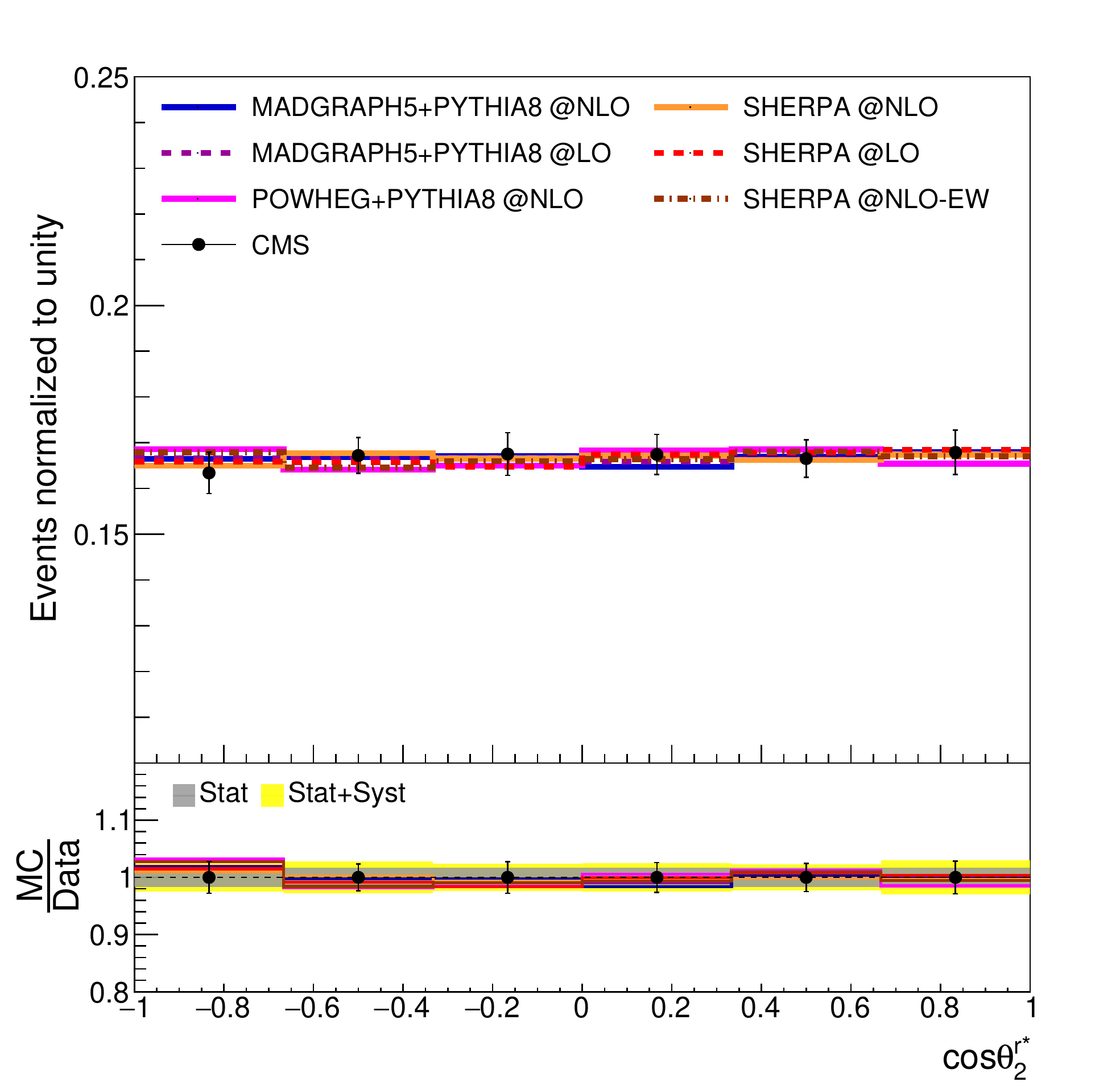}
        \caption{\label{fig:polarization_star} The normalized differential cross sections with respect to the observables $\cos\theta_{1}^{i*}$ and $\cos\theta_{2}^{i*}$, $i=k,r,n$. The ratio panels compare the MC predictions to the CMS data.}
\end{figure}

\begin{figure}[t]
     \centering
         \includegraphics[width=.445\textwidth]{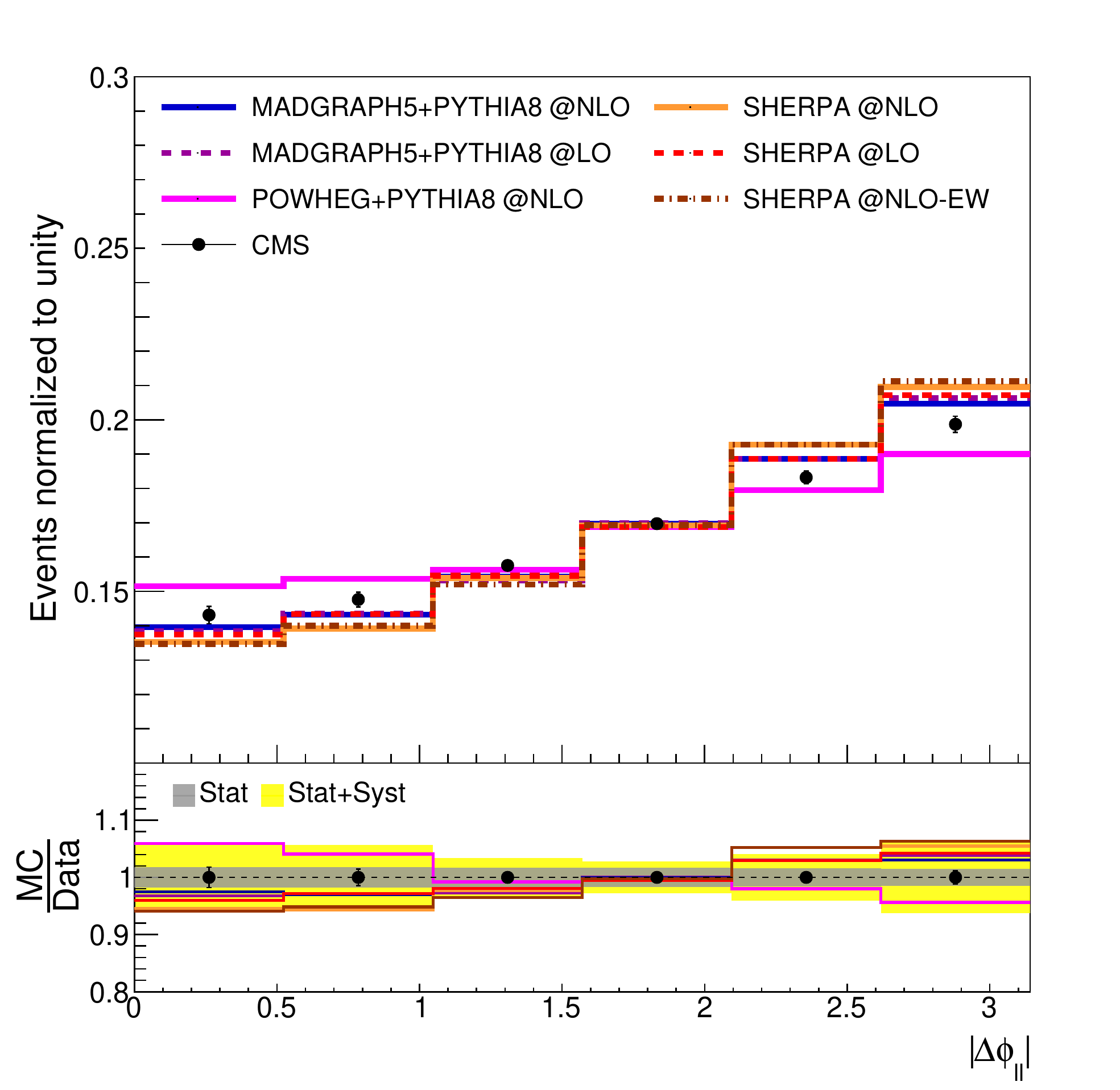}
     \hfill
         \includegraphics[width=.445\textwidth]{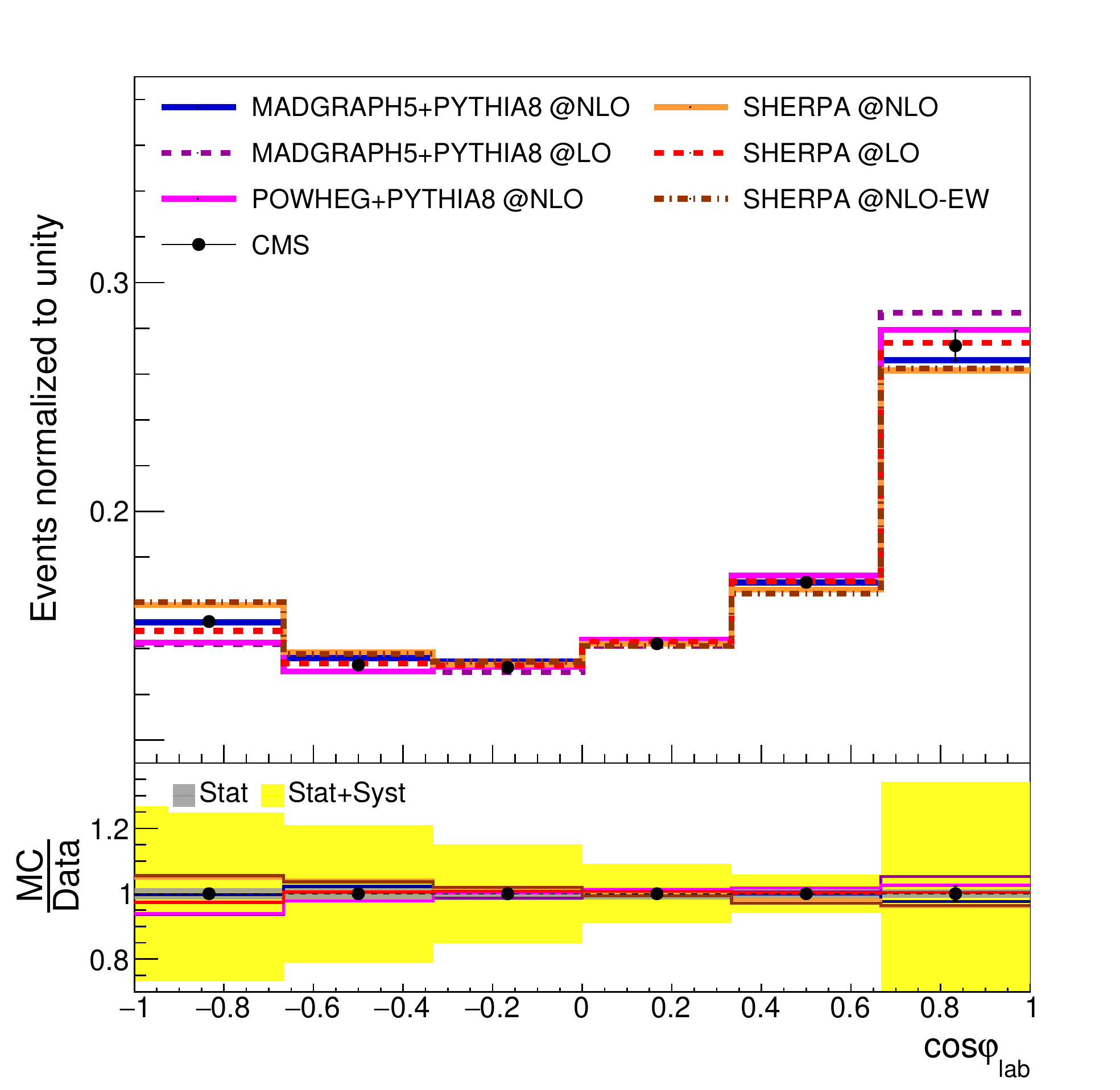}
        \caption{\label{fig:delta_phi-cos_phi} The normalized differential cross section as a function of laboratory frame observable which are $|\Delta\phi_{ll}|$ (left plot) and $\cos\varphi_{lab}$ (right plot). The ratio panels compare the MC predictions to the CMS data.}
\end{figure}

\begin{figure}[t]
    \centering
    \includegraphics[width=.445\textwidth]{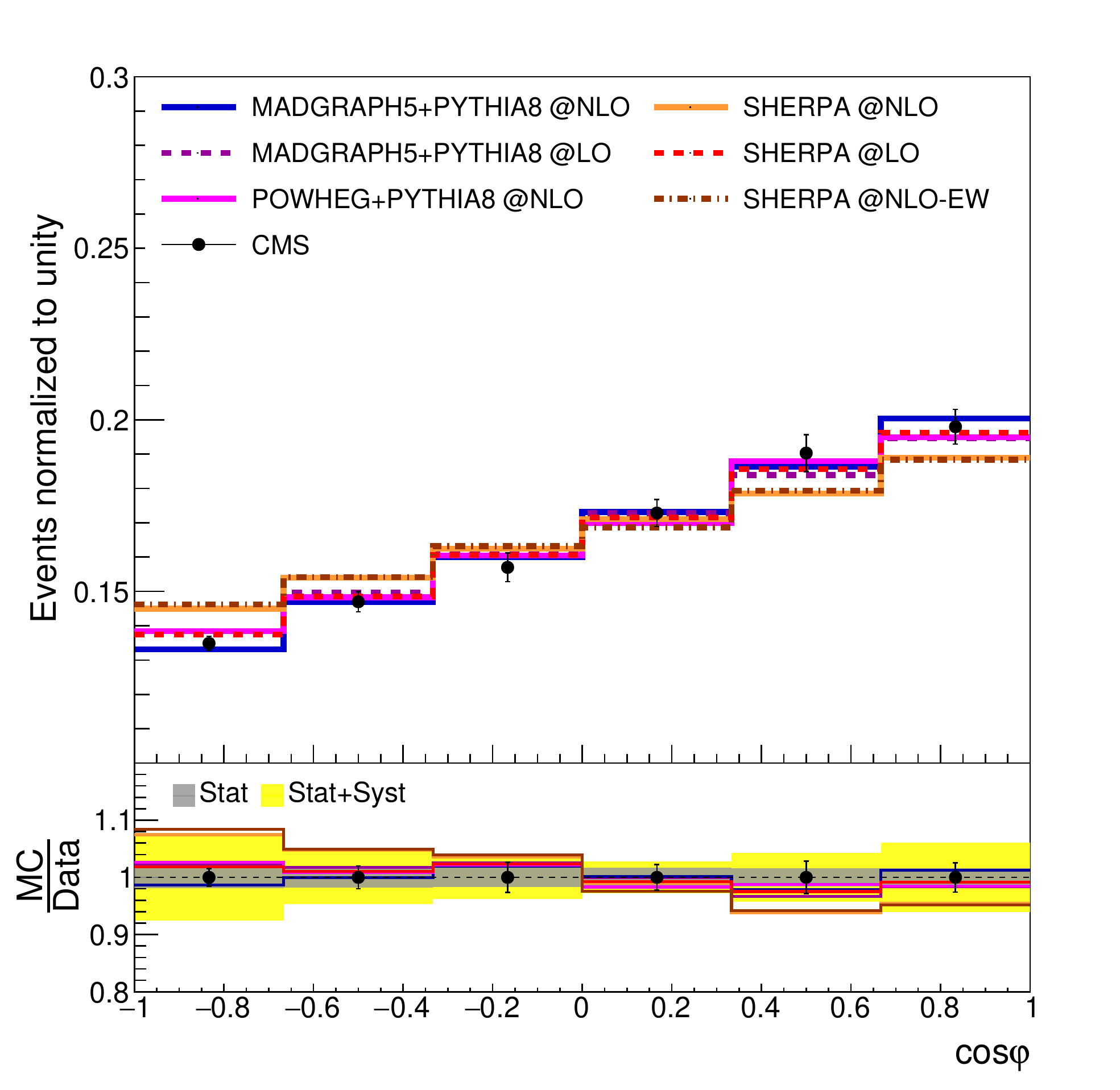}
    \caption{\label{fig:cosphi_rest_ttbar} The normalized differential cross section with respect to $\cos\varphi$ predicted by various MC event generators at LO, NLO and NLO-EW accuracy. The ratio of the MC predictions to the CMS data is shown in the lower panel.}
\end{figure}

In figures \ref{fig:diagonal_spin}-\ref{fig:cosphi_rest_ttbar}, the differential distributions predicted by \textsc{MadGraph5+Pythia8} at NLO is compared with \textsc{MadGraph5+Pythia8} at LO accuracy, with \textsc{Powheg+Pythia8} at NLO, with \textsc{Sherpa} at LO, and with \textsc{Sherpa} at NLO QCD accuracy and at NLO in QCD with EW correction. To check the compatibility with experiment, whole that theoretical predictions are compared with CMS measurements.

In figure \ref{fig:diagonal_spin}, the diagonal spin correlation observables $\cos\theta_{1}^{i}\cos\theta_{2}^{i}$ defined in eq. \ref{eq:9} are presented. There are no significant deviations in different MC approaches and different QCD corrections as well as EW correction. The experimental uncertainty comprise almost all variations of predictions.

The predictions for the differential distributions of the observables $x_{\pm}$, eq. \ref{eq:10}, are considered in figure \ref{fig:cross_spin}. Non-zero coefficients $(C_{nr}+C_{rn})$ and $(C_{nk}+C_{kn})$ are expected from mixed QCD-weak corrections correspond to ref. \cite{0h}. Due to smallness of this effect, scale, PDF and experimental uncertainties dominate it. Within the uncertainties, the experimental data and predictions for cross spin correlation observables are matched.

The normalized differential distributions in $\cos\theta_{1,2}^{i}$, eq. (\ref{eq:7}, \ref{eq:8}), which is the indicator of the polarization of top quark and anti-quark are shown in figure \ref{fig:polarization}. The polarization coefficients for all predictions behave similar in all distributions. CMS data and MC predictions are compatible within uncertainties. In the same way, the normalized differential distributions of $\cos\theta_{1,2}^{i*}$ predicted by MC, figure~\ref{fig:polarization_star}, are in agreement with data.

The differential distributions in the observables, which are calculated with kinematic variables of leptons in laboratory-frame, are in figure~\ref{fig:delta_phi-cos_phi}. When $|\Delta\phi_{ll}|$ is considered with total uncertainty which is around $6\%$, the shapes predicted by \textsc{MadGraph5} and \textsc{Sherpa} ME generators are roughly consistent. \textsc{Sherpa} and \textsc{MadGraph5} at LO accuracy show better performance than the samples simulated at NLO, this may be the result of real contributions at NLO accuracy spoiling the back-to-back alignment of the top quarks. \textsc{Powheg} is separated from the other MC predictions and the data about $6\%$ at low and high $|\Delta\phi_{ll}|$ regions. For the distributions of $\cos\varphi_{lab}$, all MC predictions and the data are in good agreement in whole range, although the systematic uncertainties coming from PDF sets and scales are more than $10\%$ in almost whole regions. The huge systematic uncertainty as seen in the lower panel of the right plot in figure \ref{fig:delta_phi-cos_phi} is coming from PDF variations in \textsc{Sherpa} samples.

In $\cos\varphi$ distribution, figure~\ref{fig:cosphi_rest_ttbar}, the predictions of MC generators are more similar compared to laboratory-frame distributions. While the distributions of different MC variations have considerably the same shape, \textsc{Sherpa} with NLO QCD accuracy and EW correction deviate from the others, especially at low $\cos\varphi$, but are still in total uncertainty.

In second part of the analysis, we have investigated various new physics models with the help of the elements of spin density matrix. For top quark pair channel, most searches for SUSY and DM particles comprise generally high missing transverse energy $E_{T}^{miss}$ region stem from undetected LSP and DM \cite{1a, 1b, 0e, 1d}. The main point for the BSM analysis in this study is the models with low $E_{T}^{miss}$ but also the same models with high $E_{T}^{miss}$, which are not excluded by CMS and ATLAS Collaborations yet, will be mentioned for the sake of broad phase space.

\begin{figure}[t]
     \centering
         \includegraphics[width=.445\textwidth]{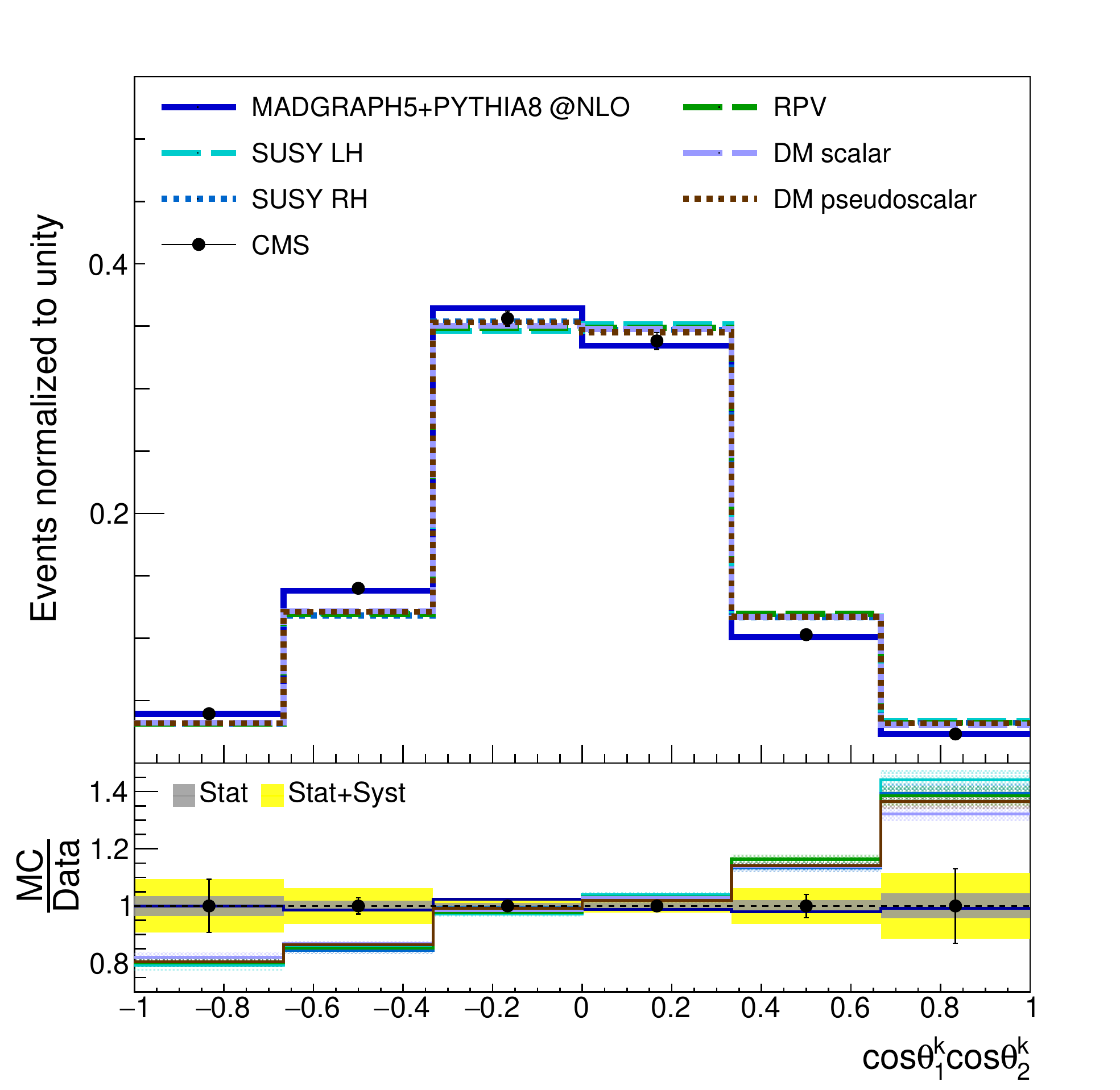}
     \hfill
         \includegraphics[width=.445\textwidth]{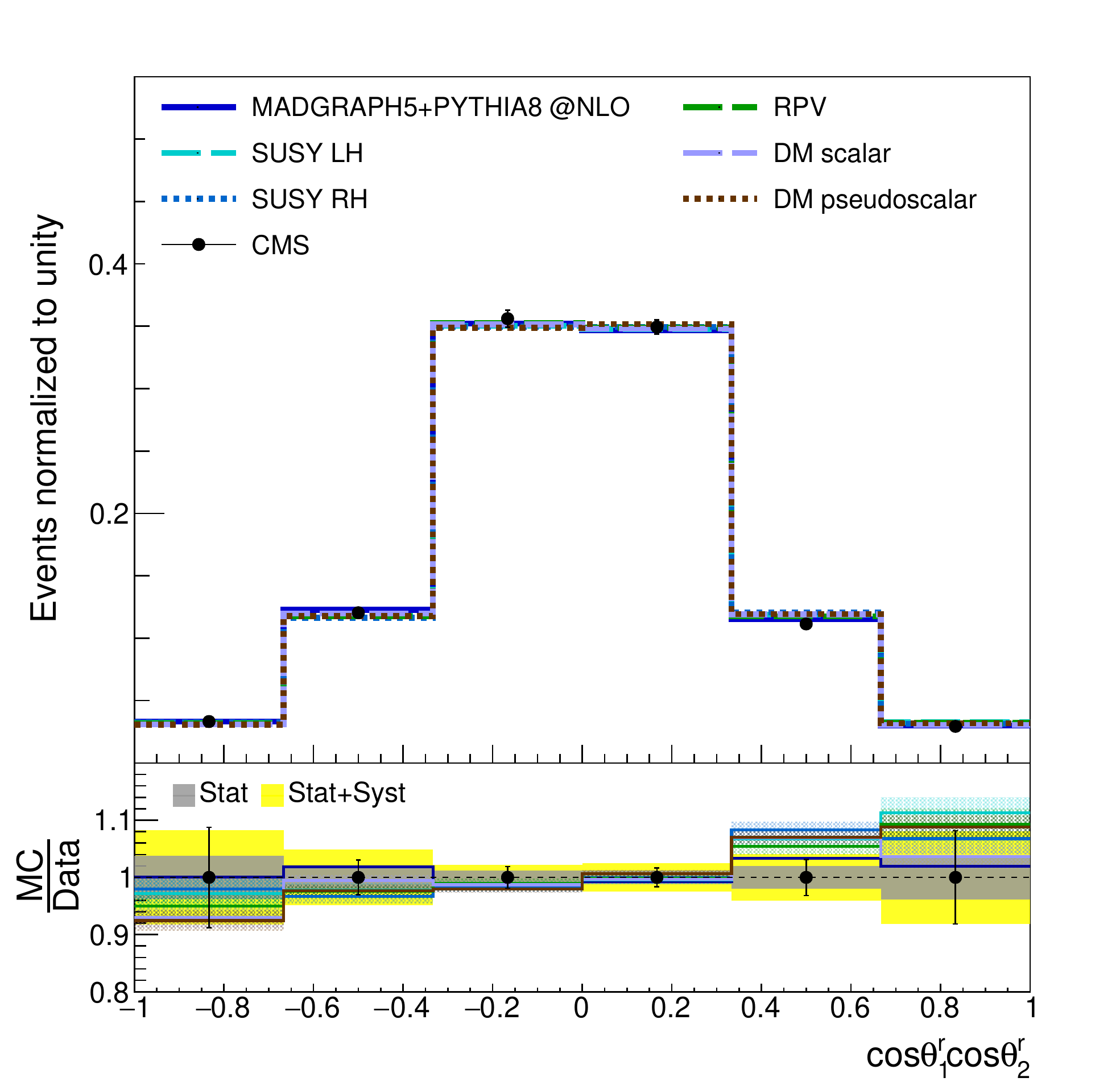}
     \hfill
         \includegraphics[width=.445\textwidth]{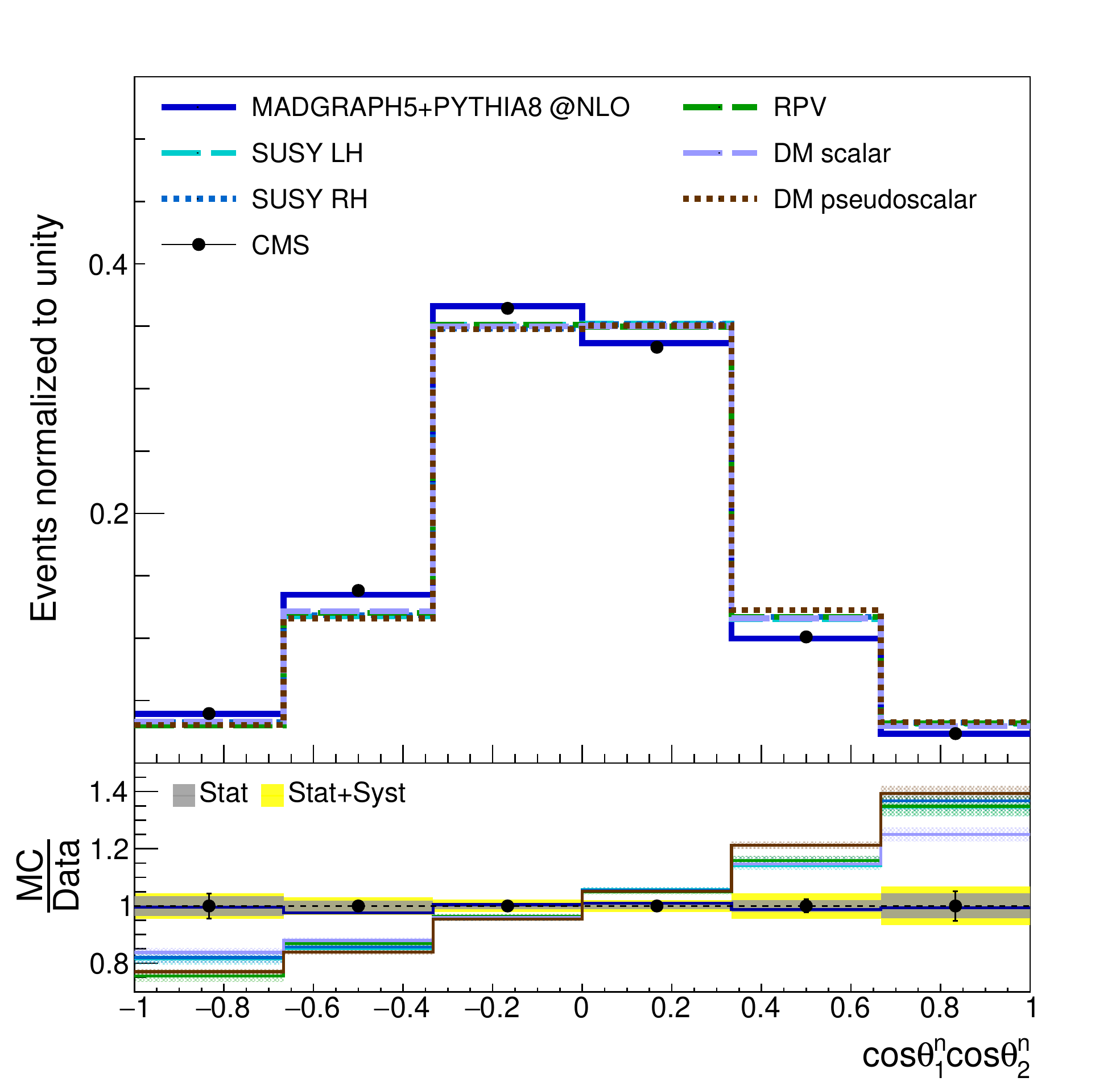}
        \caption{\label{fig:diagonal_spin_BSM} The normalized differential cross sections with respect to the diagonal spin correlation observables $\cos\theta_{1}^{i}\cos\theta_{2}^{i}$, $i=k,r,n$. The ratio panels compare the BSM predictions to the CMS data.}
\end{figure}

\begin{figure}[t]
     \centering
         \includegraphics[width=.445\textwidth]{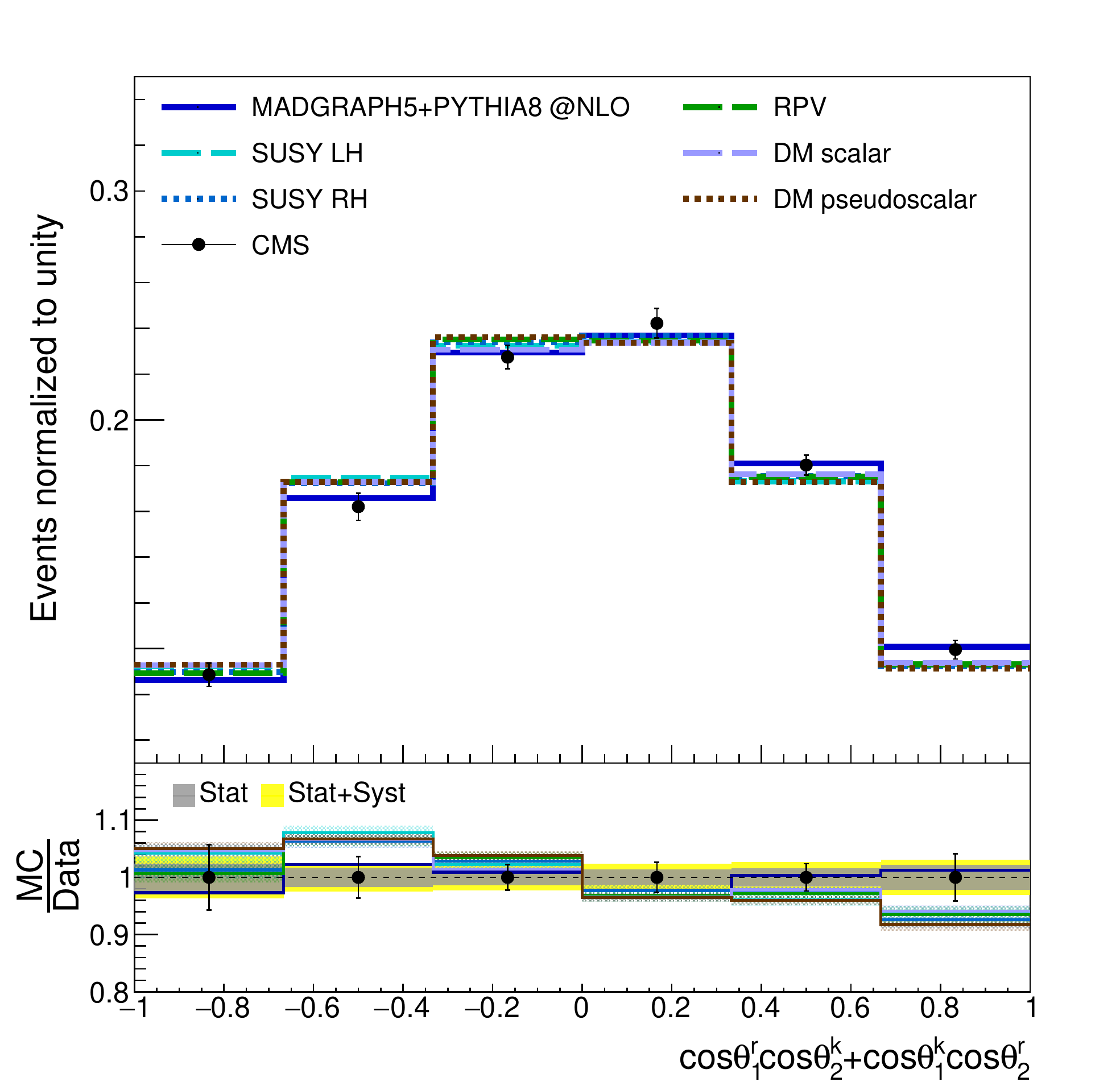}
     \hfill
         \includegraphics[width=.445\textwidth]{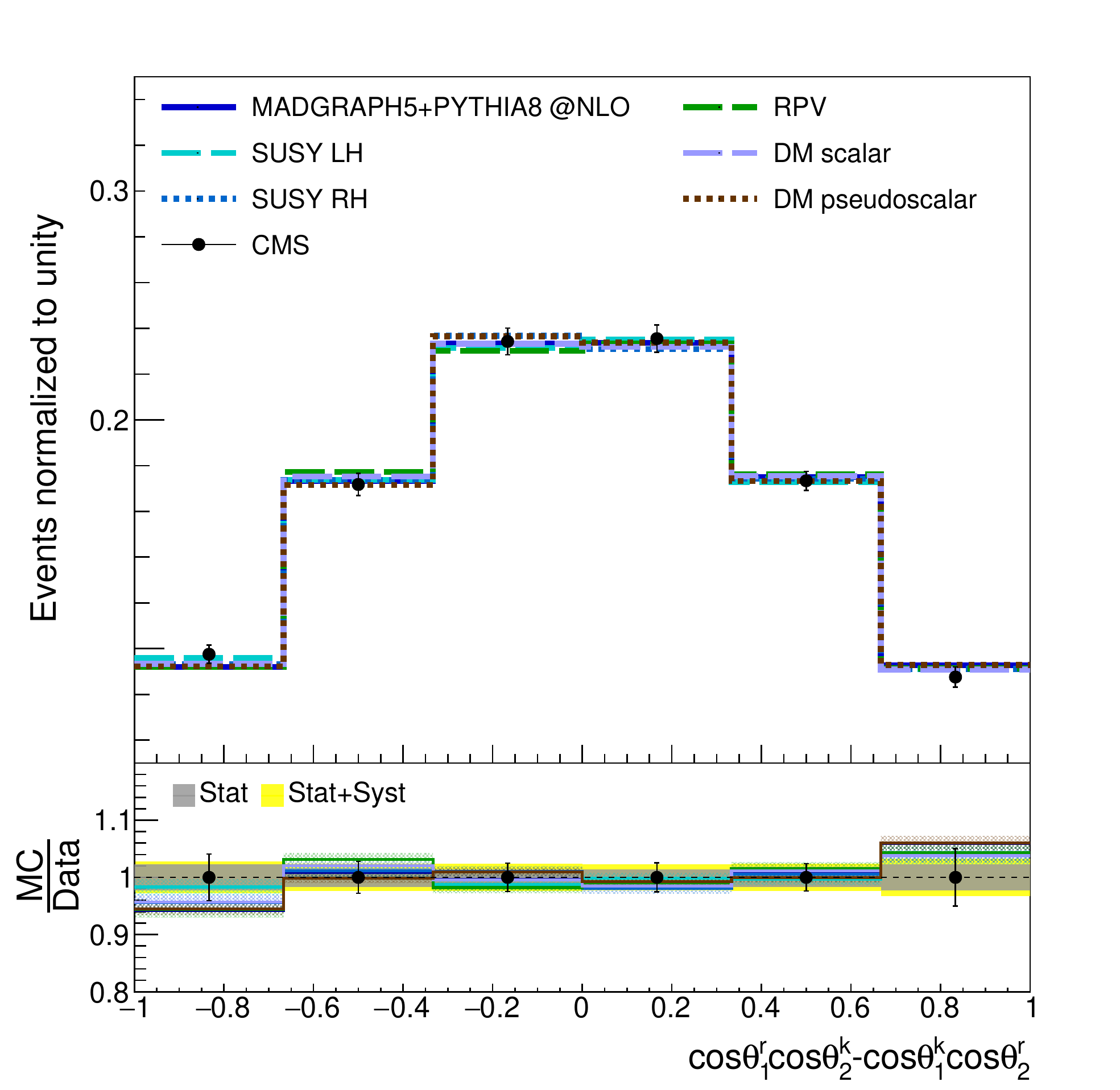}
     \hfill
         \includegraphics[width=.445\textwidth]{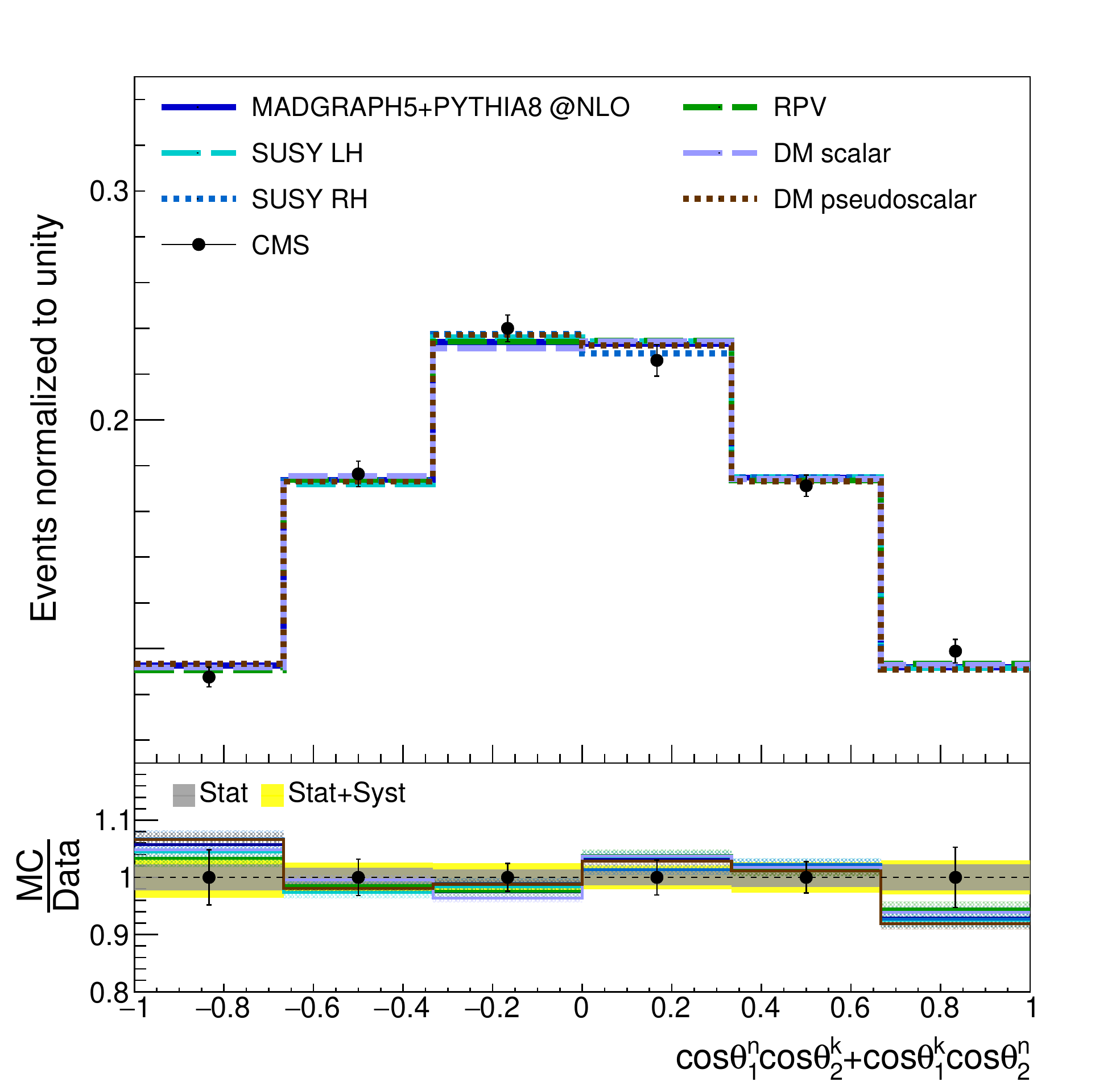}
      \hfill
         \includegraphics[width=.445\textwidth]{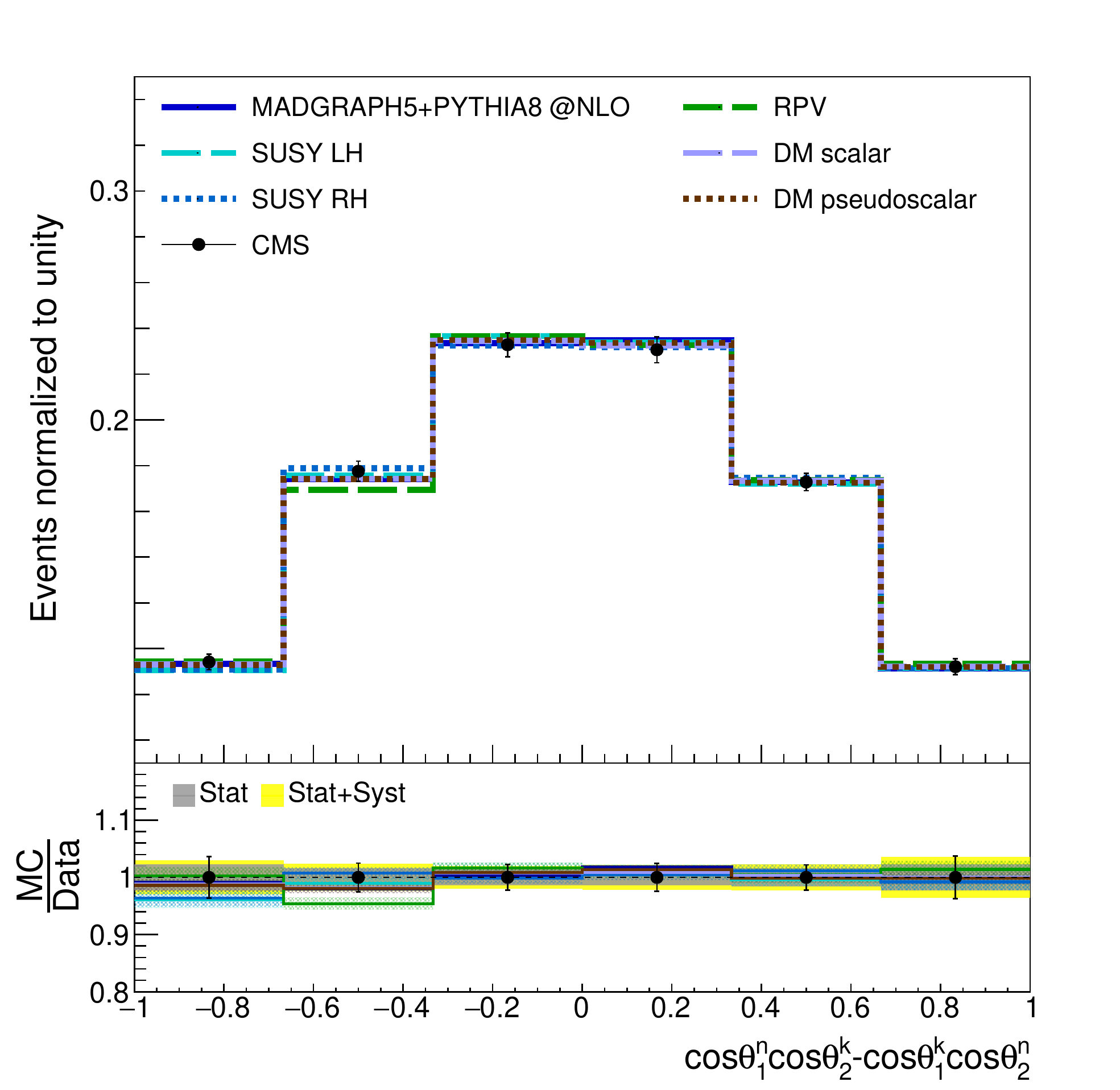}
     \hfill
         \includegraphics[width=.445\textwidth]{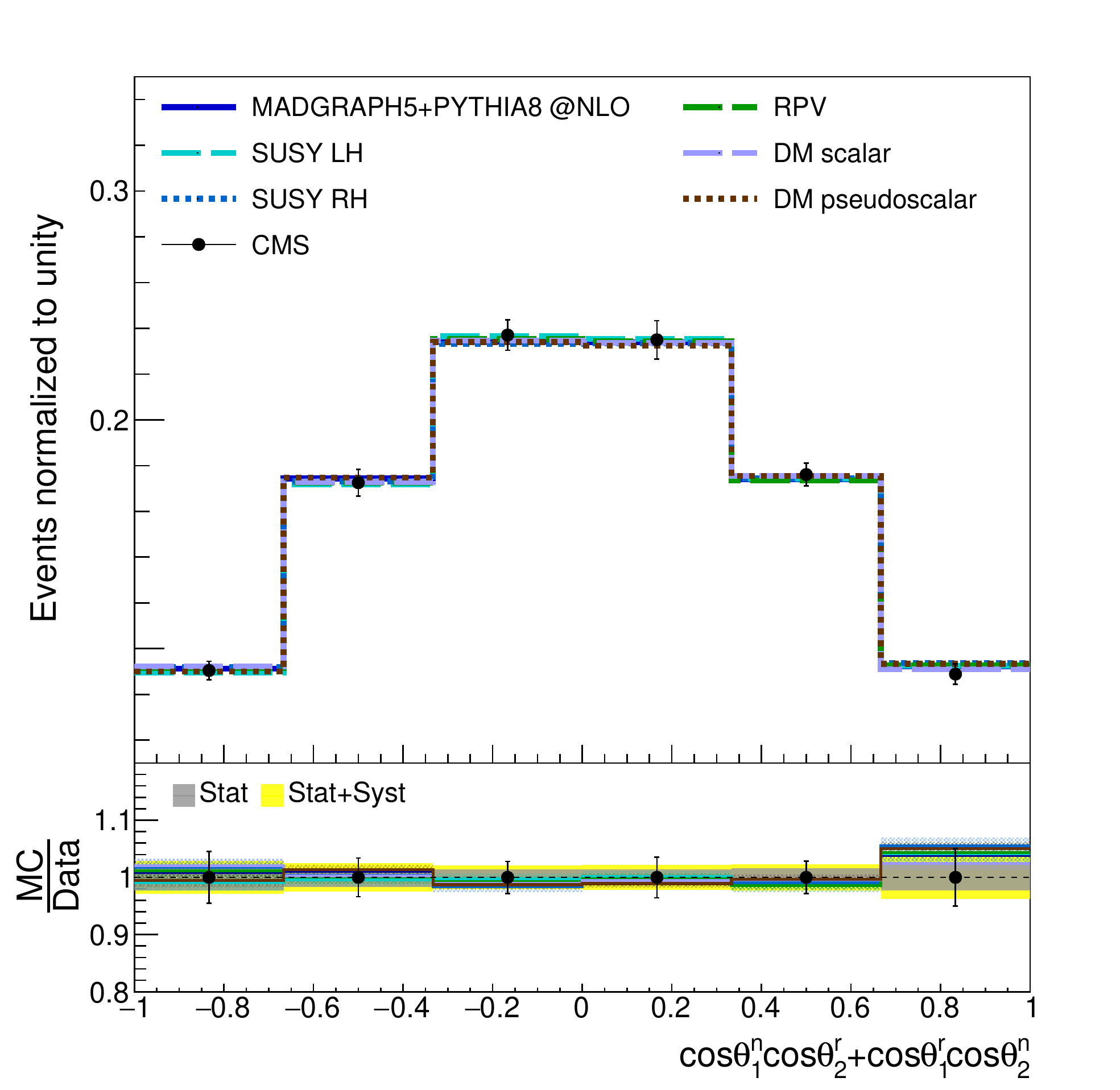}
     \hfill
         \includegraphics[width=.445\textwidth]{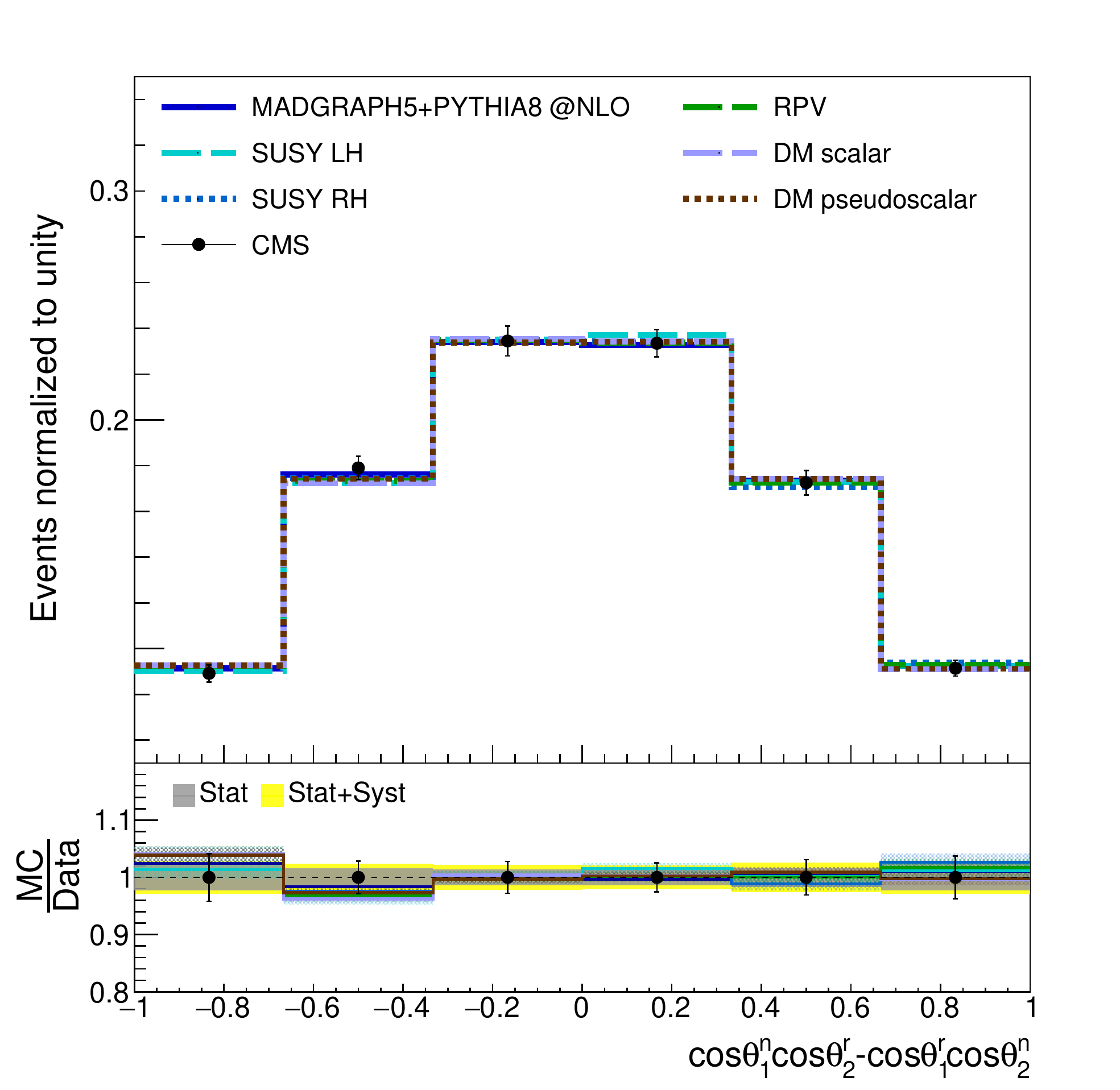}
        \caption{\label{fig:cross_spin_BSM} The normalized differential cross sections with respect to the cross spin correlation observables $\cos\theta_{1}^{i}\cos\theta_{2}^{j}\pm\cos\theta_{1}^{j}\cos\theta_{2}^{i}$, $i\neq j$. The ratio panels compare the BSM signals to the CMS data.}
\end{figure}

\begin{figure}[t]
     \centering
         \includegraphics[width=.445\textwidth]{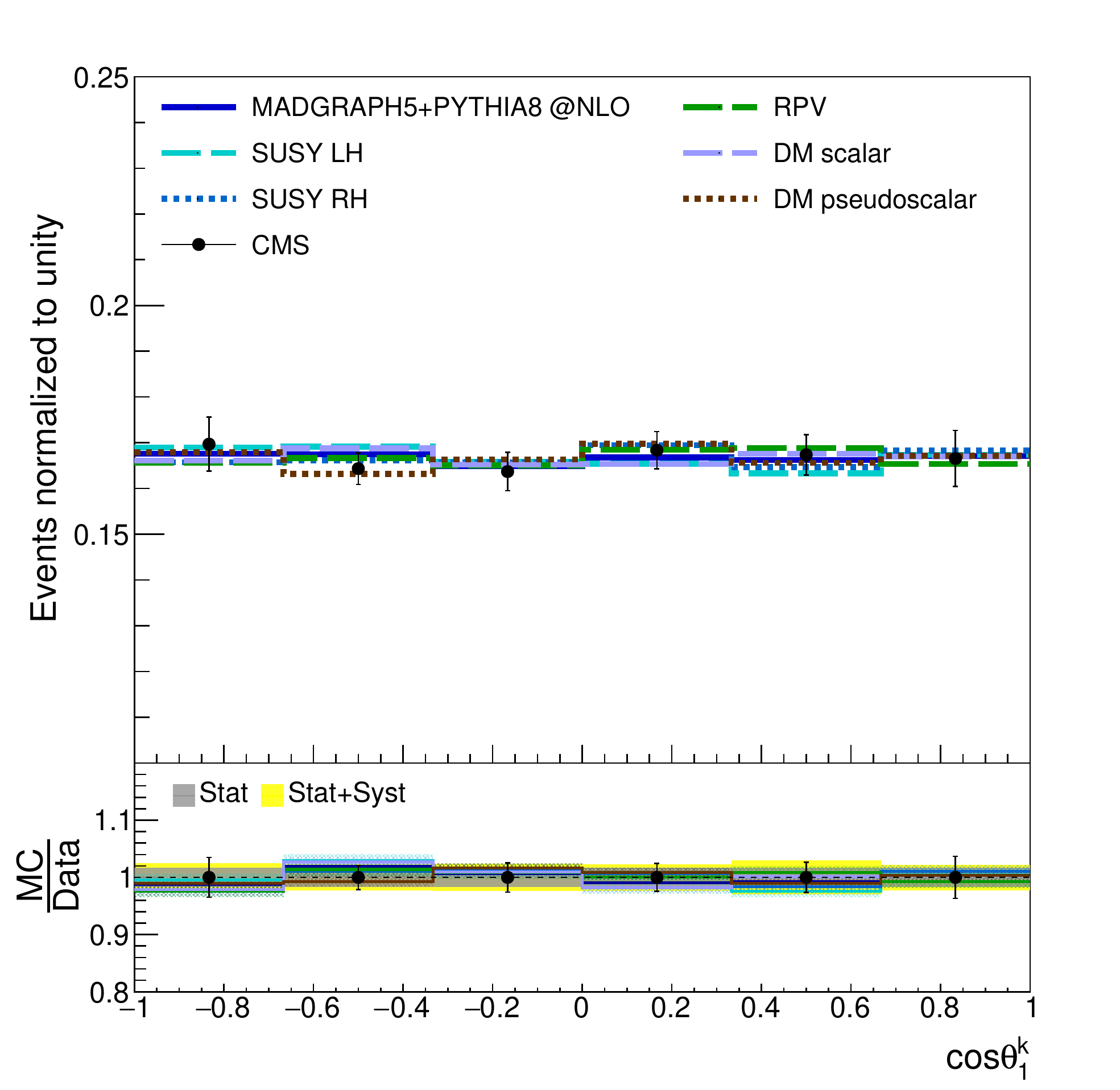}
         \hfill
         \includegraphics[width=.445\textwidth]{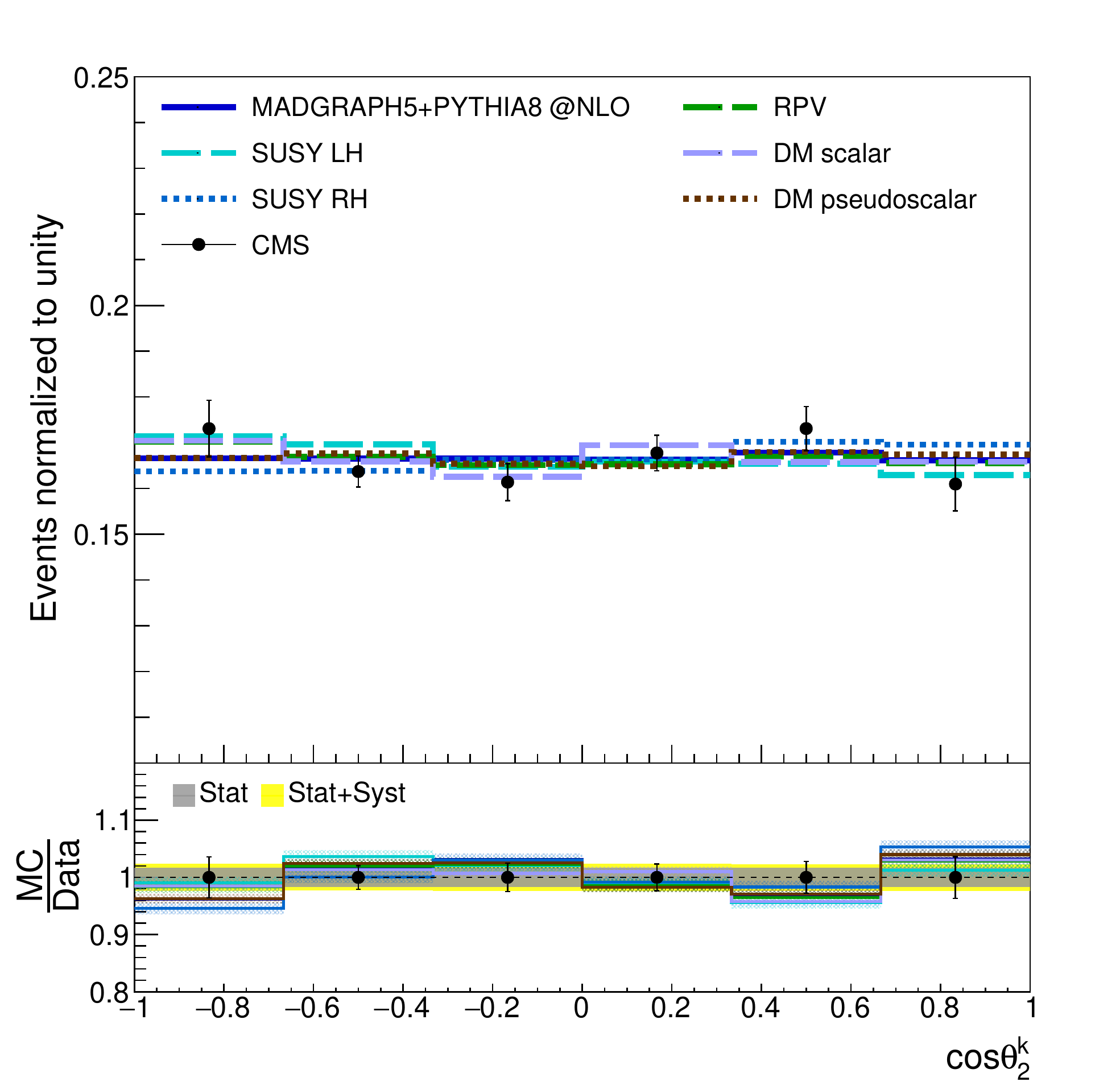}
     \hfill
         \includegraphics[width=.445\textwidth]{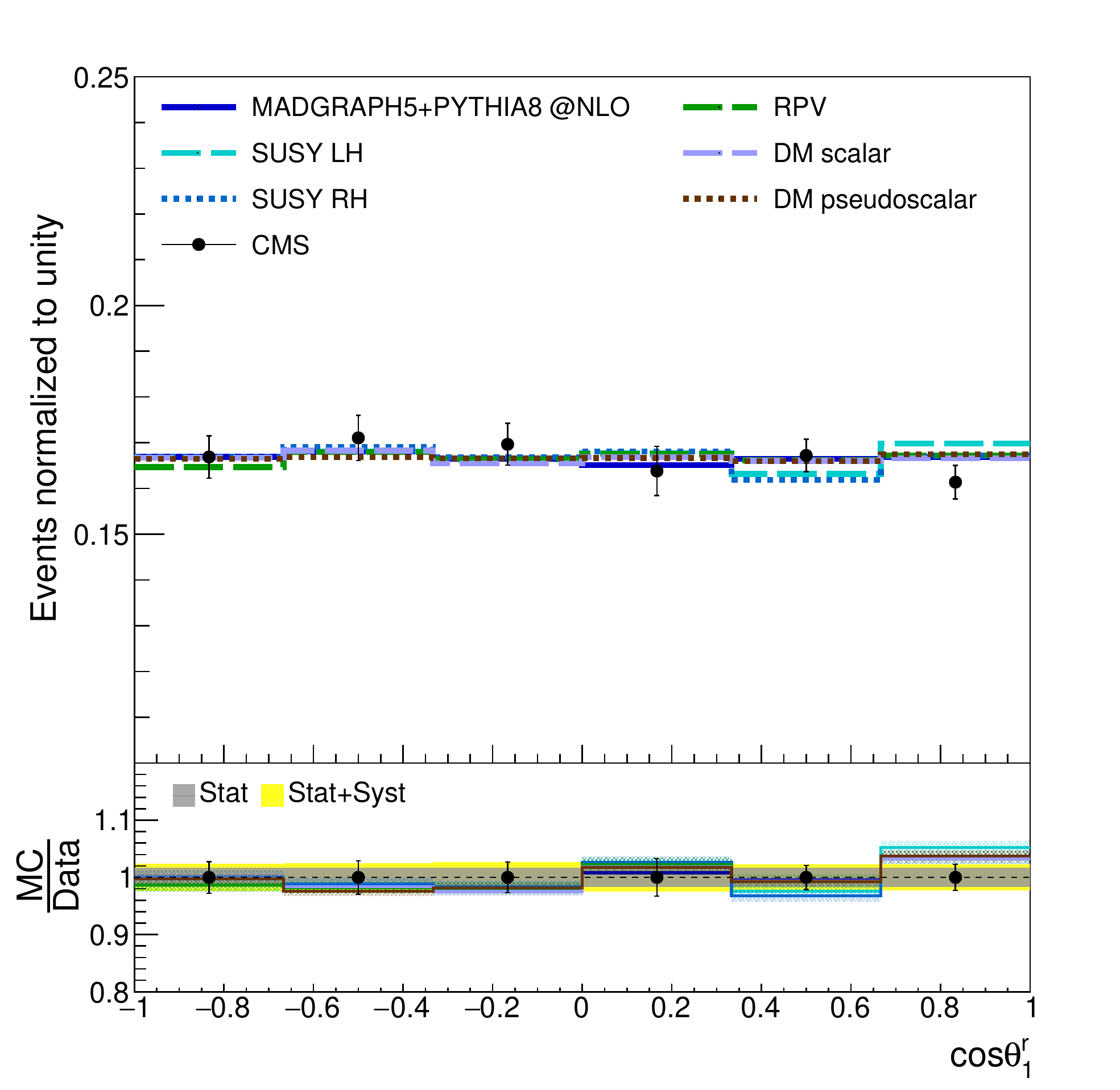}
         \hfill
         \includegraphics[width=.445\textwidth]{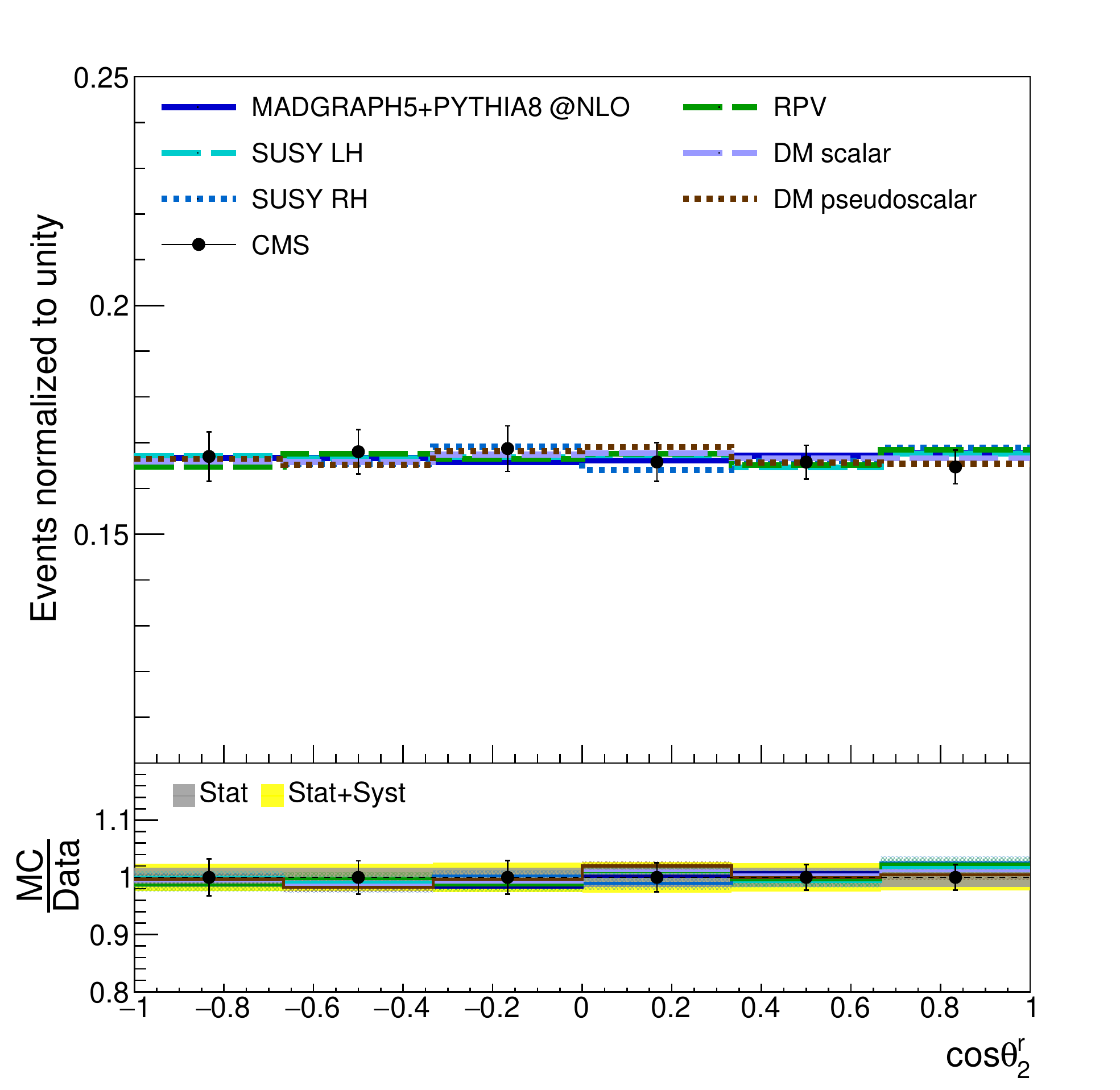}
     \hfill
         \includegraphics[width=.445\textwidth]{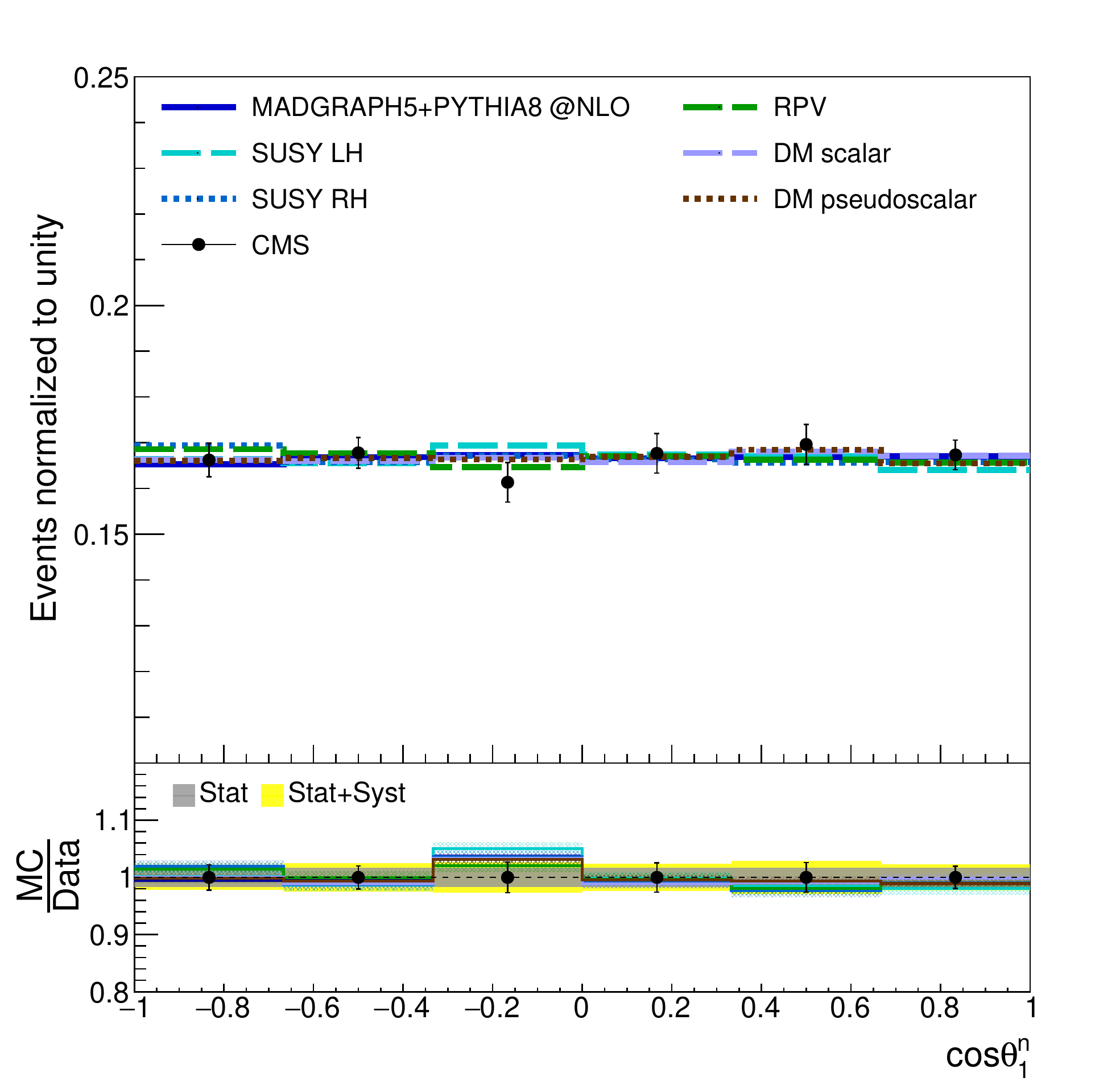}
          \hfill
         \includegraphics[width=.445\textwidth]{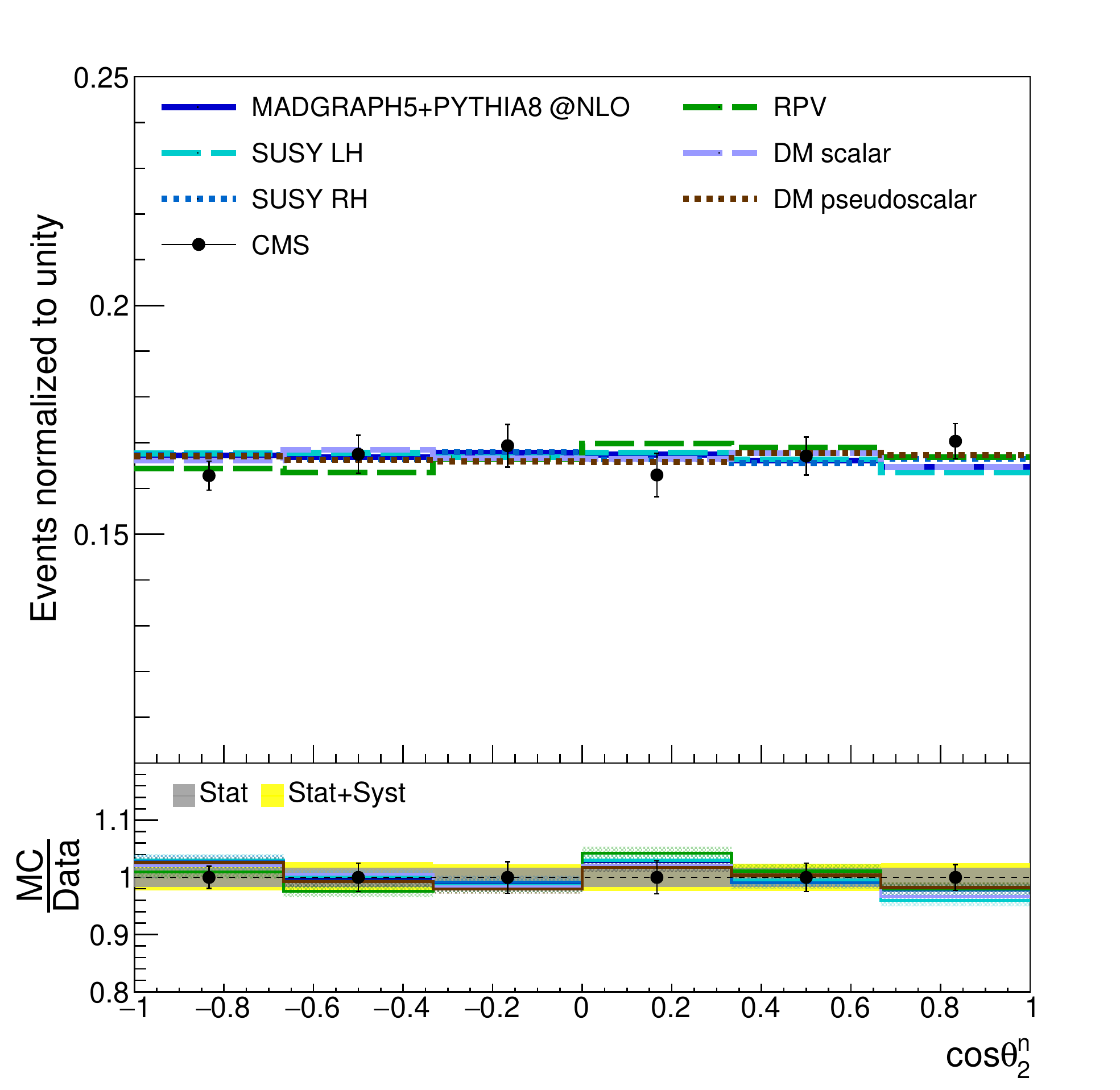}
        \caption{\label{fig:polarization_BSM} The normalized differential cross sections with respect to the observables $\cos\theta_{1}^{i}$ and $\cos\theta_{2}^{i}$, $i=k,r,n$. The ratio panels compare the BSM signals to the CMS data.}
\end{figure}

As mentioned earlier, we take into account three different BSM signal topologies to compare with SM predicted results and the data. The details of the event samples of the signals can be seen in section \ref{sec:intro}.  The SUSY samples were simulated as unpolarized.  To check importance of it, the R-Parity conserved SUSY sample is determined as two separate configurations having polarized stop quarks.  To polarize the $\tilde{t}$ as fully left- (SUSY LH) and right-handed (SUSY RH),  the distributions in the SUSY sample is multiplied by a weight obtained from the reweighting algorithm detailed in \cite{af}. The second signal is that DM model having scalar or pseudoscalar mediator associated with top quark pair.  Within this model,  the collision process has missing transverse energy $E_{T}^{miss}$ as in the model above.  For both sample, the study is focused on a narrow phase space in which mass difference between $\tilde{t}$ and $\tilde{\chi}_{1}^{0}$ is slightly bigger than top or the mediator mass is relatively small,  in this case 10 GeV.  To ensure the larger phase space in the $\tilde{t}$ mass,  the R-Parity violating (RPV) SUSY is included in the analysis framework.  The mechanism of the R-Parity violation gives us three quarks decayed from $\tilde{\chi}_{1}^{0}$.  Thus $E_{T}^{miss}$ in the $\tilde{t}$ pair production consists of only neutrinos from top quark decay chain.  The following results for RPV SUSY are performed with a signal point of $m_{\tilde{t}}=400$ GeV. Furthermore, $\tilde{t}$ masses from $300$ GeV to $1000$ GeV for RPV SUSY were analyzed and all mass points appear similarly.

\begin{figure}[t]
     \centering
         \includegraphics[width=.445\textwidth]{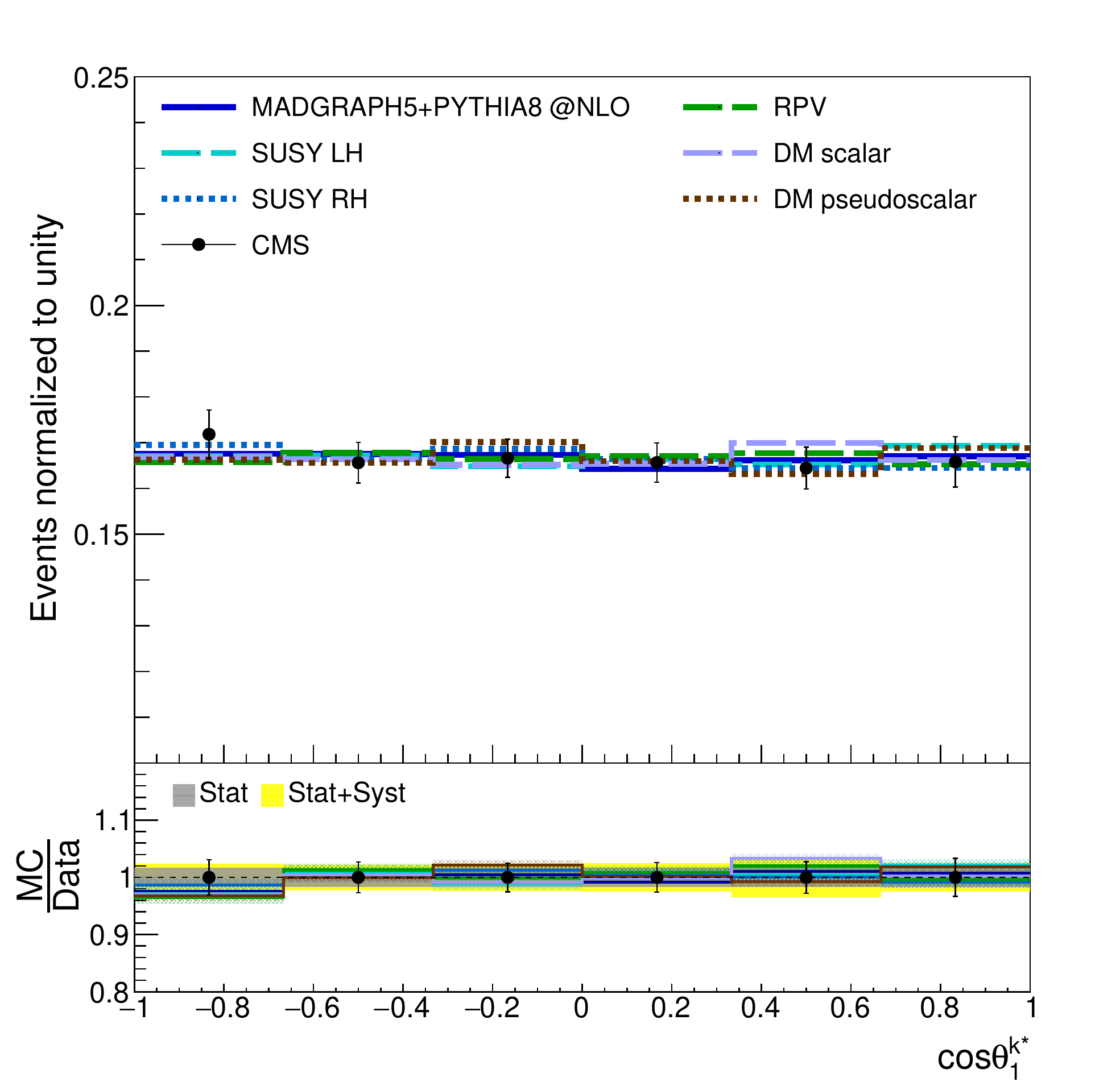}
     \hfill
         \includegraphics[width=.445\textwidth]{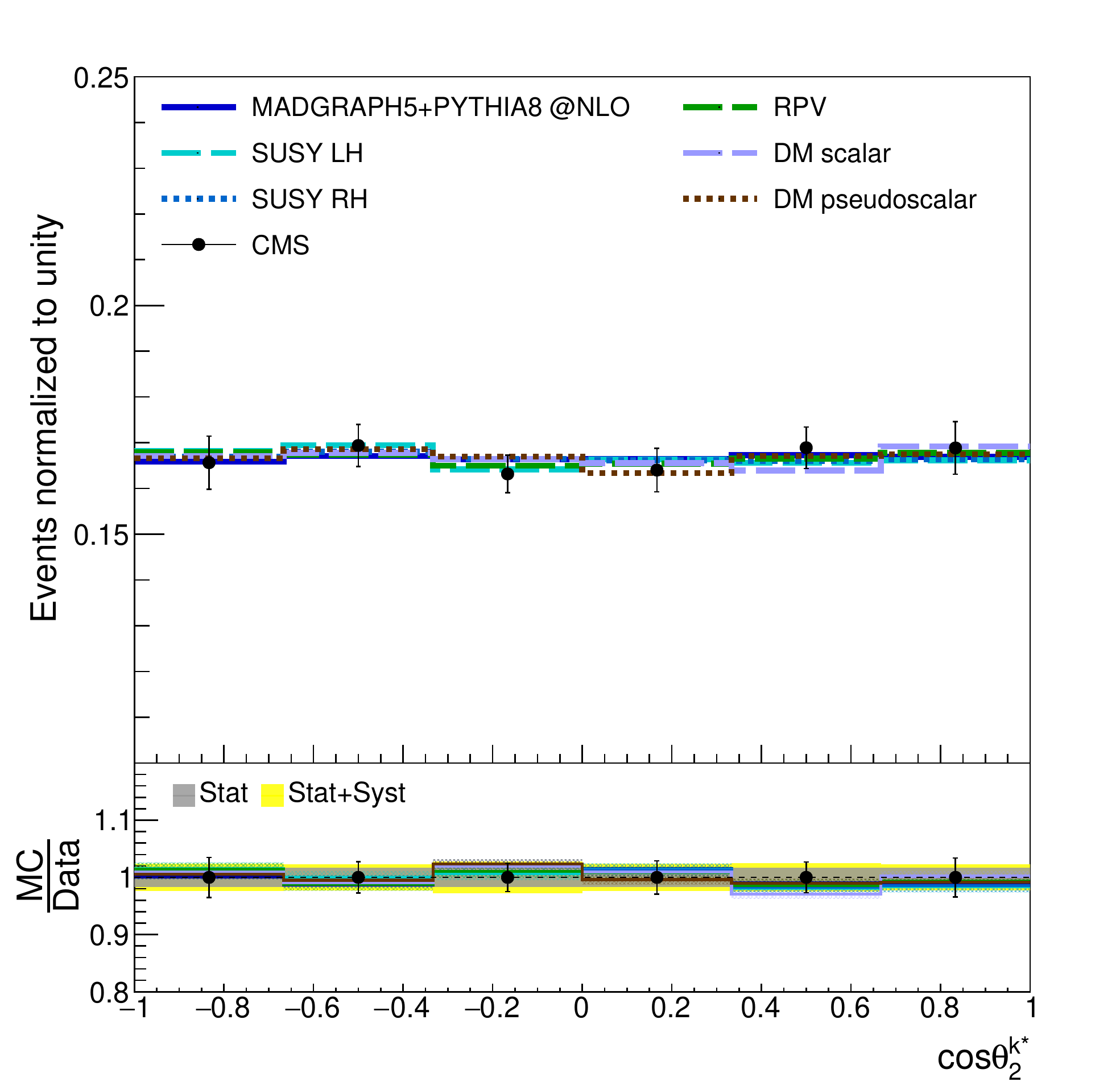}
     \hfill
         \includegraphics[width=.445\textwidth]{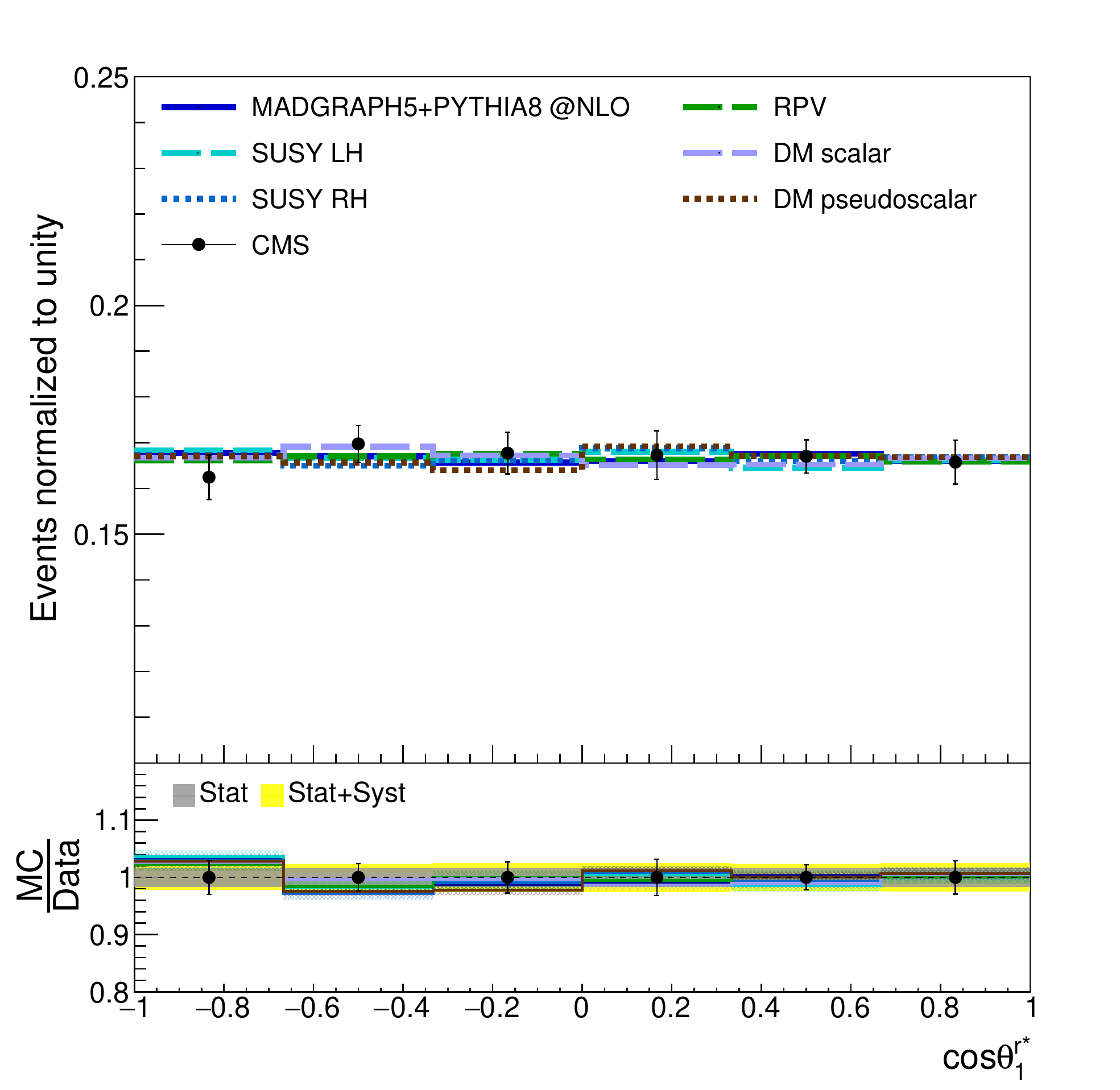}
     \hfill
         \includegraphics[width=.445\textwidth]{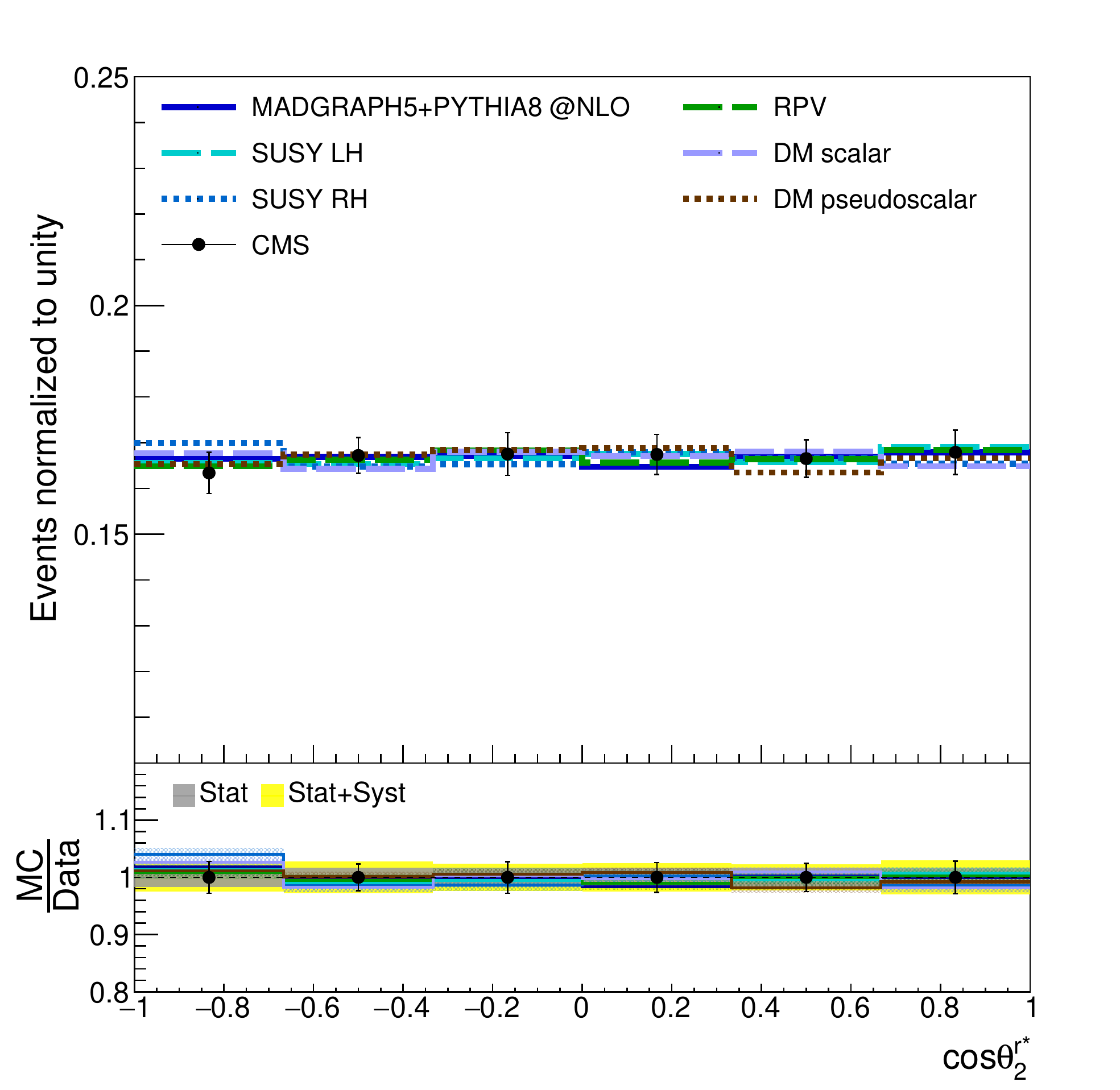}
        \caption{\label{fig:polarization_star_BSM} The normalized differential cross sections with respect to the observables $\cos\theta_{1}^{i*}$ and $\cos\theta_{2}^{i*}$, $i=k,r,n$. The ratio panels compare the BSM signals to the CMS data.}
\end{figure}

\begin{figure}[t]
     \centering
         \includegraphics[width=.445\textwidth]{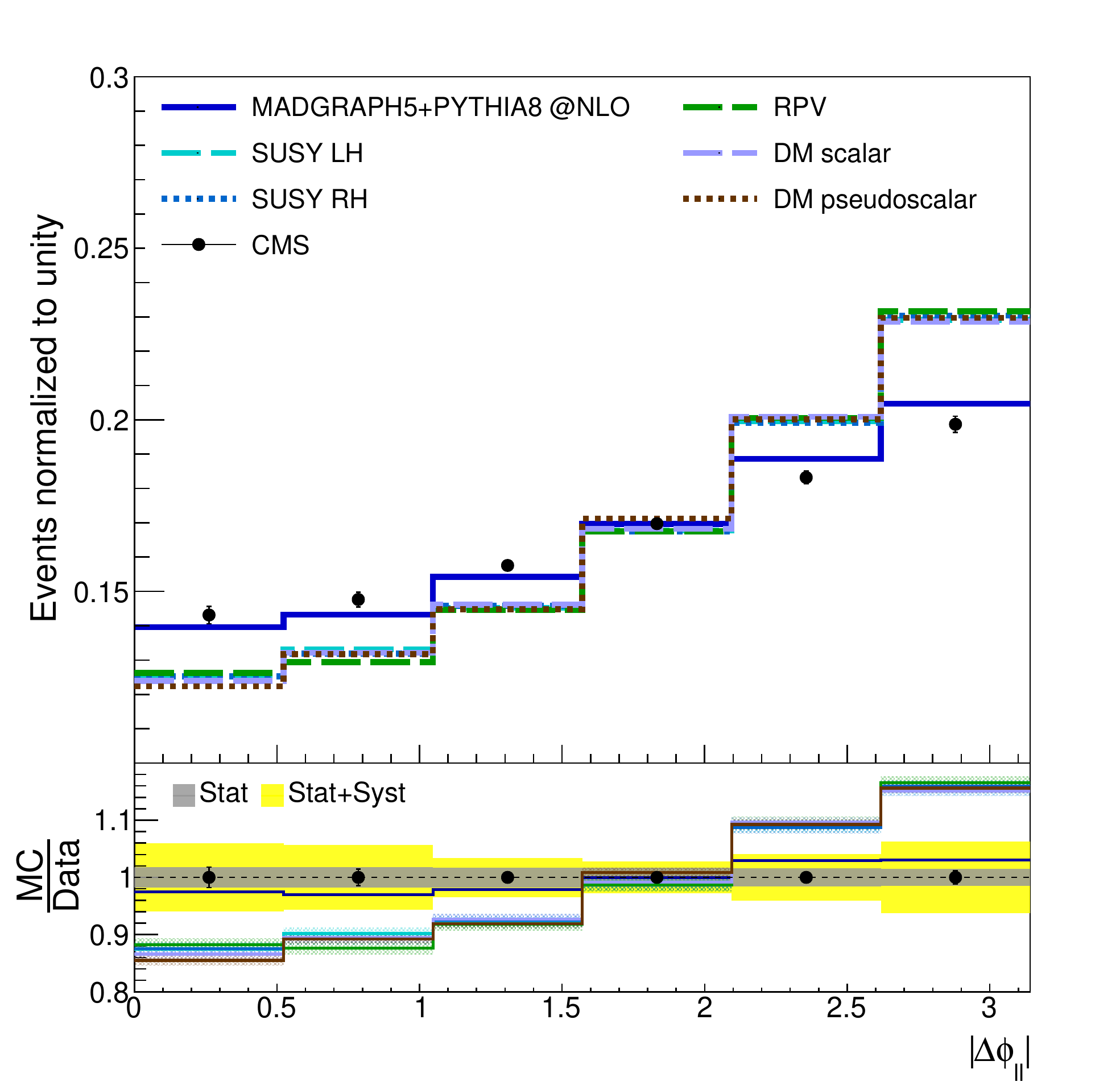}
     \hfill
         \includegraphics[width=.445\textwidth]{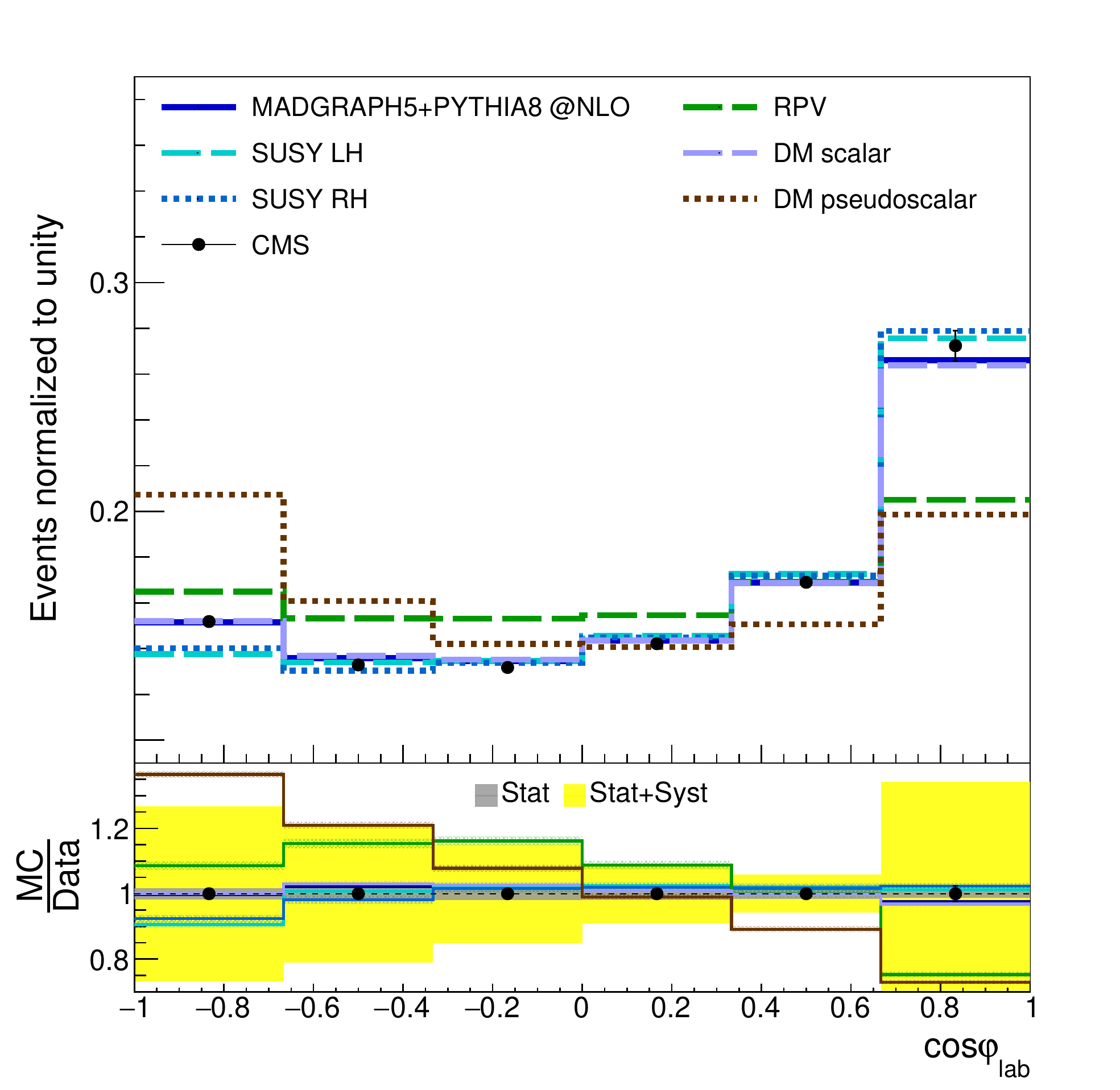}
        \caption{\label{fig:delta_phi-cos_phi_BSM} The normalized differential cross section as a function of laboratory frame observable which are $|\Delta\phi_{ll}|$ (left plot) and $\cos\varphi_{lab}$ (right plot).The ratio panels compare the BSM signals to the CMS data.}
\end{figure}

\begin{figure}[t]
    \centering
    \includegraphics[width=.445\textwidth]{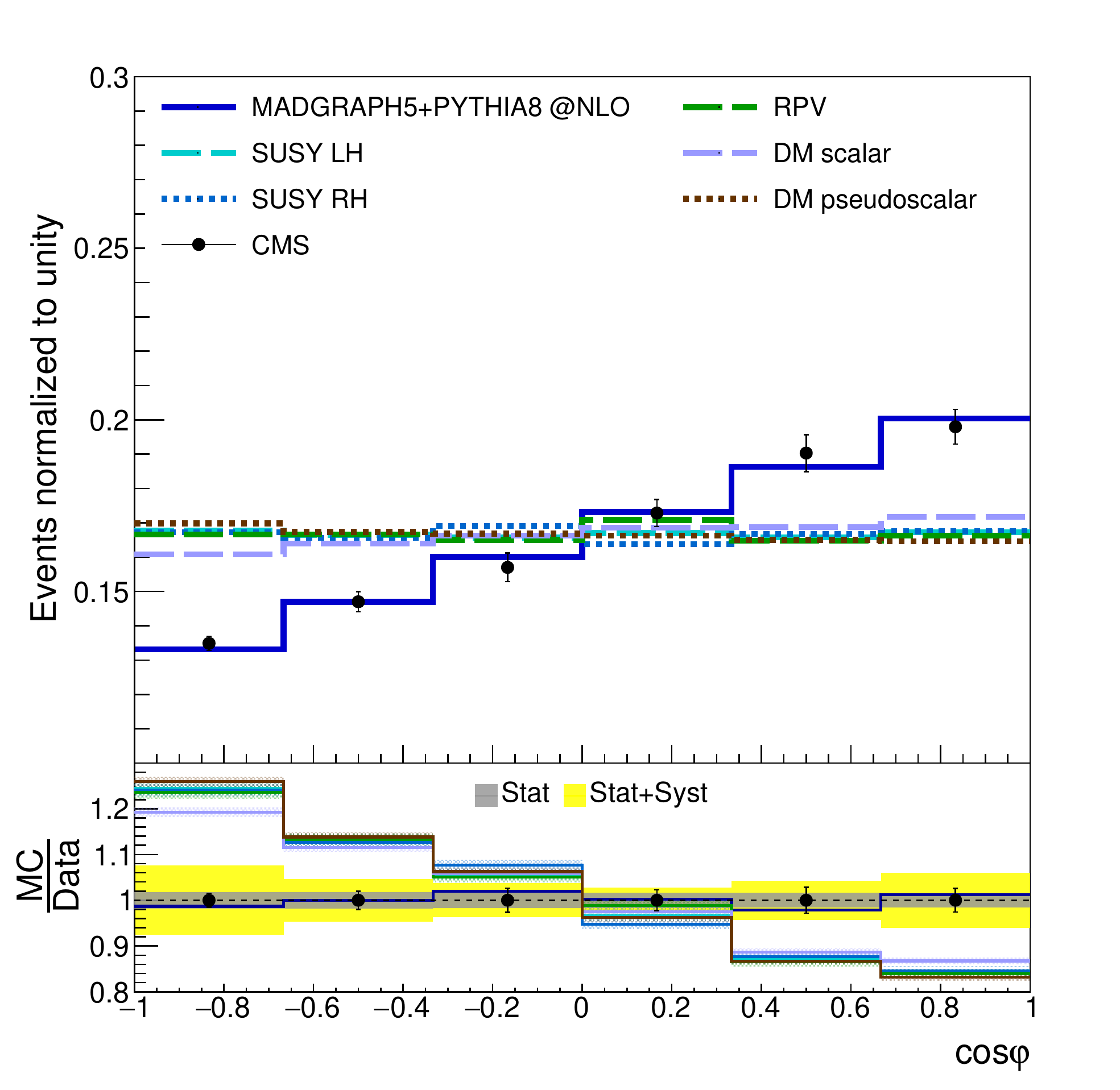}
    \caption{\label{fig:cosphi_rest_ttbar_BSM} The normalized differential cross section with respect to $\cos\varphi$. The ratio panel compares the BSM signals to the CMS data.}
\end{figure}

The distributions of spin correlation and polarization observables are presented in figures \ref{fig:diagonal_spin_BSM}-\ref{fig:cosphi_rest_ttbar_BSM}. The figures involve related BSM signals, CMS data and the nominal $t\bar{t}$ sample with systematic and statistical uncertainties calculated in the previous section. In the lower panels, the statistical uncertainties of the signals are specified on each signal sample.

The differential distributions of the diagonal spin correlation observables in BSM signals are shown in figure \ref{fig:diagonal_spin_BSM}. Almost all signals for $\cos\theta_{1}^{k}\cos\theta_{2}^{k}$ and $\cos\theta_{1}^{n}\cos\theta_{2}^{n}$ observables deviate up to $40\%$ from SM nominal $t\bar{t}$ predictions and the CMS data. For $\cos\theta_{1}^{r}\cos\theta_{2}^{r}$, the difference between signals and the nominal sample can be seen despite the domination of the uncertainties. It is seen that the scalar-mediated DM signal apart from the others has a non-zero diagonal spin correlation coefficient.

In figure \ref{fig:cross_spin_BSM}, the distributions of the observables $x_{\pm}$ defined in equation \ref{eq:10} are discussed for signals with the  nominal $t\bar{t}$ sample and the CMS data. The distribution in the upper left corner has a difference between signals and nominal sample because the coefficient of this observable is generated by P and CP conserving interactions as in the diagonal spin coefficients. For the other cross spin correlation observables, the distributions of signals and the nominal $t\bar{t}$ sample are compatible with the experimental data when considered total uncertainty.

The observables in equations (\ref{eq:7}, \ref{eq:8}) are not sensitive to small amount of top quark polarization predicted in both BSM and SM as seen in figures \ref{fig:polarization_BSM}-\ref{fig:polarization_star_BSM}. The distributions are very flat for all predictions and the data. Furthermore, polarized stops do not have any effect on the distributions as shown in fully left and right-handed separated SUSY samples.

\begin{figure}[t]
    \centering
    \includegraphics[width=\textwidth]{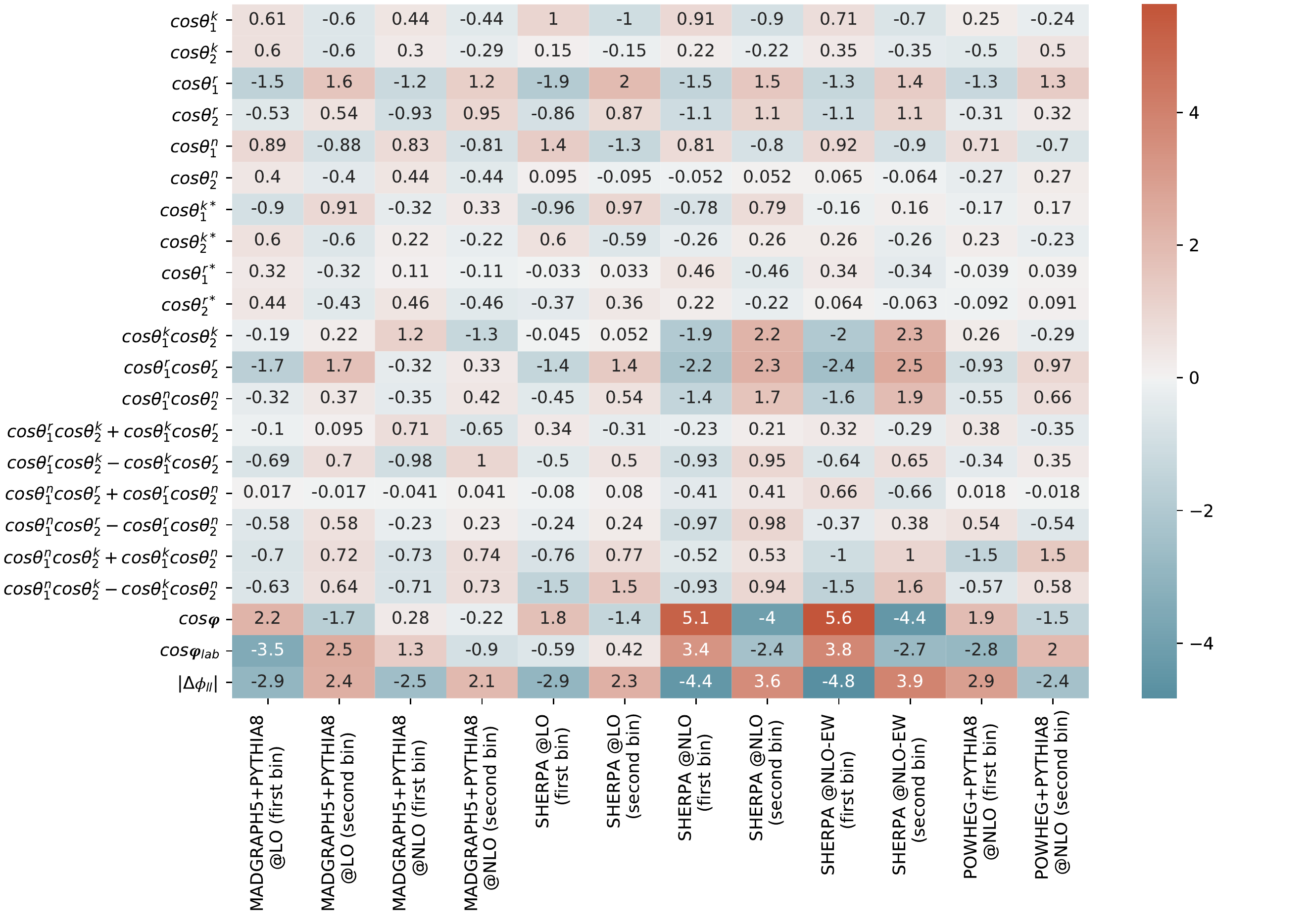}
    \caption{\label{fig:Final_SM} Deviations of the MC $t\bar{t}$ predictions from the CMS data for all observables as a matrix. Values correspond to differences in per cent between each bin of the rebinned distributions of the predictions and the data, shown in figures \ref{fig:diagonal_spin}-\ref{fig:cosphi_rest_ttbar}.}
\end{figure}

\begin{figure}[t]
    \centering
    \includegraphics[width=\textwidth]{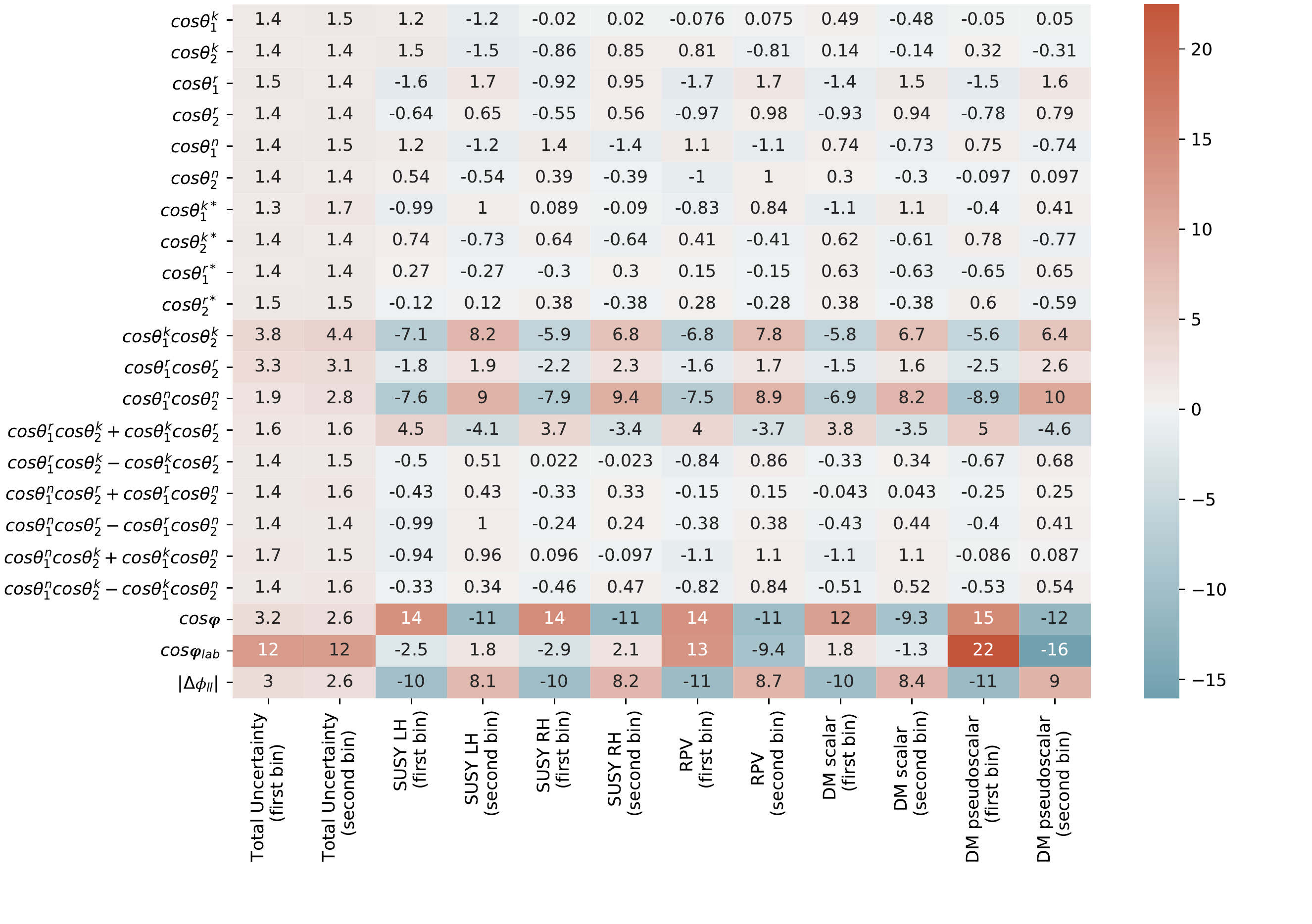}
    \caption{\label{fig:Final_BSM} Deviations of the BSM signals from the CMS data for all observables as a matrix. Values correspond to differences in per cent between each bin of the rebinned distributions of the signals and the data, shown in figures \ref{fig:diagonal_spin_BSM}-\ref{fig:cosphi_rest_ttbar_BSM}.}
\end{figure}

The laboratory frame observables are indicator of presence of spin correlations as in P- and CP-even observables. Especially the $|\Delta\phi_{ll}|$ distribution in figure \ref{fig:delta_phi-cos_phi_BSM} reveals this with the difference between BSM and SM. Moreover, it can be seen that the dark matter model with scalar mediator has spin correlation though not as much as SM. In the $\cos\varphi_{lab}$ with high systematic uncertainties, some BSM models act as in the SM while others deviate much more.

The last spin correlation sensitive observable is that the inner product of flight direction of two leptons in its own top quark center of mass frame ($\cos\varphi$). In the figure \ref{fig:cosphi_rest_ttbar_BSM}, the sensitivity to spin correlation is seen in the DM with scalar mediator while not in the other BSM samples.

To get a general picture, all distributions of various MC event generator configurations, BSM signals and the CMS data for all observables presented in figures \ref{fig:diagonal_spin}-\ref{fig:cosphi_rest_ttbar_BSM} are summarized as heat maps in figures \ref{fig:Final_SM}-\ref{fig:Final_BSM}. In these figures, each normalized distribution of each configuration is used as reduced bins from $6$ to $2$ to observe the degree of deviation of the distributions. Values on each cell in figures \ref{fig:Final_SM}-\ref{fig:Final_BSM} indicate the percentage difference between rebinned normalized distributions of $t\bar{t}$ predictions or BSM signals and of the data. In figure \ref{fig:Final_SM}, it can be explicitly seen that all MC event generator configurations explain the data, within total uncertainty, for distributions of the observables extracted from the spin density matrix. However, $\cos\varphi$ and laboratory-frame observables ($|\Delta\phi_{ll}|$ and $\cos\varphi_{lab}$) show deviations of up to $5.6\%$ in NLO QCD accuracy in \textsc{Sherpa}, as well as the inclusion of EW correction. The positive or negative deviation (in percentage) of rebinned normalized distributions of the BSM signals from the data is shown in figure \ref{fig:Final_BSM}. The first two columns indicate, for each bin, absolute total uncertainty (statistical and systematic sum) calculated in each of the above distributions. This table is a beneficial tool to see the degree of deviation of the BSM signals from the experimental data in a particular region of each observable. With this sense, it is clearly seen that the $\cos\theta_{1}^{k}\cos\theta_{2}^{k}$, $\cos\theta_{1}^{n}\cos\theta_{2}^{n}$ and $\cos\theta_{1}^{r}\cos\theta_{2}^{k}+\cos\theta_{1}^{k}\cos\theta_{2}^{r}$ observables as elements of the spin density matrix are sensitive to the considered BSM signals. In addition, all BSM signals for $\cos\varphi$ and $|\Delta\phi_{ll}|$ seem to differ significantly from the experiment. Finally, it appears that the left or right-handed polarization of the stop quarks does not cause any notable change for these observables, whereas the DM signal with the scalar mediator has more spin correlation between the top quarks compared to the pseudoscalar mediator.

\section{Conclusions and Outlook}
\label{sec:5}

In this work, we have made for the first time a comprehensive analysis of the spin correlation and polarization observable set and the direct angle-related observables between leptons in the $t\bar{t}$ production with dileptonic final states at $13$ TeV. This has provided us to investigate top quark pairs advanced properties for proton-proton collisions by using various MC event generator methods at LO, NLO QCD accuracy and NLO QCD with EW corrections in detail. For the distributions of the set of observables at the full phase space, we have compared our results with the unfolded data performed by CMS collaboration. Our study is mainly composed of two parts: the first is to compare different MC configurations with the experimental data and determine their effects on each variable, and the second is to find the degree of deviation of $\tilde{t}\tilde{t}$ and DM signals from the data for each spin correlation observable. For a clear view of deviations from the data, we have created two heat map tables of SM $t\bar{t}$ and BSM samples with the set of observables, in figures \ref{fig:Final_SM}-\ref{fig:Final_BSM}.

Considering the different theoretical approaches, the polarization observables (including terms with modified axes) and the cross terms of the spin correlation observables are in agreement within the systematic and statistical uncertainty (around the $1.5\%$). Especially samples of \textsc{Sherpa} at NLO accuracy (and with electroweak corrections) for diagonal terms in the spin correlation observables tend to have more deviation (from $1.4\%$ to $2.5\%$) compared to the others (almost all under $1\%$) but they are still in total uncertainty. The biggest contribution to the uncertainty comes from the PDF variations of predictions at LO accuracy. Separation might be seen better in NLO calculations due to the less dependency of higher order calculations to the PDF variations. Furthermore, to be more accurate in predictions, it is needed to increase the statistics in experimental data, and upcoming LHC runs could provide this. The behavior of the different configurations can be clearly seen in the $\cos\varphi$ and $|\Delta\phi_{ll}|$ distributions, the observables most sensitive to spin correlation in the top quark pair. The deviations up to $5.6\%$ in \textsc{Sherpa} samples (NLO and NLO-EW) for these observables, in figure \ref{fig:Final_SM}, could be arisen from its approximation in hard and large angle emissions.

Figure \ref{fig:Final_BSM} provides a complete map of the BSM signals according to the experiment, in the phase space of the related observables. Polarization and cross spin correlation observables (except $\cos\theta_{1}^{r}\cos\theta_{2}^{k}+\cos\theta_{1}^{k}\cos\theta_{2}^{r}$) are compatible for the BSM signals and the data, while the signals have good separation for P- and CP-even observables (up to $10\%$). Only the DM sample with a scalar mediator has spin correlation, unlike other BSM signals, but the spin correlation power in its top-quark pair is not as much as in the SM. Furthermore, polarized stop quarks of different composition have no impact observed on top quark spin correlation.

The different MC event generator configurations are compatible with each other, within total uncertainty, for the observables of spin correlation and polarization in the top quark pair. However, for some observables as mentioned above, there are small separations dominated by systematic uncertainties. Next-to-next-to leading order (NNLO) calculations can make these differences more visible by reducing systematic uncertainties, or they can reduce the difference between configurations by making more accurate calculations with higher-order ME calculations. Thereby, this gets NNLO calculations for theoretical predictions more important. Additionally, the observables may be used, especially in low missing transverse energy region, for searching BSM signals due to their sensitivity of contributions from parity and/or CP conserving or violating interactions.


\acknowledgments

We thank Alexander Josef Grohsjean for discussions on spin correlation analysis, and Gurpreet Singh Chahal for contributions about Sherpa productions. This work is supported by Istanbul Technical University Research Fund under grant number TGA-2020-42345 and TUBITAK grant 121F065.


\end{document}